\documentclass[aps,twocolumn,nofootinbib,secnumarabic,groupedaddress,showpacs,floatfix]{revtex4}

\usepackage{amssymb,amsmath}
\usepackage{graphicx}
\usepackage{color}
\usepackage[usenames,dvipsnames]{xcolor}
\usepackage[colorlinks=true,linkcolor=Maroon,citecolor=OliveGreen,urlcolor=Blue,linktoc=page]{hyperref}
\usepackage{doi}
\usepackage{enumerate}
\usepackage[normalem]{ulem}

\newcommand{\rs}{\rm\scriptscriptstyle}
\newcommand{\s}{\scriptscriptstyle}
\newcommand{\drangle}{{\rangle \hspace{-0.69 mm} \rangle}}
\newcommand{\dlangle}{{\langle \hspace{-0.69 mm} \langle}}
\newcommand{\pdag}{{\phantom\dag}}
\newcommand{\pstar}{{\phantom*}}

\begin{document}

\title{
	Scattering matrix approach to the description of quantum electron 
	transport\footnote{
		The Russian version of this review is available online,
		\href{http://dx.doi.org/10.3367/UFNr.0181.201110b.1041}{\uline{UFN {\bf 181} 1041--1096 (2011)}}. 
		Minor misprints have been corrected. Error in Eq.~(79) in the Russian 
		version [Eq.~(\ref{eq:Hp2DLL2}) in the present text] has been fixed.
	}
}

\author{
	G.B.~Lesovik,$^{\rm a}$
	I.A.~Sadovskyy$^{\rm b}$
}

\affiliation{
	$^{a}$Landau Institute for Theoretical Physics, RAS, 
	prosp. Akademika Semenova 1-A, 142432 Chernogolovka, Moscow region, Russia
}

\affiliation{
	$^{b}$Materials Science Division, Argonne National Laboratory, 
	9700 S. Cass Avenue, Argonne, Illinois 60637, USA
}

\date{\today}

\begin{abstract}
We consider the scattering matrix approach to quantum electron transport in meso- and nano-conductors. This approach is an alternative to the more conventional kinetic equation and Green's function approaches, and often is more efficient for coherent conductors (especially for proving general relations) and typically more transparent. We provide a description of both time-averaged quantities (for example, current-voltage characteristics) and current fluctuations in time~--- noise, as well as full counting statistics of charge transport in a finite time. In addition to normal conductors, we consider contacts with superconductors and Josephson junctions.
\end{abstract}


\maketitle

\tableofcontents


\section{Introduction}

Over the past 30 years, research of electrical conductors has evolved from considering macroscopic objects to the study of mesoscopic objects\footnote{That is, objects with properties intermediate between microscopic and macroscopic. {\it Mesoscopic} translated from Greek means {\it intermediate scopic} or {\it mean scopic}.} and, finally, to nanophysics objects. While in macroscopic objects the quantum nature is mainly manifested at the level of band structure formation, the mesoscopic objects are larger than the atomic objects but smaller than the characteristic length of quantum correlations. Lastly, nanophysics operates on an even smaller scale, down to the atomic one, and incorporates quantum contacts and quantum dots, molecular and atomic contacts, carbon nanotubes, graphene, etc.

Electronic transport in conductors of a size comparable to inelastic scattering length, such as the energy relaxation length or dephasing length, or even the Fermi wavelength, exhibits a number of specific features, the most important of which is a considerable nonlocality of the transport phenomena. For such conductors, there is no reason in considering such quantities as local conductivity, while the problem to be addressed is the transportation of electrons from point A (left reservoir) to point B (right reservoir). In this case, the electron transfer through a conductor is a purely quantum mechanical process. This process can be best described by means of a well known approach used in the scattering theory of particles and atoms, where given are an initial state (in our case, an electronic state), a scatterer, and a final state (in the reservoir where the electron arrives) and where the transition from one state to another is described by the scattering matrix.

Presently, the scattering matrix approach is widely and successfully applied in the quantum transport study. The main difference between this approach and more conventional methods based, for example, on the kinetic equation, the Kubo formula, Green's functions, or diagram techniques, can be put in this way. The total conductivity (or the total current) of a system is expressed in terms of the conductor's quantum mechanical transparency, generally expressed in terms of the scattering matrix, and the occupation numbers of the exact electronic scattering states, which are determined by the parameters at the boundaries (reservoirs).

At first glance, such a method for describing the electronic transport just replaces the problem of finding the local or nonlocal conductivity with the calculation of transmission, which is equally complicated. But in fact the situation is somewhat different. First, in many cases involving a simple sample geometry and simple scattering potential, transmissions can be calculated analytically, which is easier and more instructive than, e.g., calculating a Green's function. Second, it is often possible to make a reasonable assumption regarding the scattering matrix and facilitate an acceptable description of the experiment. For disordered (dirty) conductors with a complex scattering potential, the transmission probabilities can be efficiently described statistically, for example, by methods developed for random matrices.

In addition, due to the development of mesoscopics and nanophysics, new problems emerged, which either had not attracted proper attention earlier or seemed unrealistic for systems under study. One of these problems is the description of the current beyond its average value, namely, calculation of the current fluctuations and presentation of the full counting statistics in quantum meso- and nanoconductors. It was found that these particular problems could be efficiently solved by the scattering matrix method. It is important that even if the scattering matrix is unknown, i.e., has not been calculated for a particular scattering potential, the full counting statistics for large time intervals can be formally obtained, including the average current. Thus, if the scattering matrix is given then not only the conductance $G = 1/\mathcal{R}$, where $\mathcal{R}$ is the resistance, can be calculated but also the spectral density of current fluctuations $S(\omega)$ at low frequencies, and the distribution function $P(Q)$ of the charge transferred within a certain fixed time interval can be found. Besides, it proves possible to derive general relations like, for example, the fluctuation-dissipation theorem, relating the average current and nonequilibrium fluctuations. The conventional approach would require repeated calculation of $S$ and $P$ and other quantities different from the average current.


\section{Scattering matrix approach to the description of transport: Landauer formula}
\label{sec:landauer}

The scattering matrix that transforms asymptotically free incoming states into the asymptotically free outgoing states thus describing interactions with an obstacle and between particles plays an outstanding role in quantum physics. This matrix was first introduced by Born~\cite{Born26} and then by Wheeler~\cite{Wheeler37} and independently by Heisenberg~\cite{Heisenberg43a,Heisenberg43b} to describe the scattering of particles and atoms and has been extensively used since the late 20th century in the theory of electron transport in quantum conductors.

The best known result in the theory of quantum transport obtained using the scattering matrix approach is the famous Landauer formula,\footnote{Landauer~\cite{Landauer57} was the first who used scattering matrices to describe transport problems.} which is also called the Landauer-B\"uttiker formula. In fact, this formula in its conventional form first appeared in Refs.~\cite{Anderson80,Fisher81,Economou81}. The conductance of a quasi-one-dimensional (one-channel) conductor is given by the conductance quantum $G_0 = 2e^2/h$ (where $e$ is the electron charge, $h$ is Planck's constant, and the factor 2 appears due to the spin degeneracy), known from the quantum Hall effect~\cite{Klitzing80}, times the transparency $T$ of the conduction channel. In the case of several channels, the expression for the conductance
\begin{equation}
	G =\frac{2e^2}{h}\sum_{n,m} T_{nm}
	\label{eq:GN3}
\end{equation}
contains the sum of transmission probabilities $T_{nm}$ from one mode (channel) to another (see details in Sec.~\ref{sec:waveguide}).

The Landauer approach was better understood in subsequent papers, for example, Imry~\cite{Imry86} pointed out the role of a voltage drop at the input to the conductor. Later it was extended to more complicated systems with many reservoirs~\cite{Buttiker85}, the quantum Hall effect regime~\cite{Buttiker86,Buttiker88a,Buttiker88b,Buttiker88c}, hybrid superconducting systems~\cite{Takane91,Takane92,Lambert91,Lambert93,Anantram96,Beenakker91}, and was also used to describe current fluctuations in time~\cite{Lesovik89b,Buttiker90,Martin92,Levitov93}. Currently, this method has become very clear and functional. As a whole, this approach can be applied to the description of coherent mesoscopic conductors in which the characteristic size $L$ of the voltage drop region is much smaller than all inelastic lengths.

\subsection{Conductance of one-dimensional contact}
\label{sec:cond1D}

To describe a quasi-one-dimensional coherent conductor, we first consider a purely one-dimensional problem\footnote{It is this problem that Landauer initially considered in Ref.~\cite{Landauer57}. The problem was solved by using an impressively small amount of knowledge: information on the setup and solution of scattering problems in the one-dimensional case in quantum mechanics and basic concepts about the degenerate electron gas at the general physics level.} for a system in which electron reservoirs are located to the left and to the right, far away from an obstacle (scatterer) located at the center, and emit electrons in the direction of this obstacle.

Let us assume that electrons with energies up to $\mu$ move from the left reservoir to the scatterer (we forget about spin for a while). Experimentally, this may correspond to the presence of the bias voltage $V = \mu/e$. Such states are called the Lippmann-Schwinger scattering states~\cite{Lippmann50}.

One of the problems for such states in the continuous spectrum is to count their density. It can be solved by using the so-called ``box normalization'' method. This normalization method imposes the periodic boundary conditions by closing the conductor into a circle with length $L$ to make the spectrum discrete. Thereafter, in the limit $L \to \infty$ we are back to the continuous spectrum.\footnote{This method is now out of date and replaced by the equivalent method of normalization to the $\delta$-function.} But it is difficult to rigorously perform this procedure for scattering states, and here we solve this problem in a different manner, by forming normalized wave packets from continuous-spectrum states.

By dividing the energy interval $[0,\mu]$ into $N$ segments with size $\Delta = \mu/N$, we form the wave packets
\begin{equation}
	\Psi_n (x,t) = c_n \int\limits_{(n-1)\Delta}^{n\Delta}\!\!\!
	dE \, \Psi_{{\rs L}, \s E}(x) e^{-i E t/\hbar},
	\label{eq:wavePacket}
\end{equation}
where $n = 1,\ldots,N$ and $\Psi_{{\rs L}, \s E}(x)$ is the left Lippmann-Schwinger scattering state with energy $E$, having the asymptotic form
\begin{equation}
	\Psi_{{\rs L}, \s E}(x) =
	\begin{cases}
		e^{i k x} + r(E) e^{-i k x},	& x \to -\infty, \\
		t(E) e^{i k x},			& x \to \infty.
	\end{cases}
	\label{eq:scatterL}
\end{equation}
The normalization constant can be found from the relation
\begin{equation}
	\int dx \, \Psi^*_{{\rs L},\s E'}(x) \Psi^\pstar_{{\rs L}, \s E}(x) =
	2\pi \delta(k'-k),
	\label{eq:norm}
\end{equation}
where $k = \sqrt{2m E}/\hbar$. From correct normalization of wave packets $\int\!dx\,|\Psi_n(x,t)|^2 = 1$, we obtain
\begin{equation}
	c_n = \frac{1}{\sqrt{h \Delta v_n}},
	\label{eq:normalization}
\end{equation}
where $v_n = \sqrt{2n\Delta/m}$ is the velocity of the $n$th packet and $\Delta$ is assumed to be small.

The wave packets described by expressions~(\ref{eq:wavePacket}) are localized in the vicinity of $x = 0$ at $t = 0$ and have the characteristic size $h v_n / \Delta$. These packets move with the velocity $v_n$. As $\Delta \to 0$ (i.e., $N \to \infty$), the wave packets become broader, their shape approaching the shape of scattering states~(\ref{eq:scatterL}).

We now calculate the current $I$ carried by a given orthonormalized set of wave packets. The current for them is additive since, according to Pauli's principle, only one electron can occupy each state. Thus, we can first calculate the contribution 
\begin{equation}
	I_n = -i \frac{e\hbar}{2m} \Bigl[
	\Psi_n (x)^* \Psi_n '(x) - \Psi_n '(x)^*
	\Psi_n (x) \Bigr]
	\label{eq:current_n1}
\end{equation}
to the current from each $n$th packet and then sum up the contributions.
Due to the charge conservation law,
for scattering states (as for any stationary states), the current is independent of the point at which we calculate it. Hence in the limit $\Delta \to 0$, the contribution to the current from each packet at $t = 0$, can be calculated, for example, to the right of the barrier, where the wave function has the known form $\Psi_{\rs L}(x) = t(E) e^{ikx}$. This gives
\begin{equation}
	I_n = c_n^2 \Delta^2 e v_n T(n \Delta)
	= \frac{e}{h} \Delta T(n \Delta),
	\label{eq:current_n2}
\end{equation}
where $T(E) = |t(E)|^2$ is the transparency at the energy~$E$. Summing the contributions of all packets, we find
\begin{equation}
	I = \sum_{n=1}^N I_n = \frac{e}{h} \Delta \sum_{n=1}^N T(n \Delta)
	\stackrel{(\Delta\to 0)}\rightarrow
	\frac{e}{h} \int\limits_{0}^{\mu} dE \, T(E),
	\label{eq:totalCurrent}
\end{equation}
where the sum over $n$ transforms to the integral over $\mu$ in the limit $\Delta \to 0$. The conductance, defined as the ratio of the current $I$ to the voltage $V = \mu / e$, can be written in the form
\begin{equation}
	G = \frac{I}{V} = \frac{e^2}{h}
	\int\limits_0^\mu \frac{dE}{\mu} T(E).
	\label{eq:conduct}
\end{equation}
Expression~(\ref{eq:conduct}) is a simple variant of the Landauer formula for the conductance~\cite{Landauer70,Fisher81}.

Since the continuous spectrum states cannot be normalized in the usual way, like the discrete spectrum states, it is not clear in advance what current is carried by each many-particle state constructed from the arbitrary states of the continuous spectrum. This problem can be solved by considering the wave packets and passing to the limit as we did above. Such a procedure can be used in an explicit form to analyze intricate problems, for example, to describe the full transport statistics, as was done in Ref.~\cite{Hassler08}. The current can be calculated using the rule (which can also be derived by the method outlined above) allowing the summation of the contributions to the current from continuous-spectrum states: if $\psi_\xi(x)$ satisfies a normalization condition generalizing~(\ref{eq:norm}),
\begin{equation}
	\int\!dx \, \psi_\xi^{\ast} (x) \psi_{\xi'}(x) = c(\xi) \delta(\xi-\xi'),
	\label{eq:generalNormalize}
\end{equation}
then the mean of the current operator is given by
\begin{equation}
	I = \int \frac{d\xi}{c(\xi)} n(\xi) I_\xi,
	\label{eq:generalCurrent}
\end{equation}
where $I_\xi$ is the current from the particle in the state $\psi_\xi(x)$ and $n(\xi)$ is the occupation number, equal to 1 if the state with the subscript $\xi$ is present in the many-particle wave function (Slater determinant) and to 0 otherwise (at finite temperatures $\Theta>0$, the number $n(\xi)$ can take values between 0 and 1). In our case, we can choose $\xi = k$, $I_k = -e \hbar k T(E) / m$, $c(k) = 2 \pi$, and
\begin{equation}
	n(k) = \begin{cases}
		1, & \hbar^2 k^2/2m < \mu, \\
		0, & \hbar^2 k^2/2m > \mu.
	\end{cases}
	\nonumber
\end{equation}
Substituting these expressions in Eq.~(\ref{eq:generalCurrent}), we obtain
\begin{equation}
	I = \frac{e\hbar}{m} \int\limits_0^{k(\mu)} \! \frac{dk}{2\pi} k \, T(E)
	= \frac{e}{h} \int\limits_0^{\mu} dE \, T(E),
	\label{eq:currentB}
\end{equation}
which coincides with Eq.~(\ref{eq:totalCurrent}). At the last calculation step, we switched from integration over the wave vector $k$ to integration over energy $E$, using the one-dimensional density of states
\begin{equation}
	\nu(E) = \frac{dk}{dE} = \frac{m}{\hbar^2 k}.
	\label{eq:density}
\end{equation}
This leads to the cancellation of the factor $k$ in the integrand in Eq.~(\ref{eq:currentB}). This implies that in the absence of scattering each energy interval carries the same current (per spin)
\begin{equation}
	i_0=\frac{\delta I}{\delta E} = \frac{e}{h},
	\label{eq:currentperchannel}
\end{equation}
which is a characteristic feature of the one-dimensional ballistic transport.

\subsection{Two reservoirs}
\label{sec:2res}

\begin{figure*}[tb]
	\includegraphics[width=9.5cm]{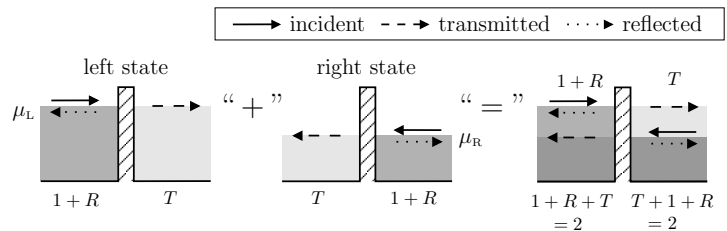}
	\caption{
Charge density caused by the left and right scattering states (we omit the details of Friedel oscillations and perform averaging over several wavelengths). Charge density is shown by image gradation: the darker shade of grey corresponds to the more populated levels while the white color indicates empty states. The total densities of states with energies smaller than $\mu_{\rs R}$ are equal. For the states with energies between $\mu_{\rs R}$ and $\mu_{\rs L}$, the charge to the left of the scatterer is greater (for $T < 1$).
	}
	\label{fig:Landauer1}
\end{figure*}

In Sec.~\ref{sec:cond1D} we discussed the case of spinless electrons emitted by one reservoir. We now consider the more realistic case where spin-$1/2$ electrons are emitted by both reservoirs. We assume that the left reservoir with the electrochemical potential\footnote{We recall that the electrochemical potential is the maximum total energy of one electron at zero temperature, which is the sum of the kinetic (Fermi) energy and the potential energy of a charge in the electrostatic potential.} $\mu_{\rs L}$ emits the ``left'' scattering states $\Psi_{\rs L}(x)$ and the right reservoir with the electrochemical potential $\mu_{\rs R}$ emits the ``right'' scattering states $\Psi_{\rs R}(x)$, see Fig.~\ref{fig:Landauer1}. Then the total current is determined by contributions from both reservoirs:
\begin{equation}
	I_{\rs L} = \frac{2e}{h} \int\limits_0^{\mu_{\rs L}} dE \, T(E)
	\label{eq:currentL}
\end{equation}
and
\begin{equation}
	I_{\rs R} = - \frac{2e}{h} \int\limits_0^{\mu_{\rs R}} dE \, T(E),
	\label{eq:currentR}
\end{equation}
where the factor 2 appears due to the spin degeneracy, and in contrast to $I_{\rs L}$, the current $I_{\rs R}$ determined by the right states acquires a minus sign because the wave vector and velocity for $\Psi_{\rs R}(x)$ are opposite to those for $\Psi_{\rs L}(x)$. Here, we used the important property of the scattering matrix following from its unitarity and symmetry under time reversal, namely, that the transmission probabilities for mutually inverse processes are equal.\footnote{In the one-dimensional case, the equality of the transmission probabilities follows from the unitarity, even in the absence of the time reversal invariance.\label{eq:note:prob}} In our case, the transmission probability from left to right, $T=|t|^2$, is equal to the transmission probability from right to left, $T=T'=|t'|^2$. In the total current
\begin{equation}
	I = I_{\rs L} + I_{\rs R} =
	\frac{2e}{h} \int\limits_{\mu_{\rs R}}^{\mu_{\rs L}} dE \, T(E)
	\label{eq:totalCurrentB}
\end{equation}
the contributions from energy intervals filled both on the left and on the right cancel while the states filled only in one reservoir make a contribution to the total current.

\subsection{Landauer voltage drop}

\begin{figure}[b]
	\includegraphics[width=6.7cm]{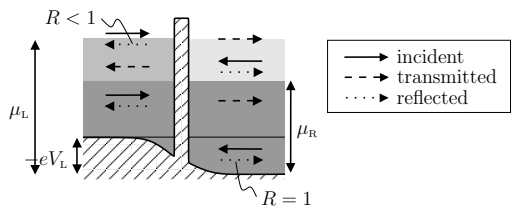}
	\caption{
Occurrence of the Landauer voltage drop $V_{\rs L}$ on a barrier. Because of the bending of the conduction band bottom caused by the voltage drop, the states emitted from the right reservoir with energies between 0 and $eV_{\rs L}$ are completely reflected. Here, $eV_{\rs L}>0$, which corresponds to the negative voltage and electric current, but to the positive flow of particles (from left to right). This difference in signs appears because the electron charge is standardly assumed negative.
	}
	\label{fig:Landauer2}
\end{figure}

Having discussed the current caused by the difference in electrochemical potentials, we now address the question about the voltage drop on a scatterer. First, we determine the electron density produced in a nonequilibrium state, assuming that $\mu_{\rs L} > \mu_{\rs R}$, see Fig.~\ref{fig:Landauer2}. The left reservoir emits states~(\ref{eq:scatterL}) and the right reservoir emits the states $\Psi_{{\rs R}, \s E}$. The total density to the right of the scatterer,
\begin{align}
	\rho_{\rs R} &= 2 \int\limits_0^{k(\mu_{\rs L})}\frac{dk}{2\pi} |\Psi_{{\rs L}, \s E}(x)|^2 +2
	\int\limits_0^{k(\mu_{\rs R})} \frac{dk}{2\pi} |\Psi_{{\rs R}, \s E}(x)|^2\nonumber \\
	&\approx
	2 \int\limits_0^{k(\mu_{\rs L})} \frac{dk}{2\pi} T(E) + 2
	\int\limits_0^{k(\mu_{\rs R})} \frac{dk}{2\pi} \bigl[ 1 + R(E) \bigr]
	\label{eq:densityR}
\end{align}
is the sum of contributions from the left and right states, the factor 2 is due to spin degeneracy. Here we do not consider the details of Friedel oscillations with the period $\pi / k(\mu)$ (see below) and perform averaging over several wavelengths $\propto \hbar / \sqrt{2m\mu}$. Calculating the density on the left gives
\begin{equation}
	\rho_{\rs L} 
	\approx 2 \int\limits_0^{k(\mu_{\rs L})} \frac{dk}{2\pi}
		\bigl[ 1+R(E) \bigr]
	+ 2 \int\limits_0^{k(\mu_{\rs R})} \frac{dk}{2\pi} T(E).
	\label{eq:densityL}
\end{equation}
In the case of nonequilibrium situation, $\mu_{\rs L} \neq \mu_{\rs R}$, and nonideal transparency, $T < 1$, the density on the right of the scatterer does not coincide with that on the left, see Fig.~\ref{fig:Landauer1}.

The difference in densities is given by
\begin{equation}
	\rho_{\rs L} -\rho_{\rs R} = 
	4 \int\limits_{k(\mu_{\rs R})}^{k(\mu_{\rs L})} \frac{dk}{2\pi} R(E),
	\label{eq:change}
\end{equation}
where we use the relation $R(E) + T(E) = 1$. If the quantum conductor is electrically neutral, then this density difference should be compensated by the voltage drop across the scatterer, which bends the conduction band bottom. This voltage drop is called the Landauer voltage drop $V_{\rs L}$ and in the stationary case it can be found from the condition of electrical neutrality, which is assumed to take place in the equilibrium. In particular, the density should be the same on both sides of the barrier as shown in Fig.~\ref{fig:Landauer2}.

In the presence of a voltage drop $V_{\rs L}$, the left states with the energy $E$ (measured from the conduction band bottom in the right reservoir) have the form
\begin{equation}
	\Psi_{{\rs L}, \s E}(x) =
	\begin{cases}
		e^{i kx} + r(E) e^{-i kx}, & x \to -\infty, \\
		\sqrt{\cfrac{k}{\tilde{k}}} \,\, t(E)\, e^{i \tilde{k}x}, & x \to \infty,
	\end{cases}
	\label{eq:PsiL}
\end{equation}
where $k(E) = \sqrt{2m(E - eV_{\rs L})} / \hbar$ and $\tilde{k}(E) = \sqrt{2mE} / \hbar$ are the wave vectors in the left and right asymptotic regions respectively. Similarly, the right scattering states are
\begin{equation}
	\Psi_{{\rs R}, \s E}(x) =
	\begin{cases}
		e^{-i\tilde{k}x} + r'(E) e^{i\tilde{k}x},	& x \to \infty, \\
		\sqrt{\cfrac{\tilde{k}}{k}} \,\, t'(E) \, e^{-i kx},	& x \to -\infty.
	\end{cases}
	\label{eq:PsiR}
\end{equation}
The factor $\sqrt{k / \tilde{k}}$ appears due to the unitarity of the scattering matrix. We also note that the scattering problem must be solved taking the bending of the conduction band bottom due to the Landauer voltage $V_{\rs L}$ into account. For example, due to the appearance of this voltage, the right scattering states with energies $E < eV_{\rs L}$ are completely reflected, $R(E) = 1$. The averaged density on the left, caused by the left scattering states, is given by
\begin{equation}
	\rho_{\rs LL} = 2 \int\limits_0^{k(\mu_{\rs L})}
	\frac{dk}{2\pi} \bigl[ 1+R(E) \bigr],
	\label{eq:rhoLL}
\end{equation}
where the factor 2 is due to spin degeneracy. The density on the left, caused by the right states, takes the form
\begin{equation}
	\rho_{\rs LR}
	= 2 \int\limits_{\tilde{k}(eV_{\rs L})}^{\tilde{k}(\mu_{\rs R})}
	\frac{d{\tilde k}}{2\pi} \, \frac{{\tilde k}}{k} \, T(E).
	\label{eq:rhoLR}
\end{equation}
Similarly, calculating the density on the right, we find
\begin{align}
	\rho_{\rs RL}
	& = 2 \int\limits_0^{k(\mu_{\rs L})}
	\frac{dk}{2\pi} \, \frac{k}{\tilde{k}} \, T(E), \\
	\rho_{\rs RR}
	& = 2 \int\limits_{\tilde{k}(eV_{\rs L})}^{\tilde{k}(\mu_{\rs R})}
	\frac{d\tilde{k}}{2\pi} \bigl[ 1+R(E) \bigr] +
	2 \int\limits_0^{\tilde{k}(eV_{\rs L})}
	\frac{d\tilde{k}}{2\pi} [1+1],
	\label{eq:rhoRL}
\end{align}
where the last term appears due to the right states completely reflected at the bottom of the conduction band.

To simplify further calculations, we switch to integrals over energies. For $\rho_{\rs LL}$, we then obtain [$dk = (m / \hbar^2 k) dE$]
\begin{equation}
	\rho_{\rs LL} = \frac{2}{\hbar} \sqrt{\frac{m}{2}}
	\int\limits_{eV_{\rs L}}^{\mu_{\rs L}} \frac{dE}{2\pi}
	\frac{1+R(E)}{\sqrt{E-eV_{\rs L}}}.
	\label{eq:rhoLL1}
\end{equation}
Similarly, for $\rho_{\rs LR}$ we have [$d\tilde{k} = (m / \hbar^2
\tilde {k}) dE$]
\begin{equation}
	\rho_{\rs LR} = \frac{2}{\hbar} \sqrt{\frac{m}{2}}
	\int\limits_{eV_{\rs L}}^{\mu_{\rs R}} \frac{dE}{2\pi}
	\frac{T(E)}{\sqrt{E-eV_{\rs L}}}.
	\label{eq:rhoLR1}
\end{equation}
Calculations for $\rho_{\rs RL}$ and $\rho_{\rs RR}$ give
\begin{align}
	\rho_{\rs RL} &= \frac{2}{\hbar} \sqrt{\frac{m}{2}}
	\int\limits_{eV_{\rs L}}^{\mu_{\rs L}} \frac{dE}{2\pi}
	\frac{T(E)}{\sqrt{E}}, 
	\label{eq:rhoRL1} \\
	\rho_{\rs RR} &= \frac{2}{\hbar} \sqrt{\frac{m}{2}}
	\int\limits_{eV_{\rs L}}^{\mu_{\rs R}} \frac{dE}{2\pi}
	\frac{1+R(E)}{\sqrt{E}} +
	\frac{2}{\hbar} \sqrt{\frac{m}{2}}
	\int\limits_0^{eV_{\rs L}} \frac{dE}{2\pi}
	\frac{2}{\sqrt{E}}.
	\label{eq:rhoRR1}
\end{align}
Summing the densities on the left, $\rho_{\rs L} = \rho_{\rs LL}+\rho_{\rs LR}$, and using the relation $T(E) + R(E) = 1$, we obtain
\begin{align}
	\rho_{\rs L} & =
	\frac{2}{\hbar} \sqrt{\frac{m}{2}}
	\int\limits_{eV_{\rs L}}^{\mu_{\rs R}}
	\frac{dE}{2\pi}
	\frac{2}{\sqrt{E-eV_{\rs L}}} \nonumber \\
	& + \frac{2}{\hbar} \sqrt{\frac{m}{2}}
	\int\limits_{\mu_{\rs R}}^{\mu_{\rs L}} \frac{dE}{2\pi}
	\frac{1 + R(E)}{\sqrt{E-eV_{\rs L}}},
	\label{eq:rhoL}
\end{align}
while the total density on the right $\rho_{\rs R} = \rho_{\rs RL}+\rho_{\rs RR}$ is given by
\begin{equation}
	\rho_{\rs R}= \frac{2}{\hbar} \sqrt{\frac{m}{2}}
	\int\limits_0^{\mu_{\rs R}} \frac{dE}{2\pi}
	\frac{2}{\sqrt{E}} +
	\frac{2}{\hbar} \sqrt{\frac{m}{2}}
	\int\limits_{\mu_{\rs R}}^{\mu_{\rs L}} \frac{dE}{2\pi}
	\frac{T(E)}{\sqrt{E}}.
	\label{eq:rhoR}
\end{equation}
Assuming the electric neutrality, we should equate the densities:\footnote{In the nonlinear case, the additional requirement of the equality of densities to their equilibrium values gives the displacement of the barrier ``pedestal'' with respect to electrochemical potentials at the boundaries.}
\begin{equation}
	\int\limits_{\mu_{\rs R}}^{\mu_{\rs L}} \frac{dE}{2\pi}
	\frac{1 + R(E)}{\sqrt{E-eV_{\rs L}}} =
	\int\limits_{\mu_{\rs R}+eV_{\rs L}}^{\mu_{\rs R}} \frac{dE}{2\pi}
	\frac{2}{\sqrt{E}} +
	\int\limits_{\mu_{\rs R}}^{\mu_{\rs L}} \frac{dE}{2\pi}
	\frac{T(E)}{\sqrt{E}}.
	\label{eq:chargeNeutrality}
\end{equation}
Equation~(\ref{eq:chargeNeutrality}) allows one to calculate the voltage $V_{\rs L}$ for an arbitrary dependence transparency on energy and an arbitrary difference of electrochemical potentials.

We consider a simple linear case and find $V_{\rs L}$ for a small difference $\Delta \mu \equiv \mu_{\rs L} - \mu_{\rs R} \ll \mu_{\rs R}$. Under such conditions, the voltage drop is also small, $|eV_{\rs L}| \ll \mu_{\rs L}$. We assume that $T(E)$ is constant on the interval $[\mu_{\rs R},\mu_{\rs L}]$. Then replacing $\sqrt{E-eV_{\rs L}}$ by $\sqrt{E}$ in Eq.~(\ref{eq:chargeNeutrality}) and taking $T$ and $R$ out of the integrand, we express the Landauer voltage as
\begin{equation}
	eV_{\rs L} = \Delta\mu R.
	\label{eq:vL}
\end{equation}
The voltage $V_{\rs L}$ is zero for an ideally transparent conductor and reaches the maximum $eV_{\rs L} = \Delta \mu$ when all the electrons are reflected. The current is [see expression~(\ref{eq:totalCurrentB})]
\begin{equation}
	I = \frac{2e}{h} T \Delta\mu,
	\label{eq:Lndr_current}
\end{equation}
which gives the Landauer resistance
\begin{equation}
	\mathcal{R}_{\rs L} = \frac{V_{\rs L}}{I} =
	\frac{h}{2e^2} \frac{R}{T}.
	\label{eq:landauerResistance}
\end{equation}

The absence of the voltage drop in an ideal conductor was the object of intensive discussions for a long time. It finally became clear that the voltage drop occurs even in this case, but in joints with reservoirs rather than in the conductor itself (see the discussion in Sec.~\ref{sec:contactcond}).

\subsection{Contact resistance}
\label{sec:contactcond}

Equating $\Delta \mu $ in Eq.~(\ref{eq:Lndr_current}) to the value specified by the bias voltage $eV$, we obtain the conductance $G = I / V$ in the form\footnote{Below, we do not explicitly indicate the energy dependence of the transparency and elements of the scattering matrix, except in the cases where this dependence is being studied.}
\begin{equation}
	G =\frac{2e^2}{h}T.
	\label{eq:Landsimple}
\end{equation}

\begin{figure}[tb]
	\includegraphics[width=6.8cm]{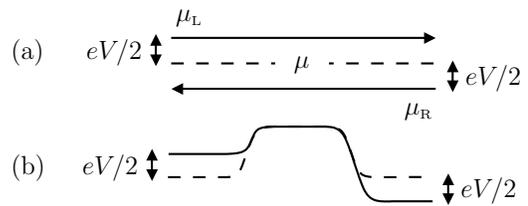}
	\caption{
(a)~Absence of the voltage drop in an ideal conductor. (b)~Initial one-dimensional electrostatic potential (dashed curve) and its modification by the bias voltage (solid curve).
	}
	\label{fig:voltDrop}
\end{figure}

Resistance~(\ref{eq:landauerResistance}) is different from the inverse of $G$ in ``Landauer formula''~(\ref{eq:Landsimple}). We can assume that~(\ref{eq:Landsimple}) is the conductance measured by the two-contact method, whereas resistance~(\ref{eq:landauerResistance}) is the resistance measured by the four-contact method.\footnote{We note that in this case, the actually measured resistance is also ill defined and depends on experimental conditions~\cite{Sukhorukov90}.} The Landauer resistance takes only the voltage drop directly across the barrier into account.\footnote{Below, we will consider the case where such voltages can be summed in the usual way, as in an ohmic conductor.} However, in a one-dimensional conductor, the voltage drop also appears in contacts with reservoirs, which is the reason for the discrepancy between the two Landauer formulas. Subtracting $V_{\rs L}$ from the bias voltage $\Delta\mu = eV$, we obtain the voltage drop $V_{\rs A}$ at the conductor entrances:
\begin{equation}
	V_{\rs A}
	= \Delta\mu-eV_{\rs L}
	= \Delta \mu T.
	\nonumber
\end{equation}
The total voltage drop can be written as the sum
\begin{align}
	V
	& = V_{\rs L} + V_{\rs A}
	= I\mathcal{R}_{\rs L} + I\mathcal{R}_{\rs A} \nonumber \\
	& = I\frac {h}{2e^2}\frac {1-T}{T} + I\frac {h}{2e^2}
	= I\frac {h}{2e^2T}
	= \frac {I}{G}.
	\nonumber
\end{align}

In the symmetric case, the voltage drop is distributed equally between contacts. The voltage drop $V_{\rs A} / 2$ at each boundary (contact) corresponds to the resistance
\begin{equation}
	\mathcal{R}_{\rs S} = \frac{h}{4 e^2},
	\label{eq:sharvin}
\end{equation}
which is the quantum analogue of the known Sharvin resistance~\cite{Sharvin65}. We can assume that this resistance is caused by the reflection of higher modes at the wire entrance (see Sec.~\ref{sec:quantcontacts} for the details).

\begin{figure}[tb]
	\includegraphics[width=3.8cm]{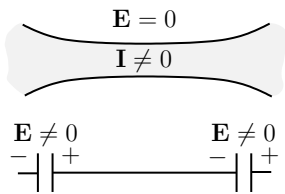}
	\caption{Input and output voltage drops in a ballistic one-channel conductor.}
	\label{fig:adiabChannel}
\end{figure}

Figure~\ref{fig:voltDrop} shows the example of a ballistic conductor $(T = 1)$. Applying a voltage, we obtain the nonzero current $I = 2e^2 / h V$, although no voltage drop occurs in the one-dimensional conductor itself due to the absence of backward reflection. The distribution of the voltage equally between contacts has been studied in detail theoretically~\cite{Glazman88,Glazman89} and verified experimentally~\cite{Patel90,Patel91b}. As a whole, the described situation is quite unusual from the standpoint of the classical local conductivity: the electric field inside the conductor is absent, although the total current is nonzero, see Fig.~\ref{fig:adiabChannel}. It is also unusual that the Joule heat dissipates far from the reservoirs due to slow energy relaxation, whereas the electromagnetic energy, from the standpoint of classical electrodynamics, enters the electron system at much smaller scales, in voltage drop regions in contacts and at the barrier.

Finally, we note that the oscillating part of the electron density (and its slowly changing part at a finite voltage), which we did not consider above, can lead to an additional scattering of electrons. Density oscillations (Friedel oscillations) are not completely screened and produce a spatially dependent electrostatic potential. The oscillating part of the potential is especially important because the oscillation period is equal to $\pi / k_{\rs F}$ and backscattering from it (by $2k_{\rs F}$ in the momentum space) is strong~\cite{Matveev89}. Therefore, the transmission probability $T(E)$ taking the total scattering potential into account can strongly differ from the bare probability (determined on a local scatterer); in addition, this probability in general case depends on the voltage $V$. Assuming that the reflection amplitude is independent of energy, we can obtain the oscillating part of the density in equilibrium in the form
\begin{equation}
	\delta n(x) =
	\frac{1}{|x|}
	\big\{
		{\rm Im}(r) \bigl[ \cos(2k_{\rs F}x)-1 \bigr]
		+ {\rm Re}(r) \sin (2k_{\rs F}|x|)
	\big\}.
\end{equation}
The case of energy-independent $r$ is realistic, for example, for almost complete reflection ($r \approx -1$), but similar oscillating dependences also appear for an arbitrary scatterer.

We emphasize once again the difference between our approach and more traditional methods: instead of calculating the nonlocal conductivity $\sigma (r, r')$ and using it in the expression
\begin{equation}
	j_{\alpha }(r)=\int \sigma (r,r')_{\alpha \beta }E_{\beta }(r')dr'
	\label{eq:nonlocal}
\end{equation}
we calculate the total conductance determining the total current $I = GV$ as a function of voltage. The convenience of such approach is obvious, because instead of calculating the self-consistent field $E$ for use in Eq.~(\ref{eq:nonlocal}), only the total voltage drop $V$ must be known. In this case, the conductance can be expressed in terms of the probability of transmission through the conductor. (Yet, to exactly solve the scattering problem in the nonlinear case, the electrostatic potential inside the conductor must also be known.)

In Sec.~\ref{sec:waveguide} we consider a multichannel conductor as a waveguide for electrons to solve a broader class of problems.


\section{Waveguides: the multichannel case}
\label{sec:waveguide}

\begin{figure}[b]
	\includegraphics[width=5.2cm]{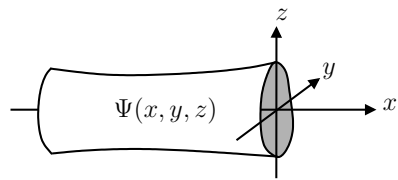}
	\caption{Waveguide elongated along the $x$ axis with an adiabatically slowly varying cross section.}
	\label{fig:waveguide}
\end{figure}

Let us describe a quantum conductor as a wire smoothly connected to reservoirs. More formally, we consider the geometry convenient for the description of such a system.

A quasi-one-dimensional system is formed as a constriction with infinitely high walls (or with a potential increasing at infinity) in transverse directions $(y, z)$ and transport is possible along the $x$ axis only, see Fig.~\ref{fig:waveguide}. Plane waves belonging to different modes can propagate along this $x$ axis. In mesoscopic physics such modes are called channels. In transverse directions each channel has a spatial structure of the bound states. The waveguide can transfer many modes. At low temperatures in a narrow waveguide only the first mode is significant and the transport becomes effectively one-dimensional (we actually discussed this situation in Sec.~\ref{sec:landauer}). In the general case the number of conducting channels involved in transport is finite.

\subsection{Mode quantization}

Let us consider now the simple case of translational invariance along the $x$ axis. We want to show how modes appear due to the transverse motion quantization. We solve the Schr\"odinger equation
\begin{equation}
	\biggl[-\frac{\hbar^2}{2m} \Delta + V(x,y,z) \biggr]
	\Psi (x,y,z) = E \Psi (x,y,z),
	\label{eq:schroedinger}
\end{equation}
where the potential $V(x, y, z) = V(y, z)$ and boundary conditions are temporarily considered independent of $x$. In this case, we can write the solution of Eq.~(\ref{eq:schroedinger}) in the form $\Psi(x, y, z) = \chi(y, z) e^{ikx}$. After the substitution of this function in Eq.~(\ref{eq:schroedinger}) the variables separate and we obtain equation for the eigenvalues
\begin{equation}
	\biggl[-\frac{\hbar^2}{2m} (\partial_y^2+\partial_z^2)
	+ V(y,z) \biggr] \chi_n (y,z) = E_n \chi_n(y,z),
	\label{eq:transversal}
\end{equation}
where $n$ is the mode (channel) index, $\chi_n(y, z)$ is the corresponding wave function, and $E_n$ is the transverse direction quantization energy. The functions $\chi_n(y, z)$ form a complete set
\begin{equation}
	\sum\limits_n \chi_n(y,z) \chi_n^*(y',z') 
	= \delta (y-y') \delta(z-z'),
	\label{eq:complete}
\end{equation}
which is also orthonormalized
\begin{equation}
	\int dy dz\, \chi_m^*(y,z) \chi_n(y,z) = \delta_{mn}.
	\label{eq:orthonormal}
\end{equation}
The general solution of Eq.~(\ref{eq:schroedinger}) can be expanded using the above set of functions
\begin{equation}
	\Psi (x,y,z) = \sum\limits_n c_n \chi_n(y,z) e^{i k_n x},
	\label{eq:solution}
\end{equation}
where $k_n = \sqrt{2m (E - E_n)} / \hbar$ is the wave vector in the $n$th channel and $c_n$ are constants. Modes with energies $E < E_n$ decay as $e^{-\varkappa_n x}$, where $\varkappa_n = \sqrt{2m(E_n -E)}/\hbar$.

\subsection{Scattering problems in waveguides}
\label{sec:unitary3d}

We consider a system that is a translation-invariant waveguide for $x \to \pm\infty$. Asymptotic solutions are described by expression~(\ref{eq:solution}). If an additional potential or a change in the boundary conditions exists in the vicinity of some finite $x$, then we can formulate a scattering problem. We assume the incident (from left or right) wave has the form
\begin{equation}
	\Psi^\text{in} (x,y,z) = \chi_n(y,z) e^{-i k_n |x|}.
	\label{eq:psiIn}
\end{equation}
Scattered waves can be written as
\begin{equation}
	\Psi^\text{out}(x,y,z) = \sum\limits_m S_{m n} \sqrt{\frac{k_n}{k_m}} \,
	\chi_m(y,z) e^{i k_m |x|},
	\label{eq:psiOut}
\end{equation}
where the sum over $m$ channels is taken for both transmitted ($S_{mn} = t_{mn}$, $m \neq n$) and reflected ($S_{nn} = r_{nn}$) states. The additional factor $\sqrt{k_n/k_m}$ is introduced to preserve the unitarity of the scattering matrix $S_{mn}$; hence each of the asymptotic states $\chi_n(y,z) e^{i k_n x}/\sqrt{k_n}$ carries the unit current.

\begin{figure}[t]
	\includegraphics[width=4.3cm]{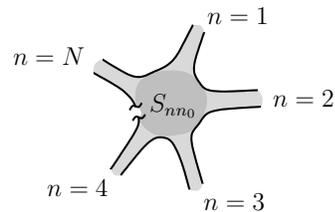}
	\caption{Multilead conductor.}
	\label{fig:multilead}
\end{figure}

We calculate the electric current in the waveguide to the right of the scattering potential. Let $\mu_1$ and $\mu_2$ be electrochemical potentials of the reservoirs so that the electron distribution functions in the reservoirs have the form
\begin{equation}
	f_{\alpha}(E) =
	\frac{1}{e^{(E-\mu_\alpha)/\Theta_\alpha} + 1}, \quad
	\alpha = 1,2,
	\label{eq:FermiDistr}
\end{equation}
where $\Theta_\alpha$ is the $\alpha$th reservoir temperature in energy units. We assume here that the temperatures $\Theta_1$ and $\Theta_2$ are equal. Electrons with energy $E$ emerging from the $n$th channel of the left reservoir $(\alpha =1)$ make a contribution to the current in the unit energy interval to the left of the scattering potential (as in purely one-dimensional problems considered in Sec.~\ref{sec:landauer}), which is proportional to $f_1(E) (2e/h)$, while to the right, after scattering into the $m$th channel, they make a contribution proportional to $f_1(E) (2e/h) T_{nm}$. Electrons emerging to the right of the $n$th channel provide an initial current of the opposite sign $-f_2(E) (2e/h)$ and, after backscattering, also make the contribution $(2e/h) f_2(E) \sum_m R_{nm}$. As a result, after summation over channels, the current is given by
\begin{equation}
	I = \frac{2e}{h}\sum\limits_{n,m}\int\limits_0^\infty dE \,
	\bigl[ f_1(E) T_{nm} - f_2(E)(\delta_{nm} - R_{nm}) \bigr],
	\label{eq:Landauer_g}
\end{equation}
where $\delta_{nm} = 1$ for $n = m$ and $\delta_{nm} = 0$ for $n \neq m$. Similarly, we can formulate the scattering problem in the multilead case shown in Fig.~\ref{fig:multilead} by replacing the mode (channel) numbers with the reservoir indices or by adding modes (channels). We now take the unitarity of the scattering matrix into account to simplify expression~(\ref{eq:Landauer_g}) for the current:
\begin{equation}
	I = \frac{2e}{h}\sum\limits_{n,m}\int\limits_0^\infty dE
	\bigl[ f_1(E) - f_2(E) \bigr] T_{nm}.
	\label{eq:Landauer_gR}
\end{equation}
The sum over the transparencies in Eq.~(\ref{eq:Landauer_gR}) can sometimes be conveniently written as the trace of the scattering amplitude matrix. In this case, we obtain the conductance in the form
\begin{equation}
	G =\frac{2e^2}{h} {\rm Tr} \{ tt^\dag \}.
	\label{eq:traceG}
\end{equation}

In what follows, with the products of the transmission and reflection amplitude matrices of types $t t^\dag$ and $1 - r r^\dag$ appearing in expressions not only for current but also for noise and more complicated quantities, it is very important that due to the unitarity of $S$, such Hermitian matrices have the same set of eigenvalues $T_1, T_2, \ldots, T_{\s N}$, and the product of matrices such as $t t^\dag t t^\dag$ has the eigenvalues $T_1^2, T_2^2, \ldots, T_{\s N}^2$, and so on. Each of these transparency eigenvalues is a real number in the interval $[0, 1]$ (see Refs.~\cite{Dorokhov82,Mello88,Martin92}). In turn, such a diagonalization of the problem implies the presence of eigenmodes (channels) representing the superposition of states like~(\ref{eq:solution}), which are no longer mixed after scattering. The conductance in the diagonal representation has the form
\begin{equation}
	G = \frac{2e^2}{h} {\rm Tr} \{ tt^\dag \} =
	\frac{2e^2}{h}\sum\limits_n T_n.
	\label{eq:Gdiag}
\end{equation}

\subsection{Adiabatically changing waveguides}
\label{sec:sma}

In general case, the boundary conditions and the potential in Eq.~(\ref{eq:schroedinger}) are inhomogeneous. Nevertheless, changes are often rather slow and small at the wavelength scale. In this case, we can use the adiabatic approximation to separate rapid transverse motion in the waveguide and slow motion along it. The eigenvalue equation for rapid motion takes the form
\begin{equation}
	\biggl[-\frac{\hbar^2}{2m}
	(\partial_y^2+\partial_z^2)
	+ V(x,y,z) \biggr] \chi_n (x,y,z) = E_n(x) \chi_n(x,y,z)
	\label{eq:transversalX}
\end{equation}
for each cross section in Fig.~\ref{fig:waveguide}. In this case, the transverse quantization energy $E_n(x)$ becomes slightly dependent on $x$. Assuming the adiabaticity, we substitute
\begin{equation}
	\Psi(x,y,z) =
	\chi_n(x,y,z) \phi_n(x),
	\label{eq:solutionX}
\end{equation}
where $\phi_n(x)$ is the solution of the equation
\begin{equation}
	\Bigl[-\frac{\hbar^2}{2m} \frac{d^2}{dx^2} + E_n(x) \Bigr]
	\phi_n(x) = E \phi_n(x)
	\label{eq:longitudinalX}
\end{equation}
for motion along the wire. We note that the transverse quantization energy $E_n(x)$ serves as the effective potential $U(x)$ for slow motion along $x$. Expression~(\ref{eq:solutionX}) is an approximate solution of the Schr\"odinger equation with the mode mixing neglected. The approximation validity conditions are
\begin{equation}
	\biggl| \frac{\partial_x \chi_n(x,y,z)}{\chi_n(x,y,z)} \biggr|
	\ll \biggl| \frac{\partial_x
	\phi_n(x)}{\phi_n (x)} \biggr| = |k(x)|
	\label{eq:adiabaticConda}
\end{equation}
and
\begin{equation}
	\biggl| \frac{\partial_x^2 \chi_n(x,y,z)}{\chi_n(x,y,z)} \biggr|
	\ll \biggl| \frac{\partial_x^2
	\phi_n(x)}{\phi_n (x)} \biggr| \approx |k(x)^2|.
	\label{eq:adiabaticCondb}
\end{equation}

In Sec.~\ref{sec:quantcontacts} we will consider the important example of a real waveguide, a microscopic constriction (a quantum point contact) in two-dimensional electron gas.


\section{Quantum contacts}
\label{sec:quantcontacts}

\subsection{Current through a quantum point contact}

\begin{figure}[tb]
	\includegraphics[width=4.7cm]{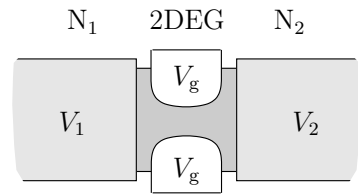}
	\caption{
The scheme of experimental realization of a point contact. Two massive electrodes are connected via a two-dimensional electron gas layer formed in a semiconductor heterostructure. A constriction is produced by the voltage $V_{\rm g}$ applied to the gates.
	}
	\label{fig:2deg_gate}
\end{figure}

Let us consider a contact between two conductors. If the contact width $W$ is comparable with a few electron wavelengths $\lambda_{\rs F}$ then such a contact is called a quantum point contact (QPC). The point contact can be realized in experiments~\cite{Wees88,Wharam88} in the following way: two massive electrodes are connected with a layer of the two-dimensional electron gas (2DEG) formed in the region of a semiconductor heterojunction, as shown in Fig.~\ref{fig:2deg_gate}. Then two gates are attached to the 2DEG layer from above.\footnote{This is the so-called {\it split gate technique} developed in Refs.~\cite{Thornton86,Wharam88}.} By applying a potential $V_{\rm g}$ to the gate we can ``expel'' electrons from the regions near the gate thus making them unavailable for electrons and thereby producing a constriction in the 2DEG (a point contact). The higher the voltage applied across the gate, the larger the region forbidden for electrons and the stronger the constriction.

\begin{figure}[tb]
	\includegraphics[width=5.5cm]{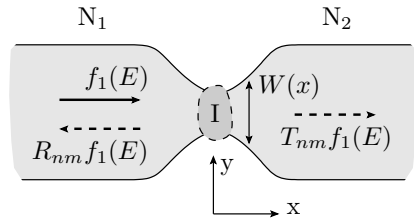}
	\caption{
Point contact in the form of a constriction. The constriction width is described by a function $W(x)$ with a minimal value $W_0$. Inside the constriction, a scatterer I (for example, an impurity) can be located.
	}
	\label{fig:2deg_constriction}
\end{figure}

\begin{figure}[bt]
	\includegraphics[width=4.5cm]{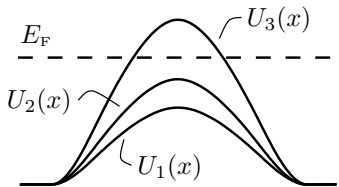}
	\caption{
Example of the effective potential $U_n(x)$ appearing due to the effect of contact walls. For each $n$, the potential has the maximum value $E_n$ determined by the smallest width. The current is provided by electrons with energies close to the Fermi energy $E_{\rs F}$. The figure corresponds to two open channels (channels 1 and 2, because the relation $E_{\rs F} > E_{1,2}$ is satisfied only for them) and one closed (channel 3 with $E_{\rs F} < E_3$).
	}
	\label{fig:channels}
\end{figure}

Now let us describe the transport in such a contact. We consider a system shown in Fig.~\ref{fig:2deg_constriction} with connected reservoirs N$_1$ and N$_2$ and assume that the system is two-dimensional, corresponding to the standard experimental situation presented in Fig.~\ref{fig:2deg_gate}.\footnote{More precisely, the size quantization along the $z$ axis is so strong that under all standard experimental conditions, only the lowest mode is always filled.} We choose the direction of the $x$ and $y$ axes as shown in Fig.~\ref{fig:2deg_constriction}. The two-dimensional electron gas lying in the $(x,y)$ plane is additionally restricted in the $y$ direction by means of voltages applied across the gates. We simulate the walls by the boundary condition $\Psi[x, \pm W(x)/2] = 0$, making motion possible only in a strip of the width $W(x)$ along the $x$ axis. Assuming that $W(x)$ varies slowly and the mean free path in 2DEG greatly exceeds all the characteristic dimensions of the contact, we obtain
\begin{equation}
	\chi_n(x,y) = \sqrt{\frac{2}{W(x)}} \sin\biggl[
		n\pi \, \frac{y+W(x)/2}{W(x)}
	\biggr]
	\label{eq:transversalMode}
\end{equation}
for the transverse modes. The wave function $\phi_n(x)$ satisfies Eq.~(\ref{eq:longitudinalX}) describing motion in the effective potential $U(x)= E_n(x) = \hbar^2 \pi^2 n^2 /2m W^2(x)$, $n \geqslant 1$. The applicability conditions~(\ref{eq:adiabaticConda}) and (\ref{eq:adiabaticCondb}) now become $W'(x)/W(x) \ll k(x)$ and $W''(x)/W(x) \ll k^2(x)$. Let $W_0$ denote the minimal value of $W(x)$. Then the effective potential (depending on the transverse quantum number $n$) in the resultant Schr\"odinger equation has the form of a potential barrier with the height
\begin{equation}
	E_n = \frac{(n\pi\hbar)^2}{2m W_0^2},
\end{equation}
decreasing to zero as $x \to \pm\infty$, see Fig.~\ref{fig:channels}.

For a wave function with the mode number $n$, the transverse motion of an electron is specified by the condition that an integer number of half waves $\lambda_{\rs F} / 2$ fit in the contact width. Therefore, for electrons flowing through the contact, either one, or two, or three, etc., half-waves fit in the contact width. These waveguide modes are called channels. For example, it is customary to say that an electron in the state with the wave function $\chi_n$ is in the $n$th channel.\footnote{The terms ``channels'' and ``leads'' should be distinguished in multilead systems.}

Since $W(x)$ changes slowly, Eq.~(\ref{eq:longitudinalX}) can be solved in the Wentzel-Kramers-Brillouin (WKB) approximation. In the leading approximation, only electrons with energies $E > E_n$ can propagate through the constriction. In the general case, the additional scattering of electrons in the constriction, for example, from the impurity potential, must be taken into account. Such a scatterer is schematically shown by the dashed contour in Fig.~\ref{fig:2deg_constriction}.

\subsection{Conductance quantization}

We now consider the linear conductance $G = dI/dV$ at $V \to 0$. We assume that scattering by impurities in the constriction is absent and channels do not mix. Then expression~(\ref{eq:Gdiag}) defines the conductance directly in terms of the transparencies $T_n$ in each channel:
\begin{equation}
	G =\frac{2e^2}h \sum\limits_n T_n(E_{\rs F}).
	\label{eq:G}
\end{equation}
The quantity $G_0 = 2e^2 / h$, which is called the conductance quantum, is the natural unit for conductance measurements in mesoscopic systems. In the zeroth-order WKB approximation described in Sec.~\ref{sec:landauer}, $T_{nm} = \delta_{nm}$ for ``open'' channels, whence
\begin{equation}
	G = N G_0, \qquad
	N = \sum\limits_n \theta(E_{\rs F}-E_n),
	\label{eq:G_q}
\end{equation}
where $N$ is the number of open channels and $\theta$ is the Heaviside function.

\begin{figure}[tb]
	\includegraphics[width=8.2cm]{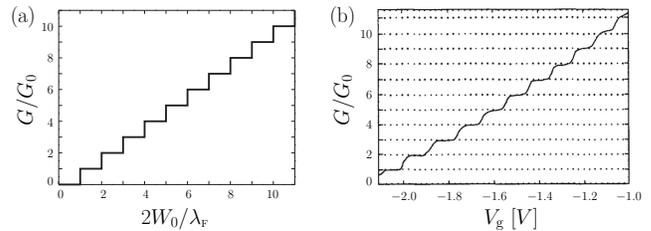}
	\caption{
(a)~Quantization of the conductance of a point contact under the variation in the constriction width $W_0$ due to a voltage applied to the gate (see Fig.~\ref{fig:2deg_gate}). (b)~Experimental dependences of the constriction conductance on the gate voltage $V_{\rm g}$. It can be assumed with good accuracy that $W_0$ is a linear function of $V_{\rm g}$. The plot is taken from the first experimental paper~\cite{Wees88}. Similar results were presented about the same time in Ref.~\cite{Wharam88}.
	}
	\label{fig:cond_quantization}
\end{figure}

Let us consider how $G$ changes when we change the QPC width $W_0$ by applying a voltage across the gate, see Fig.~\ref{fig:channels}. If $W_0 \to 0$, we obtain $E_{\rs F} < E_1$, therefore, $N = 0$ and electrons cannot pass through the QPC. This effect can be simply explained qualitatively: in a narrow QPC, due to the Heisenberg uncertainty principle, an electron should have a large quantization energy, and if this energy exceeds the specified energy, the presence of the electron in this region is forbidden. If $E_1 < E_{\rs F} < E_2$, then one channel is open and $G = G_0$. If $E_2 < E_{\rs F} < E_3$, then two channels are open, therefore, $G = 2G_0$, and so on. The QPC conductance is thus quantized in $G_0$ units (see Fig.~\ref{fig:cond_quantization}), similarly to the case of the integer quantum Hall effect (IQHE).\footnote{For a waveguide with a two-dimensional effective cross section, the quantization picture can be much more intricate, because it depends on the energy level structure in a two-dimensional box formed by the cross section. When a certain spatial symmetry exists and a two-dimensional problem is integrable (for example, if the wire cross section is nearly circular), the levels are grouped and, when the parameters are changed, several channels can be ``switched on'' at once, almost simultaneously~\cite{Falko95}.} The analogy becomes even more direct in the presence of the Zeeman splitting (which will be discussed in Sec.~\ref{sec:zeeman}), when the steps are split and the conductance is quantized in $G_0/2$ units, as in the IQHE.

\begin{figure}[tb]
	\includegraphics[width=1.2cm]{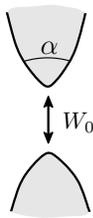}
	\caption{
Quantization is observed for angles $\alpha \gg 1 / \pi^2$.
 	}
	\label{fig:qpc_angle}
\end{figure}

The step height in the experimental plot in Fig.~\hyperref[fig:cond_quantization]{\ref{fig:cond_quantization}(b)} obeys the quantization rule with good accuracy, whereas the step edges are smeared. This can be caused by different factors, such as a finite temperature, finite probabilities of transmission below the barrier and reflection above the barrier, etc. (see Sec.~\ref{sec:condsmearing}). It is interesting that the experimental constriction was rather small, suggesting that quantization should not be so pronounced, see Fig.~\ref{fig:qpc_angle}. This puzzle was solved in paper~\cite{Glazman88b} (almost immediately after the publication of experimental results). It was found that the quantization conditions remained valid until the angle $\alpha$ was greater than $1/\pi^2$ rather than unity, as would be expected (the condition of applicability of the adiabatic approximation proved to be more strict). The problem therefore has a specific small parameter $1/\pi^2$. We consider this situation in more detail following Ref.~\cite{Glazman88b}.

\subsection{Smearing of conductance steps caused by tunneling through the effective potential}
\label{sec:condsmearing}

To perform a more detailed analysis, we describe the shape of a QPC by the model dependence (see Figs.~\ref{fig:2deg_constriction} and~\ref{fig:qpc_angle})
\begin{equation}
	W(x) = \frac{W_0}{L} \sqrt{x^2 + L^2},
	\label{eq:qpc}
\end{equation}
where $W_0$ and $L$ are the QPC width and length. The opening angle of the contact walls is $\alpha = 2\arctan(W_0 / 2 L)$. In this case, the effective potential
\begin{equation}
	U_n (x) = \frac{\hbar^2 \pi^2 L^2 n^2}{2m W_0^2 (x^2+L^2)} \approx
	U_n(0) - \frac{m}{2} \Omega_n^2 x^2
	\label{eq:effectivePotential}
\end{equation}
is approximately quadratic near the barrier top ($x=0$), with the expansion coefficients
\begin{equation}
	E_n = U_n(0) = \frac{\hbar^2 \pi^2 n^2}{2m W^2}
	\quad\text{and}\quad
	\Omega_n = \frac{\hbar \pi n}{m W L}.
	\nonumber
\end{equation}
The problem of tunneling through an inverted quadratic potential can be solved exactly. The probability of transmission through the potential~(\ref{eq:effectivePotential}) is given by the Kemble formula~\cite{Kemble35,Landau04BookV3}
\begin{equation}
	T_n(E) = \frac{1}{e^{2\pi[E_n-E]/\hbar \Omega_n} +1}
	\label{eq:kemble}
\end{equation}
in the form of a smeared step increasing from 0 for $E < E_n$ to 1 for $E > E_n$; the crossover occurs at the scale $\hbar \Omega_n / 2\pi$. To observe steps in the conductance as functions of $W_0$, the step width $\hbar\Omega_n / 2\pi$ should be much smaller than the distance between steps: $U_{n+1}(0) - U_n (0) \approx \hbar^2 \pi^2 n/m W_0^2$, i.e.,
\begin{equation}
	\frac{L}{W} \gg \frac{1}{2 \pi^2} \approx 0.051.
	\label{eq:adiabatic}
\end{equation}
Good quantization is therefore observed even for a relatively short point contact~\cite{Wees88,Wharam88}. It is also important that the region of the potential responsible for scattering is sufficiently small, and therefore quadratic approximation~(\ref{eq:effectivePotential}) can be justified and the Kemble formula well describes the behavior of the transparency in the range from low to high transparencies. The nonquadratic shape of the scattering potential is manifested only in case of very small reflection or transmission probabilities.

Let us now briefly discuss the mixing between channels.
The condition for the absence of channel mixing in the constriction region is well satisfied. Away from the throat, in the banks, the situation for the first channels is the opposite: in this region motion along the $x$ axis is faster while transverse motion is slower and distances between the transverse quantization levels are small. Therefore, even smooth inhomogeneities cause mode mixing. Yet, the mixing of transmitted modes does not affect the quantization picture, in particular, the transport remains reflectionless on a plateau. The point is that the eigenmodes that diagonalize the transmission amplitude matrix are important here. In the constriction, the eigenmodes look like usual transverse modes, which we already considered, whereas on the banks, they can be a complex mixture of transmitted modes. But if the transmitted modes are mixed with the reflected ones then the conductance in the plateau can of course change and, moreover, the entire quantization picture can be smeared.

It is interesting that for the chosen boundary conditions (impenetrable walls), variables separate in the Schr\"odinger equations if the wall shape is described by a second-order curve such as a parabola or hyperbola~\cite{Kawabata89}. In this case, the absence of channel mixing is an exact fact rather than the result of approximation. In addition, variables are separated in the saddle potential~\cite{Fertig87}, which is also used in simulations of QPCs~\cite{Buttiker90b}. Such a wall shape is also {interesting because it allows to solve the problem in the presence of magnetic field.}

The conductance quantization is observed not only in QPCs and a 2DEG but also in contacts of carbon nanotubes with metals~\cite{Frank98,Poncharal99,Martel98} and in atomic point contacts~\cite{Olesen94,Krans95,Scheer97,Rodrigues00}. Recently it was predicted~\cite{Peres06} and observed in graphene~\cite{Tombros11}.

The nature of quantization in these systems is similar to that in QPCs, however, differences also exist. For example, the number of channels is related not only to the form of orbital transverse modes (in the case of atomic point contacts, they are caused by the electron wave functions of contacting atoms) but also to the physical amount of layers in nanotubes or atoms in the constriction. The adiabaticity of the bottleneck-bank joining is also caused not by the smoothness of the conduction region opening, as in QPCs, but by a weak tunneling from a quasi-one-dimensional conductor to massive banks on a large effective contact area.


\section{Waveguide in a magnetic field}

A finite magnetic field in an electron waveguide, and in a QPC in particular, leads to two effects. First, Zeeman splitting appears. Second, orbital effects appear in two- and three-dimensional cases, which are absent in one-dimensional systems, where the vector potential leads simply to the phase accumulation and does not affect observables. In this case, the form of the wave functions of transverse modes (channels) changes considerably, and we consider these changes in Secs.~\ref{sec:zeeman} and \ref{sec:edgestates}.

\subsection{Zeeman effect in a quantum point contact}
\label{sec:zeeman}

\begin{figure}[tb]
	\includegraphics[width=8.0cm]{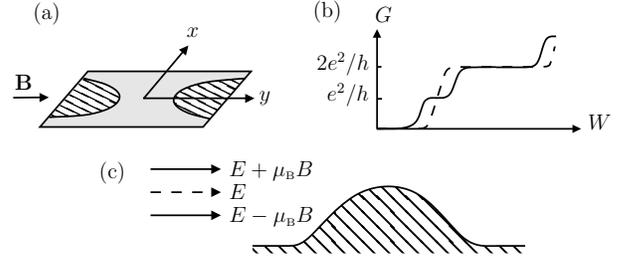}
	\caption{
(a)~Quantum point contact in a magnetic field collinear to the plane of a two-dimensional gas. (b)~The solid curve shows conductance steps in the magnetic field, and the dashed curve does so in the absence of the magnetic field. (c)~Each scattering state with an energy $E$ is split into two with energies $E \pm \mu_{\rs B} B$.
	}
	\label{fig:Ht2D}
\end{figure}

A QPC in a two-dimensional gas in the $(x,y)$ plane in a magnetic field with the vector lying in the same plane [Fig.~\hyperref[fig:Ht2D]{\ref{fig:Ht2D}(a)}] is described by the Hamiltonian
\begin{equation}
	{\hat H} = \frac{1}{2m} \Bigl(
		{\bf\hat p} - \frac{e}{c} {\bf A}
	\Bigr)^2 +
	U(x,y) + \mu_{\rs B} {\bf B} \boldsymbol{\sigma},
	\label{eq:Ht2DHam}
\end{equation}
where $e$ is the electron charge, $\mu_{\rs B}$ is the Bohr magneton, $\boldsymbol{\sigma} = (\sigma_x, \sigma_y, \sigma_z)$ are the Pauli matrices, and
\begin{equation}
	{\bf B} = B {\bf e}_y.
	\label{eq:Ht2Dmagnf}
\end{equation}
Here ${\bf e}_x$, ${\bf e}_y$, and ${\bf e}_z$ denote the unity vectors in $x$, $y$, and $z$ directions correspondingly. In-plane magnetic field~(\ref{eq:Ht2Dmagnf}) does not affect the orbital motion of particles, and we can rewrite Hamiltonian~(\ref{eq:Ht2DHam}) in the form
\begin{equation}
	{\hat H} = {\hat H}_0 + \mu_{\rs B} B \sigma_y,
	\label{eq:Ht2DHam1}
\end{equation}
i.e., represent $\hat H$ as a sum of the Hamiltonian ${\hat H}_0 = p^2/2m + U(x,y)$ without a magnetic field and the Zeeman term. Two solutions with kinetic energies $E \pm \mu_{\rs B} B$ correspond to each scattering state or bound state of the Hamiltonian $\hat H_0$ with an energy $E$, see Fig.~\hyperref[fig:Ht2D]{\ref{fig:Ht2D}(c)}. As the constriction width $W$ increases, the spin degeneracy is lifted and the conductance of the system increases by steps $e^2 / h$ as shown in Fig.~\hyperref[fig:Ht2D]{\ref{fig:Ht2D}(b)}. Such a splitting was already observed in the pioneering paper~\cite{Wharam88} and was later thoroughly studied in Refs.~\cite{Patel91,Thomas96}. This effect was considered theoretically in Ref.~\cite{Glazman89b}.

We note that in the plateau mode after odd steps, the spin-polarized current flows through the contact.

\subsection{Edge states in a magnetic field}
\label{sec:edgestates}

\begin{figure}[tb]
	\includegraphics[width=8.0cm]{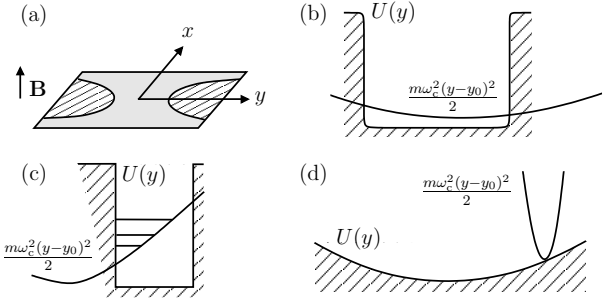}
	\caption{
(a)~Quantum point contact in a magnetic field perpendicular to a sample. (b)~Weak magnetic field in a potential box. (c)~Edge states in a strong magnetic field for a steep wall $U(y)$. (d)~Edge states in a strong magnetic field for a smooth potential $U(y)$.
	}
	\label{fig:Hp2D}
\end{figure}

If the magnetic field $\bf B$ is perpendicular to the plane shown in Fig.~\hyperref[fig:Hp2D]{\ref{fig:Hp2D}(a)}, then orbital effects appear along with the Zeeman effect. For simplicity, we consider only orbital effects in this section. They are described by the Hamiltonian
\begin{equation}
	{\hat H} = \frac{1}{2m} \Bigl(
		{\bf\hat p} - \frac{e}{c} {\bf A}
	\Bigr)^2 + U(y),
	\label{eq:Hp2DHam}
\end{equation}
where the potential $U(y)$ is independent of the coordinate $x$ along the wire. We still assume that the magnetic field is homogeneous, but this time it is perpendicular to the plane of the 2DEG,
\begin{equation}
	{\bf B} = B {\bf e}_z.
	\label{eq:Hp2Dmagnf}
\end{equation}
This magnetic field can be described by the vector potential in the form (the Landau gauge)
\begin{equation}
	{\bf A} = -By \, {\bf e}_x.
	\label{eq:Hp2DGaugCond}
\end{equation}
Then Hamiltonian~(\ref{eq:Hp2DHam}) takes the form
\begin{equation}
	{\hat H} =
	\frac{1}{2m} \Bigl(
		{\hat p}_x + \frac{eB}{c} y
	\Bigr)^2 +
	\frac{{\hat p}_y^2}{2m} + U(y).
	\label{eq:Hp2DHam1}
\end{equation}
Variables in the Schr\"odinger equation with Hamiltonian~(\ref{eq:Hp2DHam1}) can be separated by the substitution
\begin{equation}
	\Psi (x,y) = e^{ikx} \chi(y).
	\label{eq:Hp2Dform}
\end{equation}
The transverse modes $\chi(y)$ then satisfy the equation
\begin{equation}
	\chi_n''(y) + \frac{2m}{\hbar^2} \Bigl[
		E_n(k) - U(y) - \frac{m\omega_\text{c}^2}{2} (y-y_0)^2
	\Bigr] \chi_n(y) = 0,
	\label{eq:Hp2DSch1}
\end{equation}
where $\omega_\text{c} = |e|B/mc$ is the cyclotron frequency, $y_0 = \ell_B^2 k$, and $\ell_B = \sqrt{c\hbar / |e|B}$ is the magnetic length. Solving Eq.~(\ref{eq:Hp2DSch1}), we obtain the dispersion and $E_n(k)$ and the wave function in the presence of the magnetic field. In the absence of the additional potential, $U(y) = 0$, Eq.~(\ref{eq:Hp2DSch1}) reduces to the equation for a harmonic oscillator. The solution gives the Landau levels:
\begin{equation}
	E_n(k) = \hbar \omega_\text{c} \biggl( n + \frac{1}{2} \biggr),
	\label{eq:Hp2DLL1}
\end{equation}
which form a flat dispersionless band~\cite{Landau04BookV3}.

In the case of {\it a weak magnetic field} in a QPC, we can regard the quadratic potential produced by $B$ as a perturbation, see Fig.~\hyperref[fig:Hp2D]{\ref{fig:Hp2D}(b)}. The energy levels take the form
\begin{equation}
	E_n(k) = E_n + \frac{\hbar^2 k^2}{2m}
	+ \langle \chi^{(0)}_n | V(y) | \chi^{(0)}_n \rangle,
	\label{eq:Hp2Dpert}
\end{equation}
where $E_n$ is the transverse quantization energy in the $\chi^{(0)}_n(y)$ state (in the absence of a magnetic field),
\begin{equation}
	V(y) = \frac{m\omega_\text{c}^2}{2} (y^2-2 y_0y).
	\label{eq:Hp2Dpert1}
\end{equation}

Averaging over wave functions~(\ref{eq:transversalMode}) gives
\begin{equation}
	\langle \chi^{(0)}_n | V(y) | \chi^{(0)}_n \rangle=
	\frac{m\omega_\text{c}^2W^2(x)}{24} \left(1- \frac{6}{\pi^2n^2} \right).
	\label{eq:magnenergyweak}
\end{equation}
This addition to the transverse quantization energy shifts the steps and increases the plateau width~\cite{Glazman89b}. An even more substantial effect is the narrowing of the step width due to a decrease in the curvature of the effective scattering potential ${\tilde\Omega}_n^2 = \Omega_n^2 - \Omega_{\s H}^2$, where
\begin{equation}
	\Omega_{\s H}^2=
	\frac{\omega_\text{c}^2 W^2(0)}{12L^2}
	\left( 1- \frac{6}{\pi^2n^2} \right).
	\label{eq:omega}
\end{equation}
Both these effects improve quantization. But another contribution of the same order in the magnetic field exists, which can lead to the step broadening~\cite{Glazman89b}. Taking the kinetic energy variation into account in the second-order perturbation theory [with a term linear in the magnetic field in Eq.~(\ref{eq:Hp2DHam1})] complicates the picture: for the first step, it always provides a further increase in the quantization, whereas for the next steps, the effect can be the opposite due to the possible change in sign in the second-order perturbation theory.

At the same time, the magnetic field effect in Ref.~\cite{Buttiker90b} resulted only in the improvement of quantization. This difference can be caused by the use of different QPC models and the different choice of parameters (although the improvement of quantization in a magnetic field is intuitively the most natural result).

In the case of {\it a strong magnetic field and a steep wall}, transverse modes can change considerably for large $k$ and $B$. Such a situation for a magnetic field in a potential box is shown in Fig.~\hyperref[fig:Hp2D]{\ref{fig:Hp2D}(c)}, where the parabola of the quadratic potential is strongly displaced with respect to the center. The states formed at the boundaries, which are called edge states, play a key role in transport in the IQHE regime, when the magnetic film is so strong that only several modes contribute to the transport even in a wide contact, which are in fact edge states.

In a strong magnetic field for a smooth potential $U''(y_0)/m \ll \omega_\text{c}$, the wave function of the edge states is not deformed, $U(y)$ can be replaced with the potential $U(y_0)$, and the energy levels have the form
\begin{equation}
	E_n(k) =
	\hbar \omega_\text{c} \biggl( n + \frac{1}{2} \biggr) + U(y_0).
	\label{eq:Hp2DLL2}
\end{equation}
An exact solution can be obtained for parabolic walls, $U(y) = m\omega_0^2 y^2 / 2$, when the equation takes the form
\begin{equation}
	\chi''_n(y) + \frac{2m}{\hbar^2} \biggl\{
		E_n(k) - \frac{m}{2} \Bigl[
			\omega_0^2 y^2 + \omega_\text{c}^2 (y-y_0)^2
		\Bigr]
	\biggr\} \chi_n(y) = 0.
	\label{eq:Hp2Dpar}
\end{equation}
Introducing the new variables
\begin{align}
	& \tilde{\omega}^2 = \omega_\text{c}^2 + \omega_0^2,
	\\
	& \tilde{y}_0 = y_0 \frac{\omega_\text{c}^2}{\omega_\text{c}^2 + \omega_0^2},
	\\
	& \tilde{E}_n(k)
	= E_n(k) - \frac{
		m\omega_\text{c}^2 \omega_0^2
	}{
		2(\omega_\text{c}^2 + \omega_0^2)
	} y_0^2,
\end{align}
we can reduce Eq.~(\ref{eq:Hp2Dpar}) to the equation of a harmonic oscillator
\begin{equation}
	\chi_n''(y) + \frac{2m}{\hbar^2} \left[
		\tilde{E}_n (k) - \frac{m {\tilde \omega}^2 (y-{\tilde y}_0)^2}{2}
	\right] \chi_n(y) = 0
	\label{eq:Hp2Dusual0}
\end{equation}
with the spectrum
\begin{equation}
	\tilde{E}_n(k) = \hbar \tilde{\omega} \biggl( n + \frac12 \biggr).
	\label{eq:Hp2Dusual}
\end{equation}
New variable ${\tilde y}_0$ indicates the edge state position.
Returning to the usual variables, we obtain
\begin{equation}
	E_n(k)
	= \hbar \sqrt{\omega_\text{c}^2 + \omega_0^2} \left( n + \frac12 \right) +
	\frac{m\omega_\text{c}^2 \omega_0^2}{2(\omega_\text{c}^2 + \omega_0^2)} y_0^2,
	\label{eq:Hp2Den}
\end{equation}
where the dependence on $k$ enters through $y_0 = \ell_B^2 k$. We fix the energy $E$ and express ${\tilde y}_0$ in terms of $E$ and $n$:
\begin{equation}
	{\tilde y}_0^2 = 
	\frac{
		2 \omega_\text{c}^2
	}{
		m\omega_0^2(\omega_\text{c}^2+\omega_0^2)
	}
	\biggl[ E - \hbar\sqrt{\omega_\text{c}^2 + \omega_0^2} 
		\biggl( n + \frac{1}{2}\biggr)
	\biggr].
	\label{eq:Hp2Dedge}
\end{equation}
It follows from Eq.~(\ref{eq:Hp2Dedge}) that the higher the energy $E$ is, the closer the edge state is to the sample boundary. The total excess nonequilibrium current in the IQHE mode is transferred just by these states. This is explained by the fact that the edge-state energy is higher than the energy of bulk states, and hence edge states are typically the first to touch the Fermi surface (level), making the contribution to transport. It is important that, as in the case of one-dimensional motion without a magnetic field, each channel (each Landau level in the strong-field approximation) carries the same current $i_0 = e/h$ per energy interval per spin, see expression~(\ref{eq:currentperchannel}). This occurs because the current in the presence of a magnetic field can still be expressed in terms of the velocity, which cancels the velocity from the density of states, as in the normal case.

By analyzing the behavior of transverse modes, which are converted to edge states as the magnetic field increases, we can see that the quantization of the conductance both at the QPC and in the IQHE has the same nature in a certain sense, namely, the switching on of new modes when changing parameters (width or magnetic field) upon passage from plateau to plateau through steps. As regards the transport on a plateau without reflection, this property is caused in the case of QPCs by motion without reflection in the semiclassical potential, while in the case of the IQHE, it is caused by a similar phenomenon of the suppression of scattering from boundary to boundary, because the edge states with opposite momenta are located near the opposite walls.

In pure conductors, the picture described above is clear and raises no doubts. In dirty conductors, the picture is more complicated and is commonly described by using quite different approaches. However, a similarity can be seen to exist between these pictures, which we discuss in Sec.~\ref{sec:condindisord}, where we consider the transmission distribution function in dirty conductors.

The quantum Hall effect is an intricate and diverse phenomenon deserving a special discussion that is outside the scope of our review. Here, we only wanted to show that even a simple analysis of edge states based on the Landauer approach can give useful information. A more detailed analysis by means of scattering matrices was performed in Ref.~\cite{Buttiker88c} (after papers~\cite{Laughlin81,Halperin82}, in which the nature of the IQHE was considered by using edge states). The theoretical and experimental aspects of edge states are discussed in detail in review~\cite{Devyatov07}.


\section{Aharonov-Bohm effect}
\label{sec:AA-effect}

\begin{figure}[tb]
	\includegraphics[width=4.5cm]{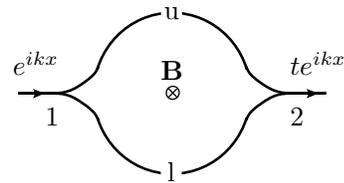}
	\caption{
Aharonov-Bohm effect. A quantum conductor with one open channel (one propagating mode) is connected at point~1 to a single-mode ring. The ring is connected at point~2 to another quantum single-mode conductor.
	}
	\label{fig:AA_effect}
\end{figure}

Let us consider now the Aharonov-Bohm effect~\cite{Aharonov59}~--- one of the most interesting effects, where the nonlocality of quantum mechanics manifests itself. The Aharonov-Bohm effect has been observed in mesoscopic quantum conductors~\cite{Yacoby95}. Let a quantum wire (Fig.~\ref{fig:AA_effect}) with one open channel (one propagating mode) be connected at point 1 to a single-mode ring connected at point 2 to another quantum single-mode conductor. We study the transmission probability $T$ from one conductor (point 1) to another (point 2) in the case where a magnetic flux $\Phi$ penetrates in the ring, for example, in a weak homogeneous magnetic field $\bf B$ perpendicular to ring's plane.

We calculate the scattering amplitude using the Feynman path integral approach~\cite{Feynman42,Feynman48} The total scattering amplitude can then be found by summing the amplitudes of transmission of a particle from one conductor to another through the ring over all possible paths. The shortest propagation paths are lying through the upper (u) or lower (l) part of the ring. We assume for simplicity that the ring and the contacts are symmetric, and hence, for ${\bf B} = 0$, the transmission amplitudes $t_{12}^{\rm u(l)}$ for the particle along these paths are the same and equal to $t_{12}$. If the magnetic field $\bf B$ is nonzero, the particle acquires different phases after propagation through the upper and lower parts of the ring:
\begin{align}
	& t_{12}^{\rm (u)} = t_{12}e^{i\chi_1}, \quad
	t_{12}^{\rm (l)} = t_{12}e^{i\chi_2}, \\
	& \chi_1 = \frac e{c\hbar} \int\limits_{\rm u} {\bf A}d{\bf l}, \quad
	\chi_2=\frac e{c\hbar} \int\limits_{\rm l} {\bf A}d{\bf l},
\end{align}
where $\bf A$ is the vector potential and the integral is taken along the particle path between points 1 and 2. The difference between these phases can be expressed in terms of the ratio of the magnetic field flux $\Phi = \oint {\bf A} d{\bf l}$ through the ring to the magnetic flux quantum $\Phi_0 = hc / e$,
\begin{equation}
	\chi = \chi_1-\chi_2 =
	\frac e{c\hbar} \oint{\bf A}d{\bf l} =
	2 \pi \frac\Phi{\Phi_0}.
	\label{eq:Phi}
\end{equation}
Then the total transmission amplitudes and transmission probability are
\begin{align}\label{eq:t_total}
	& \tilde t=t_{12}e^{i(\chi_1+{\chi_2})/2} \big[e^{i\chi/2}+e^{-i\chi/2} \big], \\
	& \tilde t'=t_{12}e^{-i(\chi_1+{\chi_2})/2} \big[e^{i\chi/2}+e^{-i\chi/2}\big], \\
	& T=|\tilde t|^2=2 T_{12}+2 T_{12}\cos(\chi).
	\label{eq:T_total}
\end{align}
We note that the amplitude $\tilde t'$ of scattering from left to right, which can be found by using rule~(\ref{eq:SB}) (see Appx.~\ref{sec:timerev}) from the expression for $\tilde t$, is not equal to $\tilde t$ in general, unlike that in problems with the symmetric ($t=t'$) scattering matrix considered in Secs.~\ref{sec:landauer}--\ref{sec:quantcontacts}. Here, this symmetry is broken [but the transmission probabilities are still equal because the scattering problem is effectively one-dimensional (see footnote~\ref{eq:note:prob} in Sec.~\ref{sec:2res})].

The periodic dependence of the transmission probability $T$ on the magnetic field represents the Aharonov-Bohm effect. When the system shown in Fig.~\ref{fig:AA_effect} is connected at the right and left to electron reservoirs, the conductance of such a contact is described by the Landauer formula $G = G_0 T$. If the motion of a particle were noncoherent, we would obtain $T=2T_{12}$. Transmission becomes zero $T = 0$ for $\chi = \pi + 2\pi n$, $n = 0,\pm 1, \pm 2, \ldots$ due to interference in the system. The vanishing of the transmission indicates the presence of the so-called Fano resonance~\cite{Fano61}, which appears because of hybridization of the continuous and discrete spectra.\footnote{The transparency never vanishes in usual purely one-dimensional problems of scattering on finite potentials.} In the case $\chi = 2\pi n$, the conductance is twice that in the noncoherent case.

\begin{figure}[tb]
	\includegraphics[width=3.5cm]{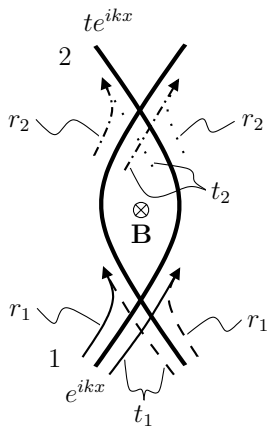}
	\caption{Reflectionless scattering in a ``four-tail'' figure.}
	\label{fig:AA_effect2}
\end{figure}

We note that we did not take all the contributions to the scattering amplitude into account in Eq.~(\ref{eq:t_total}), and considering only two amplitudes is incorrect in general case. A particle can tunnel at point 1 to the ring from the left conductor, pass several times along the ring, and only then enter the right conductor. Multiple reflections typical of a Fabry-Per\'ot interferometer can be avoided by using a Mach-Zehnder interferometer in which only two amplitudes interfere, see Fig.~\ref{fig:AA_effect2}.\footnote{Since the geometry of such an interferometer is not one-dimensional (four contacts exist), the transmission probabilities are no longer symmetric in the contact indices in a nonzero magnetic field.} In this case, generally speaking, it is necessary to fabricate a reflectionless scatterer (``beamsplitter''). This problem is quite complicated but can be solved under quantum Hall effect conditions, e.g., see Ref.~\cite{Ji03}.


\section{Double barrier: the Fabry-Per\'ot interferometer}
\label{sec:doubleBarrier}

Scattering on the real potential in meso- and nano-quasi-one-dimensional conductors can be simulated by scattering on the potentials for which the problem can be solved exactly. We considered such example in Sec.~\ref{sec:condsmearing}, where we used the Kemble formula for scattering on the quadratic potential. Another example (perhaps most frequently used) involves the Dirac delta function $\delta(x)$. The potential can be written in the form
\begin{equation}
	U(x) = \alpha\delta(x)
	\label{eq:DiracPot}
\end{equation}
if its range is shorter than the particle wavelength $\lambda$. In the case of metals, such a description is usually valid for boundaries between different materials. However, in quasi-one-dimensional conductors, where the effective wavelength can considerably exceed 1~nm, the applicability of the $\delta$-function description broadens and this approximation can sometimes be used even for QPCs.

The scattering amplitudes {of the potential~(\ref{eq:DiracPot})} are given by the known expressions
\begin{gather}
	t = t' = \frac{1}{1 + iZ},
	\label{eq:tDelta} \\ 
	r = r' = \frac{-iZ}{1 + iZ},
	\label{eq:rDelta}
\end{gather}
where
\begin{equation}
	Z = m \alpha / \hbar^2 k.
\end{equation}

\subsection{Double delta barrier}

Another very important case, which we will consider several times further, is scattering on the double barrier. The double barrier is a structure with two scatterers connected in series. Such a scatterer can successfully simulate transport through a quantum dot, for example, in a carbon nanotube. In the case of coherent transport, interference occurs due to multiple scatterings and resonances appear in the transmission amplitude and the transparency of the double barrier. Each of the barriers can be typically described by the $\delta$-function potential~(\ref{eq:DiracPot}). The transmission and refection amplitudes of this structure can be calculated in a standard way by matching the wave functions on different sides of the scatterers. However, let us consider a more illustrative calculation method based on an analogy with the optical Fabry-Per\'ot interferometer, which also gives an exact result. The method involves the summation of all possible semiclassical trajectories with successive reflections, along which the particle can propagate (the method can be formally substantiated by integrating over Feynman trajectories). In addition, this method allows to account for the phase fluctuations gained when moving between barriers.

\begin{figure}[tb]
	\includegraphics[width=8.0cm]{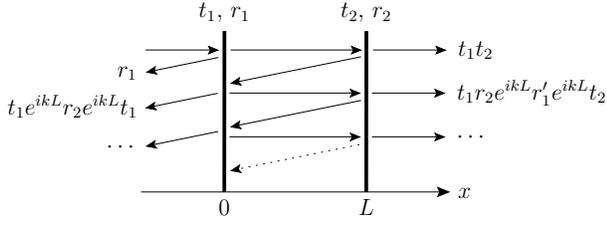}
	\caption{
A double barrier (two scattering potentials in series) can be regarded as an analogue of the Fabry-Per\'ot interferometer known in optics.
	}
	\label{fig:interferometer}
\end{figure}

Let us assume that the left scatterer has the transmission and reflection amplitudes $t_1$ and $r_1$, and the right scatterer has the corresponding amplitudes $t_2$ and $r_2$; the distance between the barriers is $L$. All possible paths of the particle are shown in Fig.~\ref{fig:interferometer}. The transmission amplitude is determined by the sum of the series
\begin{equation}
	t = t_2 t_1 + t_2 [r_1' (r_2 e^{2i kL})] t_1 + 
	t_2 [r_1' (r_2 e^{2i kL})]^2 t_1 + \ldots,
	\label{eq:ta0}
\end{equation}
where the first term corresponds to the trajectory passing through the two barriers without reflection, the second term corresponds to the trajectory with two reflections forming one loop, and so on. The summation of the (geometrical) series gives
\begin{equation}
	t = t' = \frac{t_1 t_2}{1 - r_1' r_2 e^{2ikL}}.
	\label{eq:ta}
\end{equation}
We recall that $t_1 = t_1'$, $t_2 = t_2'$, and $t' = t$ if the Hamiltonian of the system is invariant under time reversal (in the general case, $r \neq r'$ in the absence of spatial symmetry).

Similarly, we can sum over trajectories for the backward reflection amplitude:
\begin{equation}
	r = r_1 + t_1 r_2 e^{2ikL}t_1
	+ t_1 (r_2 e^{2ikL} r_1) r_2 e^{2i k L} t_1 + \ldots,
	\label{eq:ra0}
\end{equation}
which gives
\begin{equation}
	r = r_1 + \frac{t_1t_1 r_2 e^{i2kL}}{1 - r_1 r_2 e^{2ikL}} =
	\frac{r_1+r_2 e^{2ikL} (t_1t_1-r_1^2)}{1-r_1r_2 e^{2ikL}}.
	\label{eq:ra1}
\end{equation}

\begin{figure}[tb]
	\includegraphics[width=6.0cm]{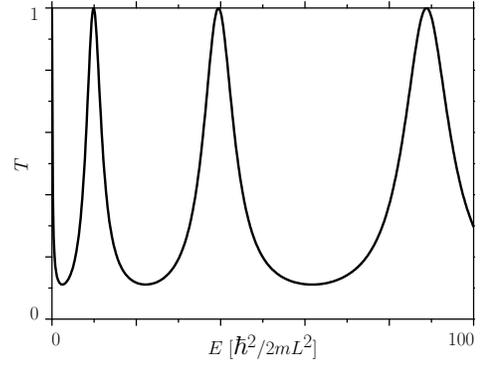}
	\caption{
Transmission probability $T$ as a function of energy; $T_1 = T_2 = 1/2$ and $\chi^r = 0$. (Figure from Ref.~\cite{Chtchelkatchev11Book}.)
	}
	\label{fig:double_bar_res}
\end{figure}

The transparency of the whole system is
\begin{equation}
	T \equiv |t|^2 =
	\frac{T_1T_2}{1 + R_1 R_2 - 2\sqrt{R_1 R_2} \cos(\theta)},
	\label{eq:Taa_T}
\end{equation}
where $\theta = 2kL + 2 \chi^r$, $T_i = |t_i|^2$ and $R_i = |r_i|^2$ are the transmission and reflection probabilities for barriers, and $\chi^r = (\chi^{r'}_1 + \chi^{r}_2) / 2$ (for example, $\chi^{r'}_1 \equiv \arg r_1'$). Relation~(\ref{eq:Taa_T}) is illustrated in Fig.~\ref{fig:double_bar_res}. The total transparency $T(E)$ reaches a maximum at $\theta = 2\pi n$, $n = 1, 2, \ldots$, which corresponds to wave vectors $k_n = (\pi n - \chi^r) / L$ with the energies
\begin{equation}
	E_n = \frac{\hbar^2}{2mL^2} \bigl( \pi n - \chi^r \bigr)^2.
	\label{eq:Taa_En}
\end{equation}
The maximum value of $T(E)$,
\begin{equation}
	T_{\rm max} = \frac{T_1 T_2}{(1 - \sqrt{R_1R_2})^2}
	\label{eq:Taa_Tmax}
\end{equation}
is equal to unity for $T_1 = T_2$ and to $4 T_1 T_2 / (T_1+ T_2)^2$ for $T_i \ll 1$, $i = 1, 2$.\footnote{We assume that $T_1$, $T_2$, and $\chi^r$ are virtually independent of energy at scales of the order of the distance between resonances.}

The obtained transmission probability demonstrates an important property consisting in ideal resonances $T = 1$ occurring for a symmetric barrier with $T_1 = T_2$. Therefore, the two-barrier structure becomes ideally transparent at resonance (neglecting the phase gain) even for very strong scattering from each of the barriers. This effect, explained by interference, can act as an indicator of particle movement coherence. If coherence is absent, the transmission probability is given by the product of probabilities $T \approx T_1T_2$, which can be much smaller than unity. The measurement of $T$ is used for the experimental verification of the coherence degree. Note that for $T < 1$ this method cannot determine whether the system is coherent or not. On the contrary, the case of $T = 1$ definitely indicates that the system is coherent.

Outside the resonance (in the destructive interference region), we have
\begin{equation}
	T = T_{\rm min} =
	\frac{T_1 T_2}{(1 + \sqrt{R_1R_2})^2}.
	\label{eq:ta1}
\end{equation}
For $T_1, T_2 \ll 1$, we obtain $T \approx T_1 T_2 / 4$, and therefore the destructive interference is stronger than the dephasing effect, which will be discussed in more detail below.

We define the spacing between resonances as
\begin{align}
	\Delta_n
	& = \frac{|E_{n+1}-E_{n-1}|}{2}
	= 2\pi \frac{\hbar^2 |\pi n - \chi^r|}{2mL^2} \nonumber \\
	& = \frac{\pi\hbar v_n}{L}
	= \pi\hbar\nu_n,
	\label{eq:Taa_Sn}
\end{align}
where $v_n = (dE / \hbar dk) |_{E = E_n}$ is the velocity of an electron moving between the potential walls of the double-barrier potential. We note that the resonance energies are not equidistant and the definition $\Delta_n = |\partial E_n / \partial n|$ gives the same result. The quantity
\begin{equation}
	\nu_n = v_n / L
	\label{eq:nu}
\end{equation}
has the dimension of frequency and its physical meaning corresponds to the number of electron attempts to leave the trap between potential barriers per unit time.

We analyze expression~(\ref{eq:Taa_T}) near the resonance energy~$E_n$ in Eq.~(\ref{eq:Taa_En}). For this purpose, we expand the cosine in the denominator in the right-hand side of Eq.~(\ref{eq:Taa_T}) to the second order in the energy deviation from the resonance $\delta E_n = E - E_n$:
\begin{align}
	& \cos\theta \approx 1-\frac12\left(\frac{d\theta}{dE_n}\right)^2(\delta E_n)^2, \quad
	\nonumber \\
	& \frac{d\theta}{dE_n} = \frac{d\theta}{dE}\biggl|_{E=E_n}=\frac{1}{\hbar \nu_n}.
	\label{eq:cos}
\end{align}
Substituting this expression in Eq.~(\ref{eq:Taa_T}), we find that the transmission probability near the $n$th resonance can be approximated by a Lorentzian function (the Breit-Wigner approximation~\cite{Landau04BookV3}):
\begin{equation}
	T(E\sim E_n)\approx T_{\rs BW}
	= \frac{\gamma_n^2}{\gamma_n^2 + (\delta E_n)^2} \, T_{\rm max}.
	\label{eq:Taa_T_G}
\end{equation}
Here, we define the resonance half-width as
\begin{equation}
	\gamma_n = \frac12{\frac{dE}{d\theta}\biggl|_{E=E_n}} \,
	\frac{1-\sqrt{R_1R_2}}{\sqrt[4]{R_1R_2}} =
	\frac{\hbar\nu_n(1-\sqrt{R_1R_2})}{2\sqrt[4]{R_1R_2}}.
	\label{eq:Taa_G1}
\end{equation}
The transmission probability $T$ can be approximated by a Lorentzian function for all energies:
\begin{equation}
	T(E) \approx
	\sum\limits_n T_{\rs BW}(\delta E_n).
	\label{eq:1941}
\end{equation}
The relative error of the approximation~(\ref{eq:1941}) does not exceed a few percent, even for $T_1, T_2 \lesssim 1/2$. For example, in Fig.~\ref{fig:double_bar_res}, if we additionally plot approximation~(\ref{eq:1941}) with the same parameters that determine the plot of $T$ shown in this figure, these plots coincide so perfectly that the difference is visually indistinguishable~\cite{Chtchelkatchev11Book}.

For a strong resonance $T_1, T_2 \ll 1$ and simpler expressions are often used. We introduce the partial resonance widths
\begin{equation}
	\Gamma_n^{(i)} =
	\frac{d\theta}{dE_n} T_i =
	\hbar \nu_n T_i,
	\label{eq:Gamma_p}
\end{equation}
where $i = 1, 2$. The ratio $\Gamma_n^{(i)} / \hbar $ gives the number of successful particle attempts to leave the trap of double-barrier potential per unit time. Expanding the right-hand side of Eq.~(\ref{eq:Taa_T_G}) in small probabilities of transitions through potential walls, we find that
\begin{align}
	T(E) \approx \frac{
		\Gamma^{(1)}_n\Gamma^{(2)}_n
	}{
		\Gamma^{(1)}_n+\Gamma^{(2)}_n
	} A_n(E-E_n),
	\label{eq:Taa_G1a} \\
	A_n(\epsilon) = \frac{\Gamma_n}{\epsilon^2+(\Gamma_n/2)^2}
	\label{eq:Taa_G1b}
\end{align}
near the resonance, where $\Gamma_n = \Gamma^{(1)}_n + \Gamma^{(2)}_n$ is the total resonance width and $A$ is the Lorentzian function. Then $\gamma_n = \hbar\nu_n (T_1+T_2) / 4 = \Gamma_n / 4$. We see that resonances become sharper as $T_1$ and $T_2$ decrease.

We next discuss the dephasing effects mentioned above. We rewrite expression~(\ref{eq:ta0}) by adding phase factors with random phases $\alpha_i$, $i=1, 2, \ldots$, to each term:
\begin{multline}
	t = t_2t_1e^{i\alpha_1} + t_2[ r_1' (r_2 e^{2i kL})] t_1e^{i\alpha_2} + \\
	+ t_2 [ r_1' (r_2 e^{2i kL})]^2 t_1e^{i\alpha_3} + \ldots
	\label{eq:ta00}
 \end{multline}
These phases can appear due to time fluctuations of the electrostatic potential in the quantum dot (i.e., in the region between barriers), which should be taken into account in multiple reflections in the resonance potential, which are described by the terms in the sum.\footnote{Electron transport in the presence of time-dependent fields can also be described by means of scattering matrices, which was discussed, e.g., in Refs.~\cite{Moskalets02,Moskalets04,Lesovik94b,Levitov96,LevitovIvanov93,LevitovIvanov97}. We do not consider this question because of the limited scope of our review.}

We now find the transmission probability by averaging it over phase realizations $\alpha_i$, assuming that the $\alpha_i$ are independent random quantities with dispersion much more than $2\pi$. Such a model corresponds to the assumption that the dephasing length is smaller than the distance between barriers. Then
\begin{multline}
	\langle T\rangle_\alpha = 
	|t_2t_1|^2 + |t_2[ r_1' (r_2 e^{2i kL})] t_1|^2 + \\
	+ |t_2 [ r_1' (r_2 e^{2i kL})]^2 t_1|^2 + \ldots =
	\frac{T_1 T_2}{1- R_1 R_2}.
	\label{eq:T_average}
\end{multline}
It is interesting to compare noncoherent tunneling described by Eq.~(\ref{eq:T_average}) with the so-called sequential tunneling~\cite{Datta95,Dem00}. Sequential tunneling is usually considered in a situation where there is the quasi-equilibrium distribution in the quantum dot. In this case, the total resistance equals to the sum of two barriers total resistances,
\begin{equation}
	\mathcal{R}_{\rm st}
	= \frac{h}{2e^2}\left(\frac{1}{T_1} + \frac{1}{T_2}\right)
	= \frac{h}{2e^2} \frac{T_1+T_2}{T_1T_2}.
	\label{eq:Resistance2barSeq}
\end{equation}
In the case of noncoherent tunneling (i.e., for the phase-averaged transparency considered above), the resistance
\begin{equation}
	\mathcal{R}
	= \frac{h}{2e^2} \frac{T_1+T_2- T_1T_2 }{T_1T_2}
	= \frac{h}{2e^2} \frac{T_1+T_2}{T_1T_2} - \frac{h}{2e^2}
	\label{eq:Resistance2barDecoh}
\end{equation}
is smaller than $\mathcal{R}_{\rm st}$ by the contact resistance formed by the internal modes of the region between barriers.

An interesting question arises about when a simple summation of Landauer resistances can be used? For example, if we assume that the relaxation in momentum with the same propagation direction occurs inside the quantum dot such that the independent Fermi surfaces (points) appear for each direction, then there is no additional voltage drop and the Landauer voltage can be summed (assuming that the delta function provides energy-independent scattering). We note that the reasoning about the summation of Landauer resistances when averaging over scattering amplitudes was used in different variants in the scaling theory of localization in well known papers~\cite{Landauer70,Anderson80}.\footnote{The transparency of a conductor containing many scatterers with random parameters was studied more accurately in Ref.~\cite{Dorokhov82}, where a transfer matrix was used to describe the effect produced by the addition of a new scatterer. The total transparency was found to behave like a particle randomly diffusing in the parameter space.}

If $T_1, T_2 \ll 1$, then
\begin{equation}
	\langle T \rangle_\alpha \approx \frac{T_1 T_2}{T_1 + T_2}.
\end{equation}
We note that the destructive interference [see Eq.~(\ref{eq:ta1})] suppresses $T$ 
much more efficiently than the phase decoherence:
in the former case, $T \approx T_1 T_2 / 4$, while in the latter case, $T \approx T_1 / 2$ for $T_1 \sim T_2 \ll 1$.

\subsection{Transport properties of contacts with the resonance potential}

We consider a quantum contact between two electron reservoirs in which a resonance potential, similar to that considered in Sec.~\ref{sec:doubleBarrier}, serves as a scattering potential, see Fig.~\ref{fig:interferometer}. For simplicity, we assume that only one channel is open, and hence Eq.~(\ref{eq:Landauer_g}) reduces to
\begin{equation}
	I = \frac{2e}{h} \int\limits_{0}^\infty dE \left[ f_1(E)- f_2(E) \right] T(E).
	\label{eq:Landauer_g1}
\end{equation}

\begin{figure}[tb]
	\includegraphics[width=7.8cm]{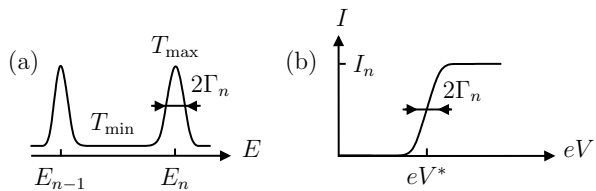}
	\caption{
(a)~Energy dependence of the transmission $T$. (b)~Current $I$ as a function of the bias voltage $V$. In the symmetric case, each resonance gives the current increment of $I_n = (2e / \hbar) \pi \Gamma_n$.
	}
	\label{fig:double_bar_t}
\end{figure}

We also assume that $T_1, T_2 \ll 1$, and therefore the Breit-Wigner approximation in the form of~(\ref{eq:Taa_G1a}) can be used. Then
\begin{gather}
	T(E) \approx \sum\limits_n \frac{
		\Gamma^{(1)}_n\Gamma^{(2)}_n
	}{
		\Gamma^{(1)}_n+\Gamma^{(2)}_n
	} A_n(E-E_n),
	\label{eq:Taa_G1aa} \\
	A_n(\epsilon) = \frac{\Gamma_n}{\epsilon^2+(\Gamma_n/2)^2},
	\label{eq:Taa_G1bb}
\end{gather}
where $\Gamma_n^{(i)}$ are the partial widths of resonances ($i = 1, 2$) and $\Gamma_n = \Gamma_n^{(1)} + \Gamma_n^{(2)}$ is the total width of the resonance [Fig.~\hyperref[fig:double_bar_t]{\ref{fig:double_bar_t}(a)}]. Substituting Eq.~(\ref{eq:Taa_G1aa}) into Eq.~(\ref{eq:Landauer_g1}), we find (for the temperature $\Theta = 0$)
\begin{gather}
	I = \frac{2e}{h}\sum\limits_n
	\frac{
		\Gamma^{(1)}_n\Gamma^{(2)}_n
	}{
		\Gamma^{(1)}_n+\Gamma^{(2)}_n
	}
	\int\limits_{E_1^\perp}^V dE \, A_n(E-E_n),
	\label{eq:I_res}
\end{gather}
where we introduce the superscript ``$\perp$'' at the transverse quantization energy $E_1^\perp$ in the constriction to distinguish it from the resonance energy of the scattering potential.

We assume that the energy interval $E_n \in [E_1^\perp,V]$ contains several transmission probability resonances. Then, according to Eq.~(\ref{eq:I_res}), the contribution from each of them to the current is
\begin{gather}
	I_n = \frac{4e}{\hbar} \pi
	\frac{
		\Gamma^{(1)}_n\Gamma^{(2)}_n
	}{
		\Gamma^{(1)}_n+\Gamma^{(2)}_n
	}
	\label{eq:I_res1}
\end{gather}
(we took into account that $\int_{-\infty}^\infty A_n(E) dE = 2\pi$). In the symmetric case, we obtain the simplest expression
\begin{gather}
	I_n = \frac{2e}{\hbar} \pi \Gamma_n.
	\label{eq:I_res1symm}
\end{gather}
In this case, the $I$-$V$ characteristic has a typical step-like profile as depicted in Fig.~\hyperref[fig:double_bar_t]{\ref{fig:double_bar_t}(b)}.


\section{Conductance in dirty conductors}
\label{sec:condindisord}

\begin{figure}[tb]
	\includegraphics[width=7.0cm]{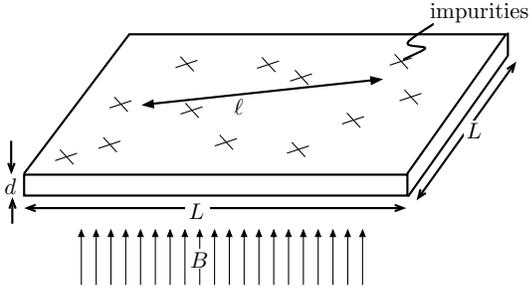}
	\caption{
Two-dimensional dirty conductor. Crosses indicate the positions of impurities fluctuating from sample to sample, $\ell$ is the mean free path. A magnetic field can be applied perpendicularly to the sample.
	}
	\label{fig:dirtySample}
\end{figure}

We now consider a multichannel dirty conductor in which electrons diffuse from one boundary to another, see Fig.~\ref{fig:dirtySample}. Some important parameters of such a sample at low temperatures, when all inelastic processes can be neglected, surprisingly resemble those of a QPC and the double-barrier system considered in Secs.~\ref{sec:quantcontacts} and \ref{sec:doubleBarrier}.

\subsection{Mesoscopic conductance fluctuations}

The question about strong fluctuations of the resistance of such mesoscopic conductors was first considered in Azbel's work~\cite{Azbel85}.\footnote{The term ``mesoscopic'' was used for such systems starting with this work.}
Quantitative studies of mesoscopic quantum effects in transport were initiated in theoretical papers by Al'tshuler~\cite{Altshuler85} and Lee and Stone~\cite{Lee85}, where large fluctuations of the conductance $G$ in a two-dimensional dirty film were predicted even for large (but still coherent) samples. A standard quantity characterizing fluctuations of the conductance from sample to sample is the mean-square deviation
\begin{equation}
	\langle \delta G ^2 \rangle_\text{im} =
	\bigl\langle (G - \langle G \rangle_\text{im})^2 \bigr\rangle_\text{im},
	\label{eq:condFluct}
\end{equation}
where the subscript ``im'' means averaging over all the possible variants of the location of impurities, and the mean conductance is
\begin{equation}
	\langle G \rangle_\text{im} = d L \sigma / L,
	\label{eq:averageCond}
\end{equation}
where $\sigma$ is the conductivity. The authors of~\cite{Altshuler85,Lee85} found that the standard deviation
\begin{equation}
	\delta G = \sqrt{ \langle
	\delta G ^2 \rangle_\text{im} } \approx G_0
	\label{eq:condFluctSol}
\end{equation}
is universal (i.e., is independent of the disorder details) and is approximately equal to the conductance quantum $G_0 = e^2/h$. The relative fluctuations
\begin{equation}
	\frac{\delta G}{\langle G \rangle_\text{im}} \approx
	\frac{e^2}{h} \frac{1}{d \sigma}
	\label{eq:condFluct2}
\end{equation}
are independent of the sample size $L$. This is a surprising result because it was usually assumed that at large scales, the conductivity even of quantum conductors is a self-averaging quantity, and its relative fluctuations decrease upon increasing the sample size. But this is not the case for a coherent quantum conductor. In addition, Lee and Stone~\cite{Lee85} as well as Al'tshuler and Khmel'nitskii~\cite{Altshuler85b} described mesoscopic fluctuations as a function of the applied magnetic field and other parameters. The fluctuations of the conductance due to the changes in magnetic field can be qualitatively explained as follows:\footnote{The explanation by D.E.~Khmel'nitskii.} The conductance is proportional to the probability $P_{\rm a \to b}$ that an electron starting from one side of the conductor will reach its opposite side. Using the path integral formalism, the probability can be represented as the square of a sum of amplitudes over all possible paths,
\begin{figure}[tb]
	\includegraphics[width=2.0cm]{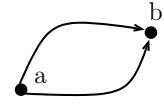}
	\caption{
Interference between two trajectories contributing to the conductance. A magnetic field induces the Aharonov-Bohm phase, which changes the relative phase between the trajectories. The sensitivity to the magnetic field (i.e., a change in the magnetic field resulting in a change in the conductance by a value of the order of $G_0$) is specified by the magnetic flux quantum $\Phi_0 = hc / e$ per sample area $L^2$.
	}
	\label{fig:Pab}
\end{figure}
\begin{equation}
	P_{\rm a \to b} =
	|A_1 + A_2|^2 =
	|A_1|^2 + |A_2|^2 + A_1 A_2^* + A_1^* A_2.
	\label{eq:condFluct3}
\end{equation}
For simplicity, here we consider only two semiclassical paths with amplitudes $A_1$ and $A_2$, see Fig.~\ref{fig:Pab}. The cross terms $A_1 A_2^*$ and $A_1^* A_2$ vanish in the mean probability $\langle P_{\rm a \to b} \rangle_{\rm im}$ due to averaging over the random phase (the exception is the contributions from paths or segments of paths repeating the motion backward and contributing to weakly localized corrections, which we do not consider here); the two probabilities are simply added, $\langle P_{\rm a \to b} \rangle_{\rm im} = |A_1|^2 + |A_2|^2 = P_1 + P_2$, and the interference terms vanish. In the calculation of the second moment,
\begin{align}
	\langle P_{\rm a \to b}^2 \rangle_\text{im}
	& \propto (P_1 + P_2)^2 + 2 |A_1|^2 |A_2|^2 \nonumber \\
	& = \langle P_{\rm a \to b} \rangle_\text{im}^2 + 2 P_1 P_2,
	\label{eq:condFluct4}
\end{align}
\begin{figure}[t]
	\includegraphics[width=5.6cm]{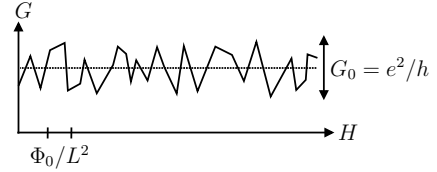}
	\caption{Conductance fluctuations as the magnetic field changes.}
	\label{fig:condFluct}
\end{figure}
\begin{figure}[b]
	\includegraphics[width=8.7cm]{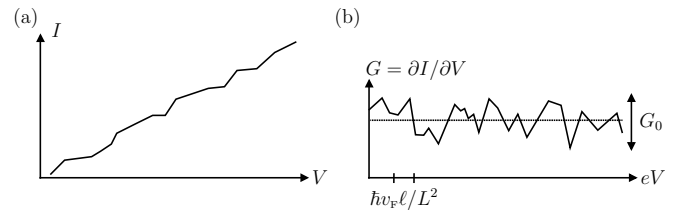}
	\caption{Conductance fluctuations as the voltage changes.}
	\label{fig:IVGV}
\end{figure}
the terms with $A_1 A_2^*$ and $A_1^* A_2$ also vanish after averaging. But additional term $2 |A_1|^2 |A_2|^2$ remains finite after averaging. The root-mean-square has the form
\begin{equation}
	\delta P_{\rm a \to b} =
	\sqrt{\langle (P_{\rm a \to b} - \langle P_{\rm a \to b}
	\rangle_\text{im} )^2 \rangle_\text{im}} = \sqrt{2 P_1 P_2}.
	\label{eq:condFluct5}
\end{equation}
If we now apply a weak magnetic field, the relative phases between all the paths change and the conductance changes accordingly. Thus, the conductance fluctuates upon changing the magnetic field in the same way as upon changing the random potential. Detailed calculations show that the fluctuation value is of the order of $G_0 = 2e^2/h$, see Fig.~\ref{fig:condFluct}. Similar fluctuations of the conductance also appear as functions of the Fermi energy (chemical potential). The characteristic energy scale at which fluctuations occur is determined by the inverse diffusion time in the sample. The phase increment on a typical path during the diffusion time is then $\delta k L \sim \pi$. Such fluctuations appear as voltage changes (Fig.~\ref{fig:IVGV})~\cite{Larkin85} and also in thermoelectric phenomena, see Sec.~\ref{sec:thermo}. Conductance fluctuations were observed in experiments~\cite{Petrashov87,Washburn88} (see also the results of subsequent experiments and the literature in Ref.~\cite{Ghosh00}).

We note that there is a possibility of some resonances existing in the transparency of dirty samples, as already discussed in the pioneering papers by Azbel~\cite{Azbel85}.

\subsection{The Dorokhov distribution function}
\label{sec:dorokhov}

Let us consider the problem of fluctuations from the standpoint of scattering matrices. The conductance represented in the basis of ``eigenchannels,'' which diagonalize the transmission matrix, has the form
\begin{equation}
	G =\frac{2e^2}{h}\sum\limits_n T_n.
	\label{eq:GN2}
\end{equation}
For a conductor with $N$ channels, Eq.~(\ref{eq:GN2}) can be written as
\begin{equation}
	G = N \frac{2 e^2}{h} \langle T \rangle,
	\label{eq:condQuantum}
\end{equation}
where $\langle T \rangle$ is the transparency averaged over all channels. The usual expression for conductivity is given by
\begin{equation}
	G = \frac{A}{L} \sigma,
	\label{eq:condClassical}
\end{equation}
where $\sigma = e^2 \nu D$ is the conductivity calculated from the Kubo formula~\cite{Kubo57,Greenwood58} at zero frequency, $\nu$ is the density of states on the Fermi surface, and $D = v_{\rs F} \ell / 3$ is the diffusion coefficient. Now Eq.~(\ref{eq:condClassical}) can be rewritten in the following way:
\begin{equation}
	G = \frac{2 e^2}{h} \frac{A k_{\rs F}^2}{\pi^2} \frac{\pi \ell}{3 L}.
	\label{eq:condClassical2}
\end{equation}
The number of channels in a wire can be estimated in the WKB approximation as $N = A k_{\rs F}^2 / \pi^2$ [i.e., one channel per the area $\pi^2/k_{\rs F}^2 = (\lambda_{\rs F}/2)^2$].

\begin{figure}[tb]
	\includegraphics[width=4.8cm]{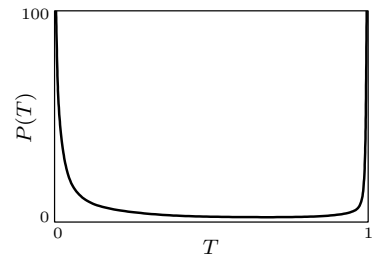}
	\caption{
Bimodal Dorokhov distribution function $P(T)$ with the most probable values of $T$ equal to 0 or 1.
	}
	\label{fig:dorokhov}
\end{figure}

Comparing expressions~(\ref{eq:condQuantum}) and (\ref{eq:condClassical2}), we obtain the mean transparency
\begin{equation}
	\langle T \rangle = \frac{\pi \ell}{3L},
	\label{eq:cond_comp}
\end{equation}
which, being proportional to $\ell /L$, tends to zero as $L \to \infty $. Does this mean that the typical transparency is approximately equal to $\ell / L$? This turns out to not be the case. A surprising property of transport in diffuse conductors is that for the eigenchannels, for which the problem is diagonal (channels are not mixed), the transparency is either very small or close to unity. In reality, most of the channels are virtually closed and $T \approx 0$, and only $n = N \ell / L $ channels are almost completely open with $T \approx 1$, providing the total conductivity. The distribution function for $T$, which was first calculated by Dorokhov, has the form~\cite{Dorokhov82,Dorokhov84}
\begin{equation}
	P(T) \propto \frac{1}{T \sqrt{1-T}},
	\label{eq:dorokhovProp}
\end{equation}
see Fig.~\ref{fig:dorokhov}. This is the general result for a quasi-one-dimensional conductor (a thick wire) with the total length $L \ll L_{\rm loc}$, where the localization length $L_{\rm loc}$ can be estimated as $L_{\rm loc} \approx N \ell$, i.e., the conductance becomes comparable to the quantum $G_0 = 2e^2/h$. Using the normalization determined by the mean conductance
\begin{equation}
	P(T)= \frac{\pi \ell}{6 L} \frac{1}{T \sqrt{1-T}}
	\label{eq:dorokhov}
\end{equation}
we obtain
\begin{equation}
	G=\frac{2 e^2}{h}N \int dT P(T)T =
	\frac{2 e^2}{h} \frac{A k_{\rs F}^2}{\pi^2} \frac{\pi \ell}{3 L}.
	\label{eq:dorokhovG}
\end{equation}
The situation resembles the case of a point contact with $n = N \ell / L $ open channels. The difference is that the eigenmodes for different energies and different magnetic field strengths in a sample are different combinations of usual propagating modes. The switching between conducting and nonconducting channels provides mesoscopic fluctuations of the conductance $\delta G \approx e^2/h$~\cite{Altshuler85}. The transparency distribution function is nontrivial. We can prove this by considering noises whose intensity is given by the sum $\sum_n T_n (1-T_n)$. Due to such nonlinearity in $T$, the result~\cite{Beenakker92b}
\begin{align}
	\Bigl\langle \sum\limits_n T_n(1-T_n) \Bigr\rangle
	& = N \int dT P(T)T(1-T) \nonumber \\
	& = \frac{1}{3} \langle \sum\limits_n T_n \rangle
\end{align}
contains additional information on the properties of $P(T)$, see Sec.~\ref{sec:noisedesc}.

As mentioned in Sec.~\ref{sec:edgestates}, the quantization of the conductance in QPCs and the IQHE in the ballistic case has a similar nature, namely, a relatively sharp switching on of new modes under a variation in the external parameters. The situation with the IQHE in dirty conductors is much more complicated and is usually described by completely different methods, in particular, by using field models~\cite{Prange89Book}. It is interesting that the authors of Ref.~\cite{Brouwer96} proved that the descriptions of a quasi-one-dimensional (multichannel) conductor in terms of a $\sigma$-model~\cite{Efetov83a,Efetov83b} and by the Dorokhov method (in particular, in the presence of a weak magnetic field) are equivalent. It seems that the analogy between a dirty conductor and a QPC described above is also valid in the presence of a strong magnetic field, and we can assume that the IQHE in dirty conductors is also provided by the presence of high-transparency eigenchannels (the number of open channels for the IQHE is obviously determined not by the ratio of the mean free path to the wire length but by the number of occupied Landau levels~\cite{Pruisken99}). The behavior of edge states in the presence of impurities was qualitatively discussed in Ref.~\cite{Buttiker88c}.


\section{Thermoelectric effects}
\label{sec:thermo}

We now show how thermoelectric effects can be described by using scattering matrices. So far, we have considered only the situation at zero temperature. The occupation numbers $f(E)$ at finite temperatures are given by Fermi distribution~(\ref{eq:FermiDistr}). The trivial effect of a nonzero temperature is manifested, for example, in the smearing of the steps of the conductance $G(W)$ or the peaks of $I(V)$ in the vicinity of resonances.

\subsection{Thermoelectric current and thermoelectromotive force}

\begin{figure}[tb]
	\includegraphics[width=6.5cm]{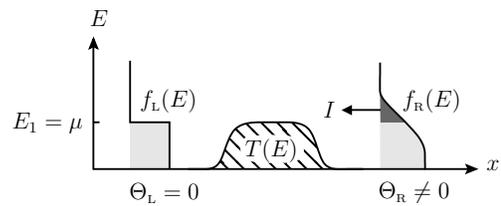}
	\caption{
Appearance of the thermoelectric current. Only electrons with energies $E > \mu$ can overcome the barrier on the right, producing the thermoelectric current.
	}
	\label{fig:thermo_curr}
\end{figure}

To study nontrivial thermoelectric effects, we consider the case of different reservoir temperatures $\Theta_{\rs L}$ and $\Theta_{\rs R}$ with their difference $\delta\Theta = \Theta_{\rs L} - \Theta_{\rs R}$ being finite. A thermoelectric current (i.e., the current caused by the difference in temperatures at a constant electrochemical potential) then appears, see Fig.~\ref{fig:thermo_curr}. In the one-dimensional case this current can be described by the expression
\begin{equation}
	I(V) = \frac{2e}{h} \int\limits_{-\infty}^\infty dE \,
	[f_{\rs L}(E)-f_{\rs R}(E)] \, T(E).
	\label{eq:TC_Land}
\end{equation}
Note, that the current is absent in the case of energy-independent transparency, $\partial_{\s E} T(E) = 0$.

As an illustration, we first consider a simple example where the transparency depends on energy, namely, a QPC with ideal quantization:
\begin{equation}
	T(E) =
	\left\{\begin{array}{ll}
		0, & E<E_1, \\
		1, & E>E_1,
	\end{array}\right.
	\label{eq:TC_IC}
\end{equation}
where the electrochemical potentials of the reservoirs are equal to the quantization energy in the first channel, $\mu = E_1$, and hence $\mu$ is the opening threshold energy for the first channel. We assume that the temperature in the left reservoir is zero, $\Theta_{\rs L} = 0$, and particles on the left cannot overcome the contact, while electrons with energies $E > \mu$ in the right reservoir can overcome the barrier resulting in non-zero current
\begin{align}
	I & = \frac{2e}{h}
	\int\limits_{-\infty}^\infty dE \, [n_{\rs L}(E)-n_{\rs R}(E)] \, T (E)
	\nonumber \\
	& = -\frac{2e}{h}
	\int\limits_0^\infty d\varepsilon \,
	\frac{1}{e^{\varepsilon/\Theta_{\rm R}}+1},
	\label{eq:TC_I_0}
\end{align}
where $\varepsilon = E-\mu$. Performing the integration over $\varepsilon$, we obtain~\cite{Lesovik89}\footnote{We use the relation $\int_0^\infty \frac{d\zeta}{e^{\zeta}+1} = \int_1^\infty \frac{d\lambda}{\lambda(\lambda+1)} = \int_1^\infty \big(\frac{1}{\lambda} - \frac{1}{\lambda+1} \big) d\lambda = \log\frac{\lambda}{\lambda+1} \big|_1^\infty = \log 2$.}
\begin{equation}
	I = \frac{2e}{h} \delta\Theta \log 2.
	\label{eq:TC_I_1}
\end{equation}

If a circuit containing the quantum wire considered here was closed then a voltage $V$ (thermoelectromotive force) would appear to compensate the thermoelectric current produced due to the difference in temperatures. For the nonzero temperature difference $\delta\Theta$ the general expression for current~(\ref{eq:TC_Land}) takes the form
\begin{align}
	I(\delta\Theta, V)
	= \frac{2e}{h} \int\limits_{-\infty}^\infty dE \, \bigg\{
		& \frac{1}{e^{(E-\mu-eV)/(\Theta+\delta\Theta)}+1} \nonumber \\
		- & \frac{1}{e^{(E-\mu)/\Theta}+1}
	\bigg\} \, T(E).
	\label{eq:TC_I_gen}
\end{align}
If the temperature difference $\delta\Theta$ is small and $T(E)$ depends on the energy $E$ relatively weakly, then the Fermi distribution function can be expanded in the vicinity of $\mu$ and the condition for the absence of the current $I(\delta\Theta,V) = 0$ gives
\begin{multline}
	\frac{2e}{h} \delta\Theta \int\limits_{-\infty}^\infty dE \, 
	\frac{\partial f(E)}{\partial E} \frac{E-\mu}{\Theta}
	\left[
		T(\mu) + (E-\mu) \frac{\partial T(\mu)}{\partial E}
	\right] \\
	+ \frac{2e^2}{h} V \int\limits_{-\infty}^\infty dE \,
	\frac{\partial f(E)}{\partial E} \, T(\mu) = 0.
	\label{eq:TC_I_exp}
\end{multline}
From~(\ref{eq:TC_I_exp}) we obtain the Katler-Mott formula
\begin{equation}
	\alpha = -\frac{\Theta}{e}
	\frac{\partial \log T(\mu)}{\partial E}
	\int\limits_{-\infty}^\infty \zeta^2 \frac{\partial n}{\partial \zeta} d\zeta =
	\frac{\pi^2}{3} \frac{\Theta}{e} \frac{\partial \log
	T(\mu)}{\partial E}
	\label{eq:TC_Vtp1}
\end{equation}
for the thermoelectric coefficient $\alpha = V / \delta\Theta$, where $\zeta = (E-\mu) / \Theta$.\footnote{The relation $\int_{-\infty}^\infty \zeta^2 \frac{\partial n}{\partial \zeta}d\zeta = -\frac{\pi^2}{3}$ is used.}

The generalization to the multichannel case is straightforward: a sum of transparencies appears instead of a transparency. Then for a dirty sample we have
\begin{equation}
	\alpha \approx
	\frac{\Theta}{e} \frac{e^2}{h} G^{-1} \frac{h L^2}{D}.
\end{equation}
A large thermoelectric coefficient for mesoscopic conductors was explicitly predicted in Ref.~\cite{Anisovich87}. The nonlinear case, which cannot be described using only the first derivative of the transparency with respect to energy (in which case the Katler-Mott formula becomes invalid), is considered in Ref.~\cite{Lesovik88}.

\subsection{Thermal flow: the Wiedemann-Franz law}

For a nonzero difference in temperatures electric current appears only when $T$ depends on energy in the vicinity of $\mu$. But the thermal flow also exists when the transparency is constant:
\begin{equation}
	I_{\s Q} =
	\frac{2}{h} \int\limits_{-\infty}^\infty dE \,
	[f_{\rs L}(E)-f_{\rs R}(E)] \, T(E) (E-\mu).
	\label{eq:WF_IQ}
\end{equation}
Here, the factor $2/h$ gives the number of electrons transmitted per unit time, while the factor $E - \mu$ in the integrand determines the amount of energy (which can dissipate) carried by each electron. For $\alpha = 0$ ($\partial_{\s E} T(\mu) = 0$), the thermal flow is
\begin{equation}
	I_{\s Q} =
	\frac{G}{e^2} \int\limits_{-\infty}^\infty dE \,
	[f_{\rs L}(E)-f_{\rs R}(E)] (E-\mu),
	\label{eq:WF_IQ1}
\end{equation}
where $G = (2e^2/h) T$ is the electric conductance.

Assuming that $\delta\Theta$ is small and expanding the difference $f_{\rs L}(E) - f_{\rs R}(E)$, we find
\begin{equation}
	I_{\s Q} =
	\delta\Theta \frac{G}{e^2} \Theta
	\int\limits_{-\infty}^\infty \zeta^2 \frac{\partial n}{\partial \zeta} \,
	d\zeta,
	\label{eq:WF_IQ1b}
\end{equation}
where $\zeta = (E-\mu) / \Theta$. Performing integration, we obtain the Wiedemann-Franz law~\cite{Franz53,Landau07BookV10}
\begin{equation}
	\varkappa = \frac{\pi^2}{3} \biggl(\frac{1}{e}\biggr)^2 G \Theta
	\label{eq:WF_law}
\end{equation}
for the heat conduction $\varkappa = I_{\s Q} / \delta\Theta$, which is also valid for usual (nonmesoscopic) conductors.

\subsection{Violation of the Wiedemann-Franz law}

The transparency of meso- and nanoconductors, unlike that in usual conductors, can strongly depend on energy in the vicinity of the electrochemical potential $\mu$, resulting in the appearance of the thermoelectromotive force
\begin{equation}
	V = \alpha \delta\Theta,
	\label{eq:vWF_tp}
\end{equation}
which also contributes to the thermal flow, and then Wiedemann-Franz law~(\ref{eq:WF_law}) can be violated. Substituting Eq.~(\ref{eq:vWF_tp}) in expression~(\ref{eq:WF_IQ}) for the thermal flow, we find
\begin{align}
	I_{\s Q} = \frac{2}{h}
	\int\limits_{-\infty}^\infty dE \, \bigg\{
		& \frac{1}{e^{(E-\mu+e\alpha \delta\Theta)/(\Theta+\delta\Theta)}+1}
		\nonumber \\
		- & \frac{1}{e^{(E-\mu)/\Theta}+1}
	\bigg\} \, T(E) (E-\mu).
	\label{eq:vWF_IQa}
\end{align}
Expansion of Eq.~(\ref{eq:vWF_IQa}) for small $\delta\Theta$ yields
\begin{equation}
	I_{\s Q} = \frac{2}{h}
	\int\limits_{-\infty}^\infty dE \, \left[
		e \alpha \frac{\partial T(\mu)}{\partial E}
		- \frac{T(\mu)}{\Theta}
	\right] \,
	(E-\mu)^2\, \frac{\partial n(E)}{\partial E} \, \delta\Theta.
	\label{eq:vWF_IQb}
\end{equation}
After integration, we obtain
\begin{equation}
	I_{\s Q} = G \Theta \left[
		- \alpha^2
		+ \frac{\pi^2}{3e^2}
	\right] \, \delta\Theta.
	\label{eq:vWF_IQc}
\end{equation}
Hence, the Wiedemann-Franz law is valid only if $\alpha \ll \pi/\sqrt{3}e$. Careful consideration shows that even in case $\alpha > \pi/\sqrt{3}e$ heat conduction $\varkappa$ also remains positive.

The possibility of the violation of the Wiedemann-Franz law in mesoscopic samples was first pointed out by Anderson and Engquist~\cite{Engquist81}, which became an important step in the understanding of specific features of quantum low-dimensional conductors that differ from usual metals.


\section{Second quantized formalism and scattering matrix approach}
\label{sec:secquant}

In the preceding sections, we discussed the mean current in coherent conductors. The method used for the calculation of current involves the summation of contributions to the current from different energy intervals. This method cannot be directly generalized to describe current fluctuations in time. Such calculations can be conveniently performed within the second quantization method. This was first done in Ref.~\cite{Lesovik89b} using the Landauer approach.\footnote{An alternative consideration can be based either on the method of wave packets developed by Landauer and Martin~\cite{Martin92,Landauer87,Landauer91,Landauer92}, which is not rigorous, or on a rigorous description in terms of wave functions~\cite{Hassler08,Schonhammer07}, which allows describing the full counting statistics but is too cumbersome, for example, for the calculation of noise.}

In this section, we describe this method, derive the Landauer formula more rigorously, and consider noises. We find, within the second quantization representation, the mean current and noise by averaging current operators over the nonequilibrium density matrix of the system, taking into account the difference in the distribution of occupation numbers in electron reservoirs.

The state of an electron in the second quantization formalism is described not by the wave function $\varphi_k(x)$ but by the creation operator ${\hat c}^\dag_k$ acting on a vacuum state~$|0\rangle$. The current density operator
\begin{equation}
	{\bf\hat j} = \frac{ie\hbar}{2m}
	\big[
		(\nabla {\hat \Psi}^\dag) {\hat \Psi} -
		{\hat \Psi}^\dag \nabla {\hat \Psi}
	\big]
	\label{eq:SCA_J_oper}
\end{equation}
is defined in terms of the $\hat\Psi$ operators
\begin{equation}
	{\hat \Psi}(x,{\bf r}_\perp) =
	\int\frac{dk}{2\pi} \sum\limits_{\alpha=1}^{N} 
	{\hat c}_{\alpha,k} \, \varphi_{\alpha, k} (x, {\bf r}_\perp),
	\label{eq:SCA_PsiC}
\end{equation}
where ${\bf r}_\perp$ is a vector in the cross section of a conductor, $k$ is the wave vector at infinity and the subscript $\alpha$ denotes a set of discrete quantum numbers, e.g., the spin, number of the channel, or reservoir index. One-particle wave functions $\varphi_{\alpha, k}$ used in second quantization form a complete orthonormalized set,
\begin{equation}
	\int dx \, d{\bf r}_\perp \,
	\varphi_{\alpha',k'}^*({\bf r}) \,
	\varphi_{\alpha, k}({\bf r}) = 2 \pi
	\delta_{\alpha\alpha'} \, \delta(k' - k)
	\label{eq:SCA_OSphi}
\end{equation}
and satisfy the Schr\"odinger equation
\begin{equation}
	{\hat H} \varphi_{\alpha, k} =
	E_{\alpha}(k) \varphi_{\alpha,k},
	\label{eq:SCA_Shr}
\end{equation}
from which the dispersion law $E_\alpha(k)$ is also determined. The commutation relation for the annihilation ${\hat c}_{\alpha,k}{\phantom\dag}$ and creation ${\hat c}_{\alpha,k}^\dag$ operators has the form
\begin{equation}
	\{ {\hat c}_{\alpha',k'}^\dag
	{\hat c}_{\alpha,k}^{\phantom\dag}\}= {\hat c}_{\alpha',k'}^\dag
	{\hat c}_{\alpha,k}^{\phantom\dag} +
	{\hat c}_{\alpha,k}^{\phantom\dag}
	{\hat c}_{\alpha',k'}^\dag =
	2\pi \delta_{\alpha' \alpha} \, \delta(k' - k),
	\label{eq:SCA_OSc}
\end{equation}
which corresponds to the normalization condition~(\ref{eq:SCA_OSphi}). The total current operator is the integral of the current density ${\bf\hat j}$ over the cross section:
\begin{equation}
	{\hat I} = \int d {\bf r}_\perp \, {\bf\hat j}(x, {\bf r}_\perp).
	\label{eq:SCA_I_oper}
\end{equation}
We express the $\hat\Psi$ operators in terms of the Lippmann-Schwinger scattering states, which form the complete orthonormalized set of eigenstates of the Hamiltonian $\hat H$ (the proof of this fact is given in Appx.~\ref{sec:scattstates}). We note that normalization~(\ref{eq:SCA_OSphi}) should match the commutation condition~(\ref{eq:SCA_OSc}). For convenience, we can redefine the normalization; for example, to obtain the delta function of energy in the right-hand side of Eq.~(\ref{eq:SCA_OSphi}), we should redefine~(\ref{eq:SCA_OSc}) correspondingly so that the same delta function appears in the right-hand side. Below, we use this renormalization.\footnote{Such a normalization is convenient, for example, in the case where scattering states are to be defined in a region with a smooth semiclassical potential.}

We now consider the problem of two electron reservoirs connected via a constriction (scatterer) with one open channel. Let us denote the states of the particle with energy~$E$ emitted from the left and right reservoirs as $\psi_{\s E,1}(x)$ and $\psi_{\s E,2}(x)$ respectively. Quantum numbers characterizing the one-particle state are the energy $E$ and the reservoir index from which the particle was emitted. For simplicity, we omit the spin subscript. The $\hat\Psi$ operator has the form
\begin{align}
	\hat\Psi(x) 
	& = \int dE \left\{
		\psi_{\s E,1}(x) {\hat a}_{\s E,1}+
		\psi_{\s E,2}(x) {\hat a}_{\s E,2}
	\right\} \nonumber \\
	& = \int dE \sum\limits_{\alpha=1,2}
		\psi_{{\s E,}\alpha}(x) {\hat a}_{{\s E,}\alpha},
	\label{eq:Psi_operator}
\end{align}
where ${\hat a}_{\s E,1}$ are electron annihilation operators in the state with quantum numbers $\{E, \alpha\}$ ($\alpha = 1,2$). These operators satisfy the commutation relations
\begin{align}
	& \!\! \{{\hat a}_{{\s E,}\alpha},{\hat a}_{{\s E',}\alpha'}^\dag\}=
	\delta_{\alpha\alpha'}\delta(E-E'),\;
	\{{\hat a}_{{\s E,}\alpha},{\hat a}_{{\s E',}\beta}\} = 0,
	\\
	& \!\! \{\hat\Psi(x),\hat\Psi^\dag(x')\}=\delta(x-x').
	\label{eq:aa_dag}
\end{align}
In the left asymptotic region we have
\begin{multline}
	\hat\Psi(x\to-\infty) =
	\int \frac{dE}{\sqrt{2\pi \hbar v_1}} \\
	\times \left\{
		\left(e^{ik_1x}+r_{\s E} e^{-ik_1x} \right) {\hat a}_{\s E,1} +
		t_{\s E} e^{-ik_1x} {\hat a}_{\s E,2}
	\right\}.
	\label{eq:Psi_L}
\end{multline}
Similarly, we can obtain the expression for $\hat\Psi$ in the right asymptotic region. It has the form
\begin{multline}
	\hat\Psi(x\to+\infty) =
	\int \frac{dE}{\sqrt{2\pi \hbar v_2}} \\
	\times \left\{
		{\hat a}_{\s E,1} t_{\s E} e^{ikx} +
		{\hat a}_{\s E,2} \left(r_{\s E} e^{ikx} + e^{-ikx}\right)
	\right\}.
	\label{eq:Psi_R}
\end{multline}
Using~(\ref{eq:Psi_L}) and (\ref{eq:Psi_R}), we can find the current operator in the asymptotic regions. For example, to the right of the scatterer, we have
\begin{align}
	{\hat I}(x)
	& = e \int dE' dE \frac{i}{2\pi m \sqrt{v_1 v'_2}} \nonumber \\
	& \times \biggl\{
		{\hat a}_{\s E',1}^\dag {\hat a}_{\s E,1}^{\phantom\dag}
		(-ik'-ik) t_{\s E'}^* t_{\s E}^{\phantom *} e^{i(k-k')x} \nonumber \\
		& + {\hat a}_{\s E',1}^\dag {\hat a}_{\s E,2}^{\phantom\dag} \Bigl[
			(-ik') t_{\s E'}^* e^{-ik'x} (e^{-ikx} + r_{\s E}^{\phantom *}e^{ikx}) \nonumber \\
			& - t_{\s E'}^* e^{-ik'x} (-ik e^{-ikx} + ik r_{\s E}^{\phantom *}e^{ikx})
		\Bigr] \nonumber \\
		& + {\hat a}_{\s E',2}^\dag {\hat a}_{\s E,1}^{\phantom\dag} \Bigl[
			(ik' e^{ik'x} - ik' r_{\s E'}^* e^{-ik'x}) t_{\s E}^{\phantom *} e^{ikx} \nonumber \\
			& - (e^{ik'x} + r_{\s E'}^*e^{-ik'x}) ik t_{\s E}^{\phantom *} e^{ikx}
		\Bigr] \nonumber \\
		& + {\hat a}_{\s E',2}^\dag {\hat a}_{\s E,2}^{\phantom\dag} \Bigl[
			(ik' e^{ik'x} - ik' r_{\s E'}^* e^{-ik'x}) (e^{-ikx} + r_{\s E}^{\phantom *}e^{ikx}) \nonumber \\
			& - (e^{ik'x} + r_{\s E'}^*e^{-ik'x}) (-ik e^{-ikx} + ik r_{\s E}^{\phantom *}e^{ikx})
		\Bigr]
	\biggr\}.
	\label{eq:I2B_I_op}
\end{align}
In the framework of second quantization formalism, the pure state of a many-particle quantum system is described by expressions like $|\psi\rangle = {\hat a}_{\s E_1}^\dag {\hat a}_{\s E_2}^\dag \ldots |0\rangle$, where $|0\rangle$ is the vacuum state. The average current is defined as
\begin{equation}
	I = \langle \psi | \hat I | \psi \rangle.
	\label{eq:pure_curr}
\end{equation}
If the state is described by the density matrix $\hat\rho$ (i.e., the state is an incoherent superposition of pure states), the mean current is given by
\begin{equation}
	I = \sum\limits_{\{\psi\}, \{\psi'\}}
		\langle \psi' | \hat \rho| \psi \rangle \langle \psi | \hat I | \psi' \rangle \equiv
	{\rm Tr} \{\hat \rho \hat I\},
	\label{eq:pure_curr1}
\end{equation}
where the current operator is multiplied by the density matrix and the trace of this product is taken. For an equilibrium system with a Hamiltonian $\hat H$, a finite temperature $\Theta$, and an electrochemical potential $\mu$, the density matrix is given by\footnote{In this case, we use the standard theoretical ``ensemble averaging.'' However, experimental averaging occurs in time. The fact that these two averaging methods give the same result is the subject of the ergodic hypothesis. Thus, we calculate one quantity, but another is measured. However, the ergodic hypothesis gives grounds to assume that they should coincide. For some particular systems the ergodic hypothesis can be proved.}
\begin{equation}
	\hat \rho = e^{-(\hat H - \mu \hat N) / \Theta}.
	\label{eq:rho_T}
\end{equation}

In the Landauer approach, the reservoirs are completely independent in the nonequilibrium case. It means that the density matrix of the total system is equal to the product of the density matrices of the left and right reservoirs, $\hat\rho = \hat\rho_1 \otimes \hat\rho_2$. The density matrix of the reservoir $\alpha$ has the form
\begin{equation}
	\hat \rho_\alpha =
	e^{-\sum_E {\hat a}_{{\s E,}\alpha}^\dag
		{\hat a}_{{\s E,}\alpha}^\pdag (E - \mu_\alpha) / \Theta_\alpha}.
	\label{eq:rho_alpha0}
\end{equation}
Then the density matrix of the total system is
\begin{equation}
	\hat \rho =
	e^{-\sum_E \hat \{ {\hat a}_{\s E,1}^\dag
		{\hat a}_{\s E,1}^\pdag (E - \mu_1) / \Theta_1 -
	{\hat a}_{\s E,2}^\dag
		{\hat a}_{\s E,2}^\pdag (E - \mu_2) / \Theta_2 \}}.
	\label{eq:rho_alpha}
\end{equation}
Using this density matrix, we can find all averages,
\begin{equation}
	\langle {\hat a}_{{\s E,}\alpha\sigma}^\dag
		{\hat a}_{{\s E',}\alpha'\sigma'}^\pdag \rangle =
	\delta(E-E')\delta_{\alpha\alpha'}\delta_{\sigma\sigma'}f_{\alpha\sigma}(E).
	\label{eq:aa_average}
\end{equation}
As an example, we indicate here the spin index $\sigma$ explicitly.
Almost all calculations presented in this review are in fact rather simple. One of the sources of this simplicity is relation~(\ref{eq:aa_average}). It implies that in the basis of scattering states, in which the current operator is written, only diagonal elements of the averages like~(\ref{eq:aa_average}) are nonzero.

The electron distribution function $f_\alpha(E)$ inside reservoirs is given by Fermi distribution~(\ref{eq:FermiDistr}).\footnote{At distances from the contact greatly exceeding the characteristic energy relaxation length $l_{\s E}$ associated with inelastic scattering of electrons on phonons or electron-electron scattering.} In general case, temperatures and electrochemical potentials in reservoirs are different.

We also note that the real bias voltage $V$ (specifying $\Delta\mu$ in the contact) can differ from the voltage $V_0$ (electromotive force) far in the reservoirs, and hence a part of the voltage drop $V_0 - V$ occurs in the lead wires of a quantum contact. This fact is taken into account experimentally quite simply, however. In addition, it may happen that a reservoir partially reflects electrons rather than absorbs them without reflection. This reflection can also be taken into account in principle as a correction to the density matrix, such that a nonzero average value $\langle {\hat a}_1^\dag {\hat a}_2^\pdag \rangle$ appears. In any case, we emphasize that approximations leading to the expressions used in this section (and above) are valid according to the experimental results. It seems that this is the main reason why the interaction with a reservoir, the possible role of reflection, the values of corrections, etc., have been so far insufficiently studied theoretically.

\subsection{Average current}

Using Eqs.~(\ref{eq:I2B_I_op}), (\ref{eq:pure_curr1}), and (\ref{eq:rho_alpha}), we obtain expressions for the average current,\footnote{It is important that the terms in Eq.~(\ref{eq:I2B_I_op}) responsible for the mixing of the reservoirs, which contain creation and annihilation operators with different subscripts $\alpha$, vanish due to Eq.~(\ref{eq:aa_average}).} coinciding in the one-dimensional case with~(\ref{eq:Landauer_gR}):
\begin{equation}
	\langle\hat I\rangle =
	\frac{2e}{h} \int
	dE\, T(E)\left\{f_{1}(E)-f_{2}(E)\right\}.
	\label{eq:I_average_simple}
\end{equation}
In the general case, in the presence of many channels and reservoirs, we have
\begin{equation}
	I_{\beta} = \frac{2e}{h}
	\sum\limits_\alpha \sum\limits_{j,l}
	\int dE \, [f_{\beta}(E) - f_{\alpha}(E)] \,
	T_{\beta \alpha, lj} (E),
	\label{eq:SCA_I_mch}
\end{equation}
where $\alpha$ and $\beta$ are the numbers of the reservoirs, and $j$ and $l$ are the numbers of the channels.

If $\mu_1=E_{\rs F} + eV$, $\mu_2 = E_{\rs F}$, and $\Theta_1 = \Theta_2 = \Theta$, then as $V \to 0$, the conductance in the one-dimensional case has the form
\begin{equation}
	G=\frac{2e^2}{h} \int dE \,
	\left( -\frac{\partial f}{\partial E} \right)
	T(E).
	\label{eq:G_g}
\end{equation}

\subsection{The Landauer approach from the standpoint of the Keldysh Green's functions}

Many efforts have been devoted to a rigorous derivation of the Landauer formula by more traditional methods, in particular, based on the Kubo formula~\cite{Stone88}. We show, omitting obvious details, how the Landauer formula can be obtained by using a more formal or, to be precise, better formalized approach. This approach was used in the Keldysh's paper~\cite{Keldysh65} to construct a diagram technique for nonequilibrium situations. The Keldysh Green's function
\begin{equation}
	iG^{\s -+}({\bf r}, {\bf r}') =
	{\rm Tr} \{ {\hat \rho} {\hat \Psi}^\dag({\bf r}') {\hat \Psi}({\bf r}) \} =
	\langle {\hat \Psi}^\dag({\bf r}') {\hat \Psi}({\bf r}) \rangle
	\label{eq:SCA_Keld_G}
\end{equation}
is an analogue of the distribution function $f(q,p,t)$ in the kinetic equation. The kinetic equation is typically solved by specifying the boundary conditions in the reservoirs, such that the distribution function be equal to the local equilibrium function. The boundary conditions for the Keldysh function at infinity, i.e., in the reservoirs, are (see, e.g.,~\cite{Larkin85})
\begin{equation}
	G^{\s -+} ({\bf r}, {\bf r}') \big|_{{\bf r}, {\bf r}' \in {\rs L(R)}}
	= G_{\rm eq}^{\s -+} ({\bf r}, {\bf r}'),
	\label{eq:SCA_Keld_G_b}
\end{equation}
where condition ${\bf r}, {\bf r}' \in {\rm L(R)}$ implies that $\bf r$ and ${\bf r}'$ belong either to the left or to the right reservoir. The current can be expressed in terms of the Keldysh Green's function as
\begin{equation}
	{\bf j} =\frac{e\hbar}{2m}
	\left[ \frac{\partial}{\partial {\bf r}} - \frac{\partial}{\partial {\bf r'}} \right]
	G^{\s -+}({\bf r}, {\bf r}')
	\Bigg|_{{\bf r} = {\bf r}'}.
	\label{eq:SCA_Keld_G_curr}
\end{equation}
Let us consider now a quasi-one-dimensional QPC with several open channels. Most of the electrons located far in the reservoirs belong to closed channels, which do not penetrate through the contact, and only a small fraction of electrons comes from the opposite reservoir. In other words, the ratio of open and closed channels in the reservoir $N_{\rm wire} / N_{\rm reservoir}$ is small. Therefore, the distribution function in the reservoirs can be treated as the equilibrium function with the specified $\mu_\alpha$ and temperature. Then the boundary conditions for the Keldysh function calculated with density matrix~(\ref{eq:rho_alpha}) are satisfied, and hence the derivation of the Landauer formula is completely verified.

Let us again describe qualitatively the transport picture. Particle flows emitted from the left and right reservoirs have distribution functions characterized by their temperatures and electrochemical potentials. In the contact region, particles experience only elastic scattering, and the distribution function is strongly nonequilibrium. Along with the states associated with the conducting channels in the contact, both reservoirs contain many other states that are not connected via the contact and do not contribute to the current. These states play a dominant role in the formation of distribution functions deep in the reservoirs, which with good accuracy turn out to be close to equilibrium functions.

\subsection{Noise description}
\label{sec:noisedesc}

Now we consider current fluctuations in time. In order to describe these fluctuations (noise), we need the time-dependent current operator in the Heisenberg representation. This operator can be expressed in terms of the time-independent Hamiltonian $H_0$ of the system as
\begin{equation}
	\hat I(x,\tau)
	= e^{i \hat H_0 \tau / \hbar} \, {\hat I}(x) \, e^{-i \hat H_0 \tau / \hbar}
	\label{eq:curr4noise}
\end{equation}
and
\begin{align}
	{\hat I}(x,\tau) & = e \int dE' dE \, \frac{ie^{-i(E-E')\tau/\hbar}}{2\pi m \sqrt{v_1 v'_2}} \nonumber \\
	& \times \bigg\{
		{\hat a}_{\s E',1}^\dag {\hat a}_{\s E,1}^{\phantom\dag}
		(-ik'-ik) t_{\s E'}^* t_{\s E}^{\phantom *} e^{i(k-k')x} \nonumber \\
		& + {\hat a}_{\s E',1}^\dag {\hat a}_{\s E,2}^{\phantom\dag} \Bigl[
			(-ik') t_{\s E'}^* e^{-ik'x} (e^{-ikx} + r_{\s E}^{\phantom *}e^{ikx}) \nonumber \\
			& - t_{\s E'}^* e^{-ik'x} (-ik e^{-ikx} + ik r_{\s E}^{\phantom *}e^{ikx})
		\Bigr] \nonumber \\
		& + {\hat a}_{\s E',2}^\dag {\hat a}_{\s E,1}^{\phantom\dag} \Bigl[
			(ik' e^{ik'x} - ik' r_{\s E'}^* e^{-ik'x}) t_{\s E}^{\phantom *} e^{ikx} \nonumber \\
			& - (e^{ik'x} + r_{\s E'}^*e^{-ik'x}) ik t_{\s E}^{\phantom *} e^{ikx}
		\Bigr] \nonumber \\
		& + {\hat a}_{\s E',2}^\dag {\hat a}_{\s E,2}^{\phantom\dag} \Bigl[
			(ik' e^{ik'x} - ik' r_{\s E'}^* e^{-ik'x}) \nonumber \\
			& \times (e^{-ikx} + r_{\s E}^{\phantom *}e^{ikx}) -
			(e^{ik'x} + r_{\s E'}^*e^{-ik'x}) \nonumber \\
			& \times (-ik e^{-ikx} + ik r_{\s E}^{\phantom *}e^{ikx})
		\Bigr]
	\bigg\}.
	\label{eq:I2B_I_op1}
\end{align}

Fluctuations can be described using the average
$\langle \Delta{\hat I}(x,\tau) \Delta{\hat I}(x',\tau') \rangle$,
where the operator $\Delta{\hat I} = {\hat I} - \langle{\hat I}\rangle$ determines a deviation from the mean current. This average, called the {\it irreducible correlator}, is denoted as $\dlangle \hat I(x,\tau)\hat I(x',\tau') \drangle$ and is given by expression
\begin{align}
	\dlangle & {\hat I}(x,\tau) {\hat I}(x',\tau') \drangle \nonumber \\
	& \equiv \langle \hat I(x,\tau)\hat I(x',\tau') \rangle -
	\langle \hat I(x,\tau) \rangle \langle \hat I(x',\tau') \rangle \nonumber \\
	& = \langle\Delta \hat I(x,\tau)\Delta\hat I(x',\tau') \rangle.
	\label{eq:erverv}
\end{align}
Current operators evaluated at different instants do not commute. Therefore the operator $\Delta \hat I(x,\tau)\Delta\hat I(x',\tau')$ is not Hermitian and, in general, quantity~(\ref{eq:erverv}) is complex. This means that this quantity cannot be directly measured in experiments. In Landau and Lifshitz's book~\cite{Landau10BookV5}, the symmetrized correlator
\begin{equation}
	\frac12 \Bigl[
		\dlangle \hat I(x,\tau)\hat I(x',\tau') \drangle +
		\dlangle \hat I(x',\tau')\hat I(x,\tau) \drangle
	\Bigr]
	\label{eq:symcorr}
\end{equation}
is considered to be a measurable quantity. Another standard quantity characterizing noise is the Fourier transform of current correlators: the {\it spectral noise density}. In Ref.~\cite{Landau10BookV5}, it was proposed to take the Fourier transform of symmetrized correlator~(\ref{eq:symcorr}). However, as follows from the analysis of the measurement process, the measurable quantity is typically the Fourier transform of the nonsymmetrized correlator\footnote{Which, unlike the nonsymmetrized current correlator at different times, is always a real quantity.}~\cite{Lesovik97b,Gavish00,Aguado00}
\begin{equation}
	S(\omega) =
	\int d\tau \, e^{i\omega\tau} \dlangle \hat I(x,0)\hat I(x,\tau) \drangle.
	\label{eq:asymmcorr}
\end{equation}
This fact was confirmed in recent experiments~\cite{Basset10}. In previous experiments~\cite{Onac06,Gustavsson07,Deblock03}, only the excess noise was measured, and, as a result, it was impossible to rigorously distinguish the symmetrized correlator from the nonsymmetrized one~\cite{Gavish02}.

In the absence of time-dependent external fields the correlation function in Eq.~(\ref{eq:asymmcorr}) must depend on the time difference only. Therefore, the Fourier transform of Eq.~(\ref{eq:erverv}) in $\tau$ and $\tau'$ can be written in the form
\begin{equation}
	\dlangle \hat I(x,\omega)\hat I(x,\omega') \drangle =
	S(\omega)2\pi\delta(\omega+\omega').
\end{equation}
The most often studied quantity is the spectral noise density at zero frequency:
\begin{align}
	S(0)
	& = \frac{2e^2}{h} \int\limits_0^{+\infty} dE \,
	\Bigl[
		f_1(E) [1-f_1(E)] T^2(E) \nonumber \\
		& + f_2(E) [1-f_2(E)] T^2(E)
		+ T(E) [1-T(E)] \nonumber \\
		& \times \big\{ f_1(E) (1-f_2(E)) + f_2(E) (1-f_1(E)) \big\}
	\Bigr].
	\label{eq:noise1chann}
\end{align}
This quantity does not depend on coordinates, which is the general property that follows from the stationarity of the random process of charge transfer.

Expression~(\ref{eq:noise1chann}) was first obtained by one of us~\cite{Lesovik89b}. Its generalization to the multichannel case~\cite{Buttiker90} looks surprisingly simple in the representation of eigenchannels in Ref.~\cite{Martin92}. In this representation the transparency is diagonal and we can write
\begin{align}
	S(0)
	& = \frac{2e^2}{h} \sum\limits_n \int\limits_0^{+\infty} dE \,
	\Bigl[
		f_1(E) [1-f_1(E)] T_n^2(E) \nonumber \\
		& + f_2(E) [1-f_2(E)] T_n^2(E)
		+ T_n(E)[1-T_n(E)] \nonumber \\
		& \times \big\{f_1(E) (1-f_2(E)) + f_2(E) (1-f_1(E))\big\}
	\Bigr].
	\label{eq:noisemultichann}
\end{align}
The latter expression coincides with the expression for a QPC with no mixing of channels~\cite{Lesovik89b}.

Now let us consider the spectral noise density at the zero frequency $S(0)$ in the equilibrium case ($f_1 = f_2 = f$). Then we have the relation $f(1-f) = -\Theta\partial_{\s E} f$ and obtain
\begin{gather}
	S(0) = \frac{4e^2\Theta}{h}\int dE \,
	\left(-\frac{\partial \, f}{\partial E}\right) T(E)
	= 2\Theta G.
	\label{eq:1223}
\end{gather}
This is the equilibrium Johnson-Nyquist noise appearing due to {\it temperature} fluctuations of the electron occupation numbers in the reservoirs.

We now consider noise in the zero temperature (quantum) limit. Occupation number has the steplike form $f_\alpha(E) = \theta(E - \mu_\alpha)$ and
\begin{align}
	S(0) & =
	\frac{2e^2}{h}\int dE [1-T(E)]T(E) \left\{f_2(E)-f_1(E)\right\}
	\nonumber \\
	& \approx \frac{2e^3|V|}{h} (1-T)T=e \langle \hat I \rangle (1-T),
	\label{eq:noise}
\end{align}
where we set $\mu_1 = \mu + eV$ and $\mu_2 = \mu$.

The approximate equality in Eq.~(\ref{eq:noise}) is valid if the transmission probability $T(E)$ weakly depends on energy. In this case the expression~(\ref{eq:noise}) (the Khlus-Lesovik formula) was obtained in Ref.~\cite{Khlus87} and then independently in Ref.~\cite{Lesovik89b}, as a particular case of general expression~(\ref{eq:noise1chann}).\footnote{One has to use a sufficient attention to hound down the semi-classical limit of formula~(\ref{eq:noise}) in the text of Ref.~\cite{Khlus87}.}

Equation~(\ref{eq:noise}) shows that
the quantum shot noise intensity is determined by scattering on a potential barrier. If a scatterer is absent, $T=1$, the noise is also absent. Noise also disappears if $T = 0$ because the electron transfer is then completely absent. In the intermediate situation, the wave packets describing electrons split into transmitted and reflected fractions during tunneling through the barrier. During measurements, electrons can be detected both in the left reservoir (``reflected electrons'') and in the right reservoir (``transmitted electrons''), and this occurs absolutely unpredictably and randomly. This principal quantum mechanical unpredictability is the main source of the quantum shot noise. It is important that in the quantum case, the electrons obeying the Fermi-Dirac statistics leave the reservoir in an almost ordered way, and therefore, in the absence of the uncertainty caused by scattering from the barrier ($T = 1$), the low-frequency shot noise is suppressed. Expression~(\ref{eq:noise}) was confirmed in the excellent experiments of two groups~\cite{Reznikov95,Kumar96} studying noise in QPCs. At a plateau, where $T_n = 1$ or $T_n = 0$ for all channels, the noise was suppressed, while in the region of steps it was found to be finite and, in accordance with Eq.~(\ref{eq:noise}), having correct dependence on transparency.

Expressions for noise as a function of the transparency make the theory, in certain sense, closed. In order to describe the conductance of the QPC we compare theoretical results based on the calculations of $T$ with experimental data. But having theoretical results for the average current and noise, it is possible to determine transparency $T$ {\it experimentally} from current measurements, and then to compare these measurements with independent experimental data on noise.

The measurements of noise in dirty samples lead to conclusion that the transparency distribution function [the Dorokhov function~(\ref{eq:dorokhovProp})] is actually nontrivial and the simple estimates of transparencies discussed in Sec.~\ref{sec:dorokhov} are incorrect. If all the transparencies are small then it follows from the general expression that
\begin{equation}
	S(0) \approx e \langle {\hat I} \rangle
	\label{eq:schottkylimit}
\end{equation}
and the Fano factor $F = S(0) / eI$ is unity, $F = 1$, as for the classical shot noise (see the end of this section). By averaging the sum $\sum_n T_n (1-T_n)$ that enters the expression for noise with the Dorokhov distribution function, it is possible to obtain the relation $\langle\sum_n T_n (1-T_n)\rangle = (1/3) \langle\sum_n T_n\rangle$ and the Fano factor $F = 1/3$~\cite{Beenakker92b}. Experiments~\cite{Steinbach95,Liefrink94} confirmed these calculations.

The energy dependence of transparency gives rise to some additional effects. In the case of ideal resonance at a voltage exceeding the width of the resonance, i.e., in the plateau of the current-voltage characteristic [Fig.~\hyperref[fig:double_bar_t]{\ref{fig:double_bar_t}(b)}], the Fano factor $F$ is $1/2$. This result follows from nontrivial dependence of transparency distribution function on energy.\footnote{See also Sec.~\ref{sec:noiseNININS}, where noise in the hybrid INIS junction is considered.} We also note that for a certain energy dependence of the transparency, noise can decrease at a nonzero voltage. In other words, the ``excess noise'' can be negative~\cite{Lesovik93}.

Finally, we see that expression~(\ref{eq:noise}) for noise contains the electron charge, and therefore the discreteness of the charge carried by quantum particles is also significantly manifested in the shot noise. Schottky was the first to point out this circumstance in 1918 and to derive the famous formula
\begin{equation}
	S(0) = e \langle I \rangle
	\label{eq:schottky}
\end{equation}
for the classical shot noise. Equation~(\ref{eq:schottky}) assuming that the random electron transfer process is Poissonian (i.e., all electrons escape independently of each other) with the escape probability for $m$ electrons $P_m=(\bar N^m/m!) \exp(-{\bar N})$, where ${\bar N} = It / e$. The mean-square deviation for the transferred charge in this process is $\langle (\delta Q)^2 \rangle = e^2 \langle(\delta N)^2 \rangle = e^2 \langle N \rangle$. Using the relation
\begin{equation}
	\lim_{t \to \infty } \langle (\delta Q(t))^2 \rangle / t = S(0)
	\label{eq:QQS}
\end{equation}
Schottky obtained formula~(\ref{eq:schottky}).

The ratio $S(0)/I$ of the noise intensity to the average current is used for experimental measurements of the charge of an elementary current carrier, which is not always identical to an isolated electron. Important measurements of a fractional charge in the fractional quantum Hall effect were performed by two groups~\cite{Saminadayar97,Picciotto97,Lefloch03,Reznikov99}. A more complete bibliography on noises is presented in reviews~\cite{Blanter00,Martin05}.


\section{Full counting statistics}
\label{sec:fcs}

Typical quantities that have been studied in the quantum transport until recently were the time-averaged current and noise, i.e., a pair current correlator.
However, it is known from the theory of random processes that in order to characterize a random process completely, one should also analyze higher-order correlators and distribution function of the transferred charge, which requires the knowledge of current correlators of all orders at low frequencies. This knowledge provides maximum information on the system, taking into account that the process is nondeterministic. One of the sources of uncertainty, as pointed out in Sec.~\ref{sec:noisedesc}, is the probabilistic nature of quantum mechanics while another source of uncertainty is inaccurate knowledge of reservoir states. Hence, along with the mean current $\langle I \rangle$, transferred charge $\langle Q(t) \rangle = \langle I \rangle t$, or mean-square deviation $\langle (\delta Q)^2 \rangle$ [and, correspondingly, noise $S(0)$], it is also interesting to study the higher-order moments
\begin{equation}
	\langle Q^n \rangle =
	\int\limits_0^t dt_1\ldots dt_n \,
	\langle I(t_1) \ldots I(t_n)\rangle
	\label{eq:higher_mom}
\end{equation}
and the characteristic function, as was first done in Refs.~\cite{Levitov92,Levitov93}. The characteristic function for the transferred charge distribution defined for a dimensionless number $Q/e$ of transferred particles
\begin{equation}
	\chi(\lambda) =
	\sum\limits_n \frac{\langle (Q/e)^n \rangle}{n!} (i\lambda)^n =
	\langle e^{i \lambda Q/e} \rangle,
	\label{eq:gen_fun}
\end{equation}
being a generating function, contains information about all moments, and these moments can be derived by differentiating the characteristic function:
\begin{equation}
	\langle (Q/e)^n \rangle =
	\frac{d^n}{d(i\lambda)^n} \chi(\lambda) \biggr|_{\lambda=0}.
	\label{eq:moments}
\end{equation}
A more convenient way to characterize a random process is to use cumulants instead of moments. The cumulant is defined by the expression
\begin{equation}
	\dlangle (Q/e)^n \drangle =
	\frac{d^n}{d(i\lambda)^n}
	\log \chi(\lambda) \biggr|_{\lambda=0}.
	\label{eq:cumulants}
\end{equation}
Cumulants have the following important properties:
\begin{enumerate}[(i)]
\item cumulants with $n>1$ do not change when a random variable is shifted by $c$, $\dlangle (Q+c)^n \drangle = \dlangle Q^n \drangle$,
\item they are homogeneous with regard to degree $n$, $\dlangle (cQ)^n \drangle = c^n \dlangle Q^n \drangle$, and
\item they are additive, $\dlangle (Q+\tilde{Q})^n \drangle = \dlangle Q^n \drangle +\dlangle \tilde{Q}^n \drangle$, if $Q$ and $\tilde Q$ are independent variables.
\end{enumerate}
It follows from the last property that $\dlangle Q^n \drangle \propto t$ for large time $t$ exceeding correlation time in the system. The argument is as follows: the whole process at a large time can be divided into independent subprocesses contributing to the net result. Since the number of subprocesses increases linearly with $t$, the total cumulant behaves similarly.

Knowing all the cumulants, we can, for example, describe frequency shift of the Josephson generation ~\cite{Lesovik94} and accurately describe the influence of noise in a wire on a near quantum system, for example, a qubit, without the usual assumption that the noise distribution is Gaussian~\cite{Falci03,Galperin06,Neder07}. In addition, it becomes possible to accurately describe the properties of a QPC as a detector related to a quantum bit~\cite{Averin05}. A third-order correlator can indicate asymmetry in a two-level system, affecting conduction electrons~\cite{Lesovik94,Falci03}, and the presence of other effects.\footnote{We do not present here the list of all possible effects in which the non-Gaussian distribution of fluctuations is manifested, since this question is outside the scope of our review.}

We now determine the number $n$ of electrons transferred in time $t$, which are related to the charge as $Q = en$. The random process is defined by the probabilities $P_n$ that exactly $n$ particles are transferred in time $t$, i.e.,
\begin{equation}
	\chi(\lambda) =
	\langle e^{i \lambda Q/e} \rangle =
	\sum\limits_n P_n e^{i\lambda n}.
	\label{eq:chi_pn}
\end{equation}
We note that the assumption 
that $n$ is an integer leads to the periodicity of $\chi(\lambda)$ with a period $2\pi$.

The probabilities $P_n$ can be obtained from the characteristic function via the Fourier transform
\begin{equation}
	P_n = \int\limits_0^{2\pi} \frac{d\lambda}{2\pi}
	e^{-i\lambda n} \chi(\lambda).
	\label{eq:pn}
\end{equation}

In quantum case, the relation between current correlators and moment observables~(\ref{eq:higher_mom}) and charge cumulants is not as simple as in classical case. Different definitions are found to lead to different results, and in order to obtain unambiguous results, it is necessary to describe not only a wire but also a detector and a measurement scheme. When calculating the characteristic function defined similarly to the classical expression as $\chi(\lambda) = \langle \exp[i\lambda \int_0^t \hat{I}(t')dt'] \rangle$, the problem of time ordering of current operators appears. If we follow this definition literally, current operators in the expressions for moments and cumulants should be symmetrized. This definition was used in 1992 in the first paper~\cite{Levitov92} on the full counting statistics (FCS). The result obtained for a one-channel conductor with a transparency $T$ at a finite voltage $V$ and zero temperature has the form
\begin{align}
	\chi(\lambda)
	& = \Bigl\langle e^{i \lambda \int\limits_0^t dt' {\hat I(t')}/e}
	\Bigr\rangle \nonumber\\
	& = \Bigl[
		\cos(\lambda \sqrt{T})
		+i \sqrt{T} \sin(\lambda \sqrt{T})
	\Bigr]^N,
	\label{eq:fcsll}
\end{align}
where $N= 2e V t / h \gg 1$ is the ``number of attempts.'' This expression is periodic with a period $2\pi / \sqrt{T}$. It can be explained by the following: the distribution function exists for a fractional charge $e^* = e \sqrt{T}$, which appears in some way in a system, but is manifested neither in the mean current nor in the noise. Although result~(\ref{eq:fcsll}) is formally correct and follows from the definition of the characteristic function, a further analysis has shown that in all the measurement schemes considered such a distribution function was never directly realized.

In order to decide how to determine the characteristic function in the quantum case, one should analyze the measurement scheme. In Refs.~\cite{Levitov94,Levitov96} the authors proposed an analog to a classic galvanometer which measures charge~--- the quantum galvanometer represented by a spin $1/2$ located near the wire and precessing in a magnetic field induced by a current. The precession angle allows to measure the passed charge $Q = \int_0^t I(t') dt'$. The interaction between the spin and an electron in the wire is described by the Hamiltonian
\begin{equation}
	{\hat H}_{\rm int} =
	-\frac{1}{c} \int\!dx\, {\hat I(x)} A(x),
	\label{eq:hint}
\end{equation}
where $A(x)$ is a component of a vector potential induced along the wire by the spin $1/2$ in the quantum conductor. In a general case, such an interaction is long-range, but to simplify calculations it can be replaced with a local interaction by representing $A(x)$ in the form\footnote{Strictly speaking, a potential of this form can give rise to certain difficulties due to the fact that the interaction Hamiltonian should take into account not only linear terms but also terms quadratic in $A(x)$. This leads to some peculiarities in the description of statistics in the many-particle perturbation theory, which we do not consider here.}
\begin{equation}
	A(x) = A_0 \, \delta(x - x_0) \sigma_z,
	\label{eq:aint}
\end{equation}
where $\sigma_z$ is the Pauli matrix, $x_0$ is the position of the measured spin, and $A_0$ specifies strength of the interaction with electrons in the wire. Correspondingly, the interaction Hamiltonian takes the form ${\hat H}_{\rm int} = {\hat H}_{{\rm int},+} |{\uparrow}\rangle \langle{\uparrow}| + {\hat H}_{{\rm int},-} |{\downarrow}\rangle \langle{\downarrow}|$, where
\begin{equation}
	{\hat H}_{{\rm int},\pm} =
	\mp \lambda \frac{\hbar I(x_0)}{2e},
	\label{eq:hintpm}
\end{equation}
$\lambda = 2e A_0 / \hbar c$, and $|{\uparrow}\rangle$ and $|{\downarrow }\rangle$ are spin states.

We assume that the initial state of the measured spin is specified by a density matrix ${\hat\rho}^{\rm s}(0)$ at the moment $t = 0$. The transfer statistics can be ``rewritten'' in terms of the spin rotation angle in time $t$, which can be obtained from nondiagonal elements of the spin density matrix. The time evolution of nondiagonal elements of the spin density matrix [assuming that at the instant $t = 0$ it is independent of the density matrix ${\hat\rho}^{\rm e}(0)$ of the electron system] is described by
\begin{align}
	{\hat\rho}^{\rm s}_{\uparrow\downarrow} (t)
	= \text{Tr}_{\rm e}
	\Bigl\{
		& e^{-i ({\hat H}_{\rm e}+ {\hat H}_{{\rm int},+}) t/\hbar} \, {\hat\rho}^{\rm e}(0)
		\nonumber \\ & \times
		e^{i ({\hat H}_{\rm e} + {\hat H}_{{\rm int}, - })t/\hbar}
	\Bigr\}
	{\hat\rho}^{\rm s}_{\uparrow\downarrow}(0)
	\nonumber \\ =
	\text{Tr}_{\rm e} \Bigl\{ &
		\mathcal{T} \bigl( e^{i\lambda \int\limits_0^t dt' I(x_0,t')/2e}\bigr)
		\,{\hat\rho}^{\rm e} (0)
		\nonumber \\ & \times
		\mathcal{\tilde T} \bigl( e^{i \lambda\int\limits_0^t dt' I(x_0,t')/2e} \bigr)
	\Bigr\}
	{\hat\rho}^{\rm s}_{\uparrow\downarrow}(0),
	\label{eq:timeevol}
\end{align}
where the trace ${\rm Tr}_{\rm e}$ is taken over the electron degrees of freedom and ${\hat H}_{\rm e}$ is the Hamiltonian of the electron subsystem. In the second expression the interaction is presented by the free energy operator ${\hat H}_{\rm e}$, which determines the dependence of the current operator $I(x_0,t)$ on time. $\mathcal{T}$ and $\mathcal{\tilde T}$ denote time ordering and antiordering respectively.\footnote{If the measured spin is located near the scatterer, the more complicated, so-called Matthew time ordering is required~\cite{Bachmann00}.} Defining $\chi(\lambda)$ as ${\hat\rho}^{\rm s}_{\uparrow\downarrow}(t) / {\hat\rho}^{\rm s}_{\uparrow\downarrow}(0)$,\footnote{If the spin is located near a classical current this quantity depends exponentially on the charge passed, $e^{i\lambda N}$, expressed in electron charge units.} we obtain the characteristic function of the transfer statistics:
\begin{equation}
	\chi(\lambda) = \Bigl\langle
		\mathcal{\tilde T}\bigl( e^{i \lambda \int\limits_0^t dt' I(t')/2e} \bigr) \,
		\mathcal{T} \bigl( e^{i \lambda \int\limits_0^t dt' I(t')/2e} \bigr) 
	\Bigr\rangle.
	\label{eq:spin_counter}
\end{equation}
We see that this definition differs from Eq.~(\ref{eq:fcsll}) by the presence of time ordering of current operators.

The characteristic function~(\ref{eq:spin_counter}) at zero temperature and a finite voltage has the form~\cite{Levitov93}
\begin{equation}
	\chi(\lambda) = \bigl[ 1- T + T e^{i\lambda} \bigr]^N.
	\label{eq:fcsll2}
\end{equation}
Since it is a function of $\lambda$, it is periodic with a period~$2\pi$ which leads to a charge quantization (in units of~$e$). However, we should note that even though the characteristic function is $2\pi$-periodic in this particular case, there is no reason to believe that this result is general for an arbitrary ${\hat H}_{\rm e}$ and ${\hat\rho}_{\rm e}(0)$. Moreover, an explicit example is presented in Ref.~\cite{Shelankov03} where the initial state is a superposition of the left and right scattering states and the characteristic function has the period $4\pi$ (which means the charge is quantized in units of $e/2$). Nevertheless, although the chosen definition~(\ref{eq:spin_counter}) does not always give an integer charge quantization, the quantity $\chi(\lambda)$ is measurable and, in particular, can be used to describe decoherence of a qubit (spin) coupled to a quantum wire. Indeed, according to the accepted definition, $\chi(\lambda)$ is a nondiagonal (normalized) element of the spin density matrix, and the absolute value $|\chi(\lambda)|$ specifies the decoherence degree. From relation~(\ref{eq:fcsll2}), we obtain
\begin{align}
	|\chi(\lambda)|
	= & \left|1-T+Te^{i\lambda}\right|^N
	\nonumber \\
	= & \left[1-4T(1-T)\sin^2 \frac{\lambda}{2} \right]^{N/2}.
	\label{eq:decoher}
\end{align}
Since the value of $\lambda$ is determined by the interaction strength, the decoherence rate is a nonmonotonic function of the coupling between the conductor and the measuring spin. In reality, a phase or a charge qubit can play the role of a spin (see the discussion in Ref.~\cite{Lesovik06}).

Shelankov and Rammer~\cite{Shelankov03} proposed an alternative definition of $\chi(\lambda)$, which always gives the period $2\pi$ and positive probabilities $P_n$. This definition corresponds to the approach in which $P_n$ is measured directly, as proposed in Ref.~\cite{Muzykantskii03}. The same definition was used in Ref.~\cite{Schonhammer07,Avron08} (also see the discussion in Ref.~\cite{Bachmann00}). By performing the measurement corresponding to the operator ${\cal Q} = \int_{x_0}^\infty dx \, |x\rangle \langle x|$, which determines a charge to the right of the detector at $t=0$, and comparing the result with a charge at time $t$, we can obtain the number of electrons that have passed in time $t$. The formulation of the problem in this way leads to the characteristic function
\begin{equation}
	\chi(\lambda) =
	\bigl\langle
		e^{i\lambda U^\dag \mathcal{Q} U/e} \,
		e^{-i\lambda\mathcal{Q}/e}
	\bigr\rangle,
	\label{eq:fcs}
\end{equation}
where $U = \exp(-i{\hat H}_{\rm e}t)$ is the unitary evolution operator; the angular brackets denote an averaging over the~${\cal Q}$ eigenstate, in which particles are initially located with certainty either to the left or to the right of the scatterer. In particular, such a definition allows to avoid the states leading to periodicity with a period of~$4\pi$.

\subsection{One-electron example}
\label{sec:fcs1electron}

In order to better understand the description of the transport statistics by means of the formalism presented (or rather outlined) above, we consider a simple problem for one electron. We assume that a wave packet with a wave function $f(k)$ in the $k$ space is concentrated near some $k_0>0$,
\begin{equation}
	\Psi_\mathrm{in}(x,t) \equiv \Psi_f(x,t)
	= \int \frac{d k}{2\pi} \, f(k) \, e^{i(kx-\omega_k t)}
	\label{eq:Psi_in}
\end{equation}
located on the left for $t \to -\infty $, moves to the right and falls on a scatterer having the transmission amplitude $t_k$ and reflection amplitude $r_k$. The function $f(k)$ is normalized by the condition $\int (dk/2\pi) |f(k)|^2 = 1$; $\omega_k = \hbar k^2 / 2m$. We locate a measuring spin near the scatterer. Then the transmitted part of the wave packet acquires an additional phase shift due to the interaction with the spin: in the case of magnetic interaction, the additional phase at the point $x$ has the form $\delta \phi_{\s A}(x) = 2 \pi \int_{-\infty}^x dx' \, A_x(x') / \Phi_0$ and does not depend on $k$; as $x \to \infty $, we obtain the full phase $\lambda/2 = 2\pi \int_{-\infty}^{\infty} dx\, A_x(x)/\Phi_0$. We note that $\phi_{\s A}$ has opposite signs for particles moving in opposite directions ($k \to -k$). Scattered waves, which have the form (for $t \to \infty$)
\begin{multline}
	\Psi_\mathrm{out}^\pm (x,t)=
	\int \frac{d k}{2\pi} f(k) e^{-i \, \omega_k t}
	\bigl[
		r_k e^{-i k x} \Theta(-x) \\
		+ e^{\pm i \lambda/2} t_k e^{i \, k x}\Theta(x)
	\bigr]
	\label{eq:psi_out}
\end{multline}
acquire an additional phase, depending on the spin state $|\pm\rangle$ (or, in reality, a qubit state). The characteristic function for the FCS is described by the expression
\begin{align}
	\chi(\lambda,t) & = 
	\int dx {\Psi_\mathrm{out}^-}^*(x,t) \Psi_\mathrm{out}^+(x,t)
	\nonumber\\
	&= \int \frac{d k}{2\pi} (R_k + e^{i \lambda} T_k) |f(k)|^2
	\nonumber\\
	&=\langle R\rangle_f + e^{i \lambda} \langle T\rangle_f,
	\label{eq:fid_behind}
\end{align}
where $R_k = |r_k|^2$ and $T_k = |t_k|^2$. We neglected the nondiagonal term $\int dk \, f^*(-k) f(k)$, which is typically exponentially small. The Fourier transform of the characteristic function gives the probabilities $P_0 = \langle R \rangle_f$ and $P_1 = \langle T \rangle_f$, coinciding, as expected, with the reflection and transmission probabilities for particles. We see from this simple example that the definition (with the measuring spin) works correctly. Of course, this method offers no advantages over standard probability calculations in this simple case, the advantages being manifested only for a large or infinite number of particles.

\subsection{Two electrons}
\label{sec:fcs2electrons}

Following~\cite{Hassler07} and using the wave-packet formalism, we calculate the characteristic function for the FCS for two particles~--- the simplest case where the Fermi statistics of particles is already manifested. Incident particles are described by the wave packets
\begin{equation}
	\psi_{\text{in},m}(x;t) = \int\limits_0^\infty
	\frac{dk}{2\pi} f_m(k) e^{i k (x - v_\text{F} t)}
	\label{eq:wave_packet}
\end{equation}
with wave functions $f_1(k)$ and $f_2(k)$ in the momentum space satisfying the normalization condition $\int (dk / 2\pi) | f_m(k) |^2 = 1$. Because we eventually consider electrons at low temperatures in the vicinity of the Fermi energy, it is convenient to linearize the spectrum $\epsilon = v_{\rs F} |k|$, where $v_{\rs F}$ is the Fermi velocity, $\hbar k$ is the momentum, and $\hbar\epsilon$ is the energy. After propagating through the scatterer, wave packet~(\ref{eq:wave_packet}) is split into reflected and transmitted parts:
\begin{multline}
	\psi^\sigma_{\text{out},m} (x,t) = \int\limits_0^\infty
	\frac{dk}{2\pi} f_m(k) e^{- i k v_\text{F} t}
	\Bigl[
		r_k e^{-ikx} \Theta(x_\text{s}-x) \\
		+ e^{i \sigma \lambda/2} t_k e^{ikx} \Theta(x-x_\text{s})
	\Bigr],
	\label{eq:wf_out1}
\end{multline}
where we introduce the phase $\exp(i\sigma \lambda/2)$ in the transmitted part of the wave packet; the sign $\sigma = \pm 1$ corresponds to the spin state, as in Sec.~\ref{sec:fcs1electron}. The two-particle wave function symmetrized (antisymmetrized) in the proper way has the form
$
	\Psi_{{\rm x},\pm}(x_1,x_2;t)\propto
	\psi_{{\rm x},1}(x_1;t)\psi_{{\rm x},2}(x_2;t) \pm (x_1\leftrightarrow x_2)
$, 
${\rm x} = {\rm in},~{\rm out}$;
here, we use the sign ``$\pm$'' to distinguish the triplet and singlet states of two electrons. 

From this, we obtain the characteristic function
\begin{multline}
	\chi_\pm (\lambda) = \frac{
		\bigl[ 1 + (e^{i \lambda} - 1 ) \langle 1 | T | 1 \rangle \bigr]
		\bigl[1 + (e^{i \lambda} - 1 ) \langle 2 | T | 2 \rangle \bigr]
	}{
		1 \pm |S|^2
	} \\
	\pm \frac{
		\bigl[ S + (e^{i \lambda} - 1 ) \langle 1 | T | 2 \rangle \bigr]
		\bigl[ S^* + (e^{i \lambda} - 1 ) \langle 2 | T | 1 \rangle\bigr] 
	}{
		1 \pm |S|^2
	}
	\label{eq:fcs_2wp}
\end{multline}
with the matrix element $\langle n | T | m \rangle = \int (dk / 2\pi) f_n^*(k)T_k$ $f_m(k)$, where $T_k = |t_k|^2$; the overlap integral is
\begin{equation}
	S = \int \frac{dk}{2\pi} f_1^*(k) f_2(k).
	\nonumber
\end{equation}
The transmission probability for $n$ particles is determined by the Fourier transform of the characteristic function $P_n = \int (d\lambda/2\pi) \chi(\lambda) e^{-i\lambda n}$.

For the amplitude $t_k \equiv t$ independent of energy, the denominator in Eq.~(\ref{eq:fcs_2wp}) cancels with the factor in the numerator that depends on the exchange term, and, as a result, the transfer statistics is independent of the exchange symmetry of the two-particle wave function. This property is typical for the one-dimensional case, whereas such a cancelation does not occur, generally speaking, in the multichannel case, and the interference (exchange) term is not zero even if the transmission amplitude is independent of energy.

If the transmission amplitude depends on energy, the exchange terms lead to significant effects in the transfer statistics. For simplicity, we consider two packets of the same form separated by a distance $\delta x$. The Fourier components of the packets satisfy the relation $f_2(k) = f_1(k) e^{- i k \delta x}$, whence $\langle 1 | T | 1 \rangle = \langle 2 | T | 2 \rangle \equiv \langle T \rangle = \int (dk/2\pi) T_k |f_1(k)|^2$. The overlap integral $S = \int (dk / 2\pi) |f_1(k)|^2 \exp (ik \delta x)$ is the Fourier transform of the packet distribution function in the momentum space. The transmission probabilities $P_{n,\pm}$, which have the form
\begin{align}
	P_{0,\pm} &= \frac{ ( 1 - \langle T \rangle )^2
		\pm | S - \langle 1 | T | 2 \rangle |^2 }
		{1 \pm |S|^2}, \nonumber\\
	P_{1,\pm} &= 2 \frac{\langle T \rangle ( 1 - \langle T \rangle)
		\pm \bigr[\text{Re} ( \langle 1 | T | 2 \rangle S^*)
		- | \langle 1 | T | 2 \rangle |^2 \bigr]}
		{1 \pm |S|^2}, \nonumber\\
	P_{2,\pm} &= \frac{ \langle T \rangle ^2
		\pm | \langle 1 | T | 2 \rangle |^2 }
		{1 \pm |S|^2}
	\label{eq:prob}
\end{align}
(with $\sum_n P_{n,\pm } = 1$), depend on the exchange symmetry if $\langle 1 |T| 2 \rangle \neq S \langle T \rangle$.

Probabilities~(\ref{eq:prob}) can be easily transformed into the cumulants of the transmitted charge $Q = \int dt \, I(t)$. The first two cumulants for two particles incident on the scatterer have the form
\begin{align}
	\langle & n \rangle_\pm
	= P_{1,\pm} + 2 P_{2,\pm}, \nonumber \\
	\dlangle & n^2 \drangle_\pm
	= P_{1,\pm} ( 1- P_{1,\pm} ) + 4 P_{2,\pm} P_{0,\pm}
	\label{eq:cum}
\end{align}
(with $n = Q/e$), and they both depend on the exchange symmetry. Surprisingly, the effect caused by the exchange symmetry in the mean charge $\langle Q \rangle$ was discovered~\cite{Hassler07} considerably later than in the noise and charge fluctuations $\dlangle Q^2 \drangle$~\cite{Buttiker92,Lesovik97Thesis,Burkard00}.

To analyze the effect quantitatively, it is necessary to specify the form of the wave packet $f_1$ and the dependence of $T_k$ on the momentum. We do not resort to the standard approximation of the Gaussian wave packet, but consider a more realistic example.

Recently, a method has been developed for sending isolated electrons on demand in quantum wires~\cite{Feve07,Mahe10}. In this case, electrons move not very high over the Fermi surface. Otherwise, the electron transport becomes incoherent due to the emission of phonons and photons (plasmons) by electrons. The presence of the Fermi sea blocks these inelastic processes, and the coherence length can reach a few micrometers, exceeding the size of a conductor, for example, a QPC.

Strictly one-particle excitations over the Fermi surface can be produced by applying~\cite{Levitov96} a Lorentzian voltage pulse~\cite{Keeling06}\footnote{In general a pulse with an arbitrary profile excites an infinite number of electron-hole pairs, which is a phenomenon quite similar to the so-called Anderson catastrophe (see the discussion in Ref.~\cite{Levitov96}).} $V_{t_1}(t) = -(2v_{\rs F} \xi \Phi_0/c) / [v_{\rs F}^2(t-t_1)^2 + \xi^2]$, where the pulse duration $\xi / v_{\rs F}$ is expressed in terms of the length parameter $\xi $, and $\Phi_0 = hc/e$. Such a voltage pulse gives rise to a wave packet with the amplitude
\begin{equation}
	f_1(k) = \sqrt{4\pi\xi} e^{-\xi (k-k_{\rs F})-i k x_1}
	\Theta(k-k_{\rs F})
	\label{eq:lorentz}
\end{equation}
($x_1 = v_{\rs F}t_1$) and the Lorentzian profile in the usual space,
\begin{equation}
	|\psi_1|^2 =
	\frac{\xi/\pi}{(x-x_1-v_{\rs F}t)^2+\xi^2}.
	\label{eq:wf_lorentz}
\end{equation}
The overlap integral for the wave packets separated by a distance $\delta x$ has the form $S = e^{-i k_{\rs F} \delta x} / (1 + i \delta x/2 \xi)$.

We consider scatterers of two types:
\begin{enumerate}[(i)]
\item {\it with a transparency resonance}, which we write in the form
\begin{equation}
	T_k^{\rm res} =
	\frac{\alpha}{1 + \beta^2 (k - k_{\rs F} - k_0)^2},
	\nonumber
\end{equation}
where $\alpha \leqslant 1$ is the amplitude of the resonance and $k_0 > 0$ is its position relative to the Fermi wave vector $k_{\rs F}$. The resonance width $\beta^{-1}$ should be much smaller than the wave-packet width $\xi^{-1}$ in the $k$-space, $\beta^{-1} \ll \xi^{-1}$. The transparency $\langle T^{\rm res} \rangle$ for one packet with the amplitude $f_1(k)$ for $\beta ^{\,-1}\ll k_0$ takes the form $\langle T^{\rm res} \rangle \approx (2\pi \alpha \xi / \beta) e^{-2 \xi k_0}$, which is a resonance away from the Fermi level. Although the small parameter $k_0\xi$ provides a strong total signal, it leads to the suppression of exchange effects because the transparency is already maximal; we therefore consider intermediate and large values of $k_0\xi$;
\item {\it with a sharp transparency step} ($\beta ^{-1} \ll \xi ^{-1}$), for example, in a QPC. In this case, we use the Kemble formula considered above, which can now be conveniently written in the form
\begin{equation}
	T_k^{\rm qpc}=
	\frac{\alpha}{1+e^{-\beta (k-k_{\rm \s F} -k_0)}}.
	\nonumber
\end{equation}
The mean transparency here is $\langle T^{\rm qpc}\rangle\approx \alpha e^{-2\xi k_0}$; a small factor $\xi / \beta$ is absent in this case.
\end{enumerate}

For a pronounced (narrow) resonance, the exchange term takes the simple form
\begin{equation}
	\langle 1| T^{\rm res} | 2 \rangle \approx
	e^{-i (k_{\rs F}+k_0) \delta x} \langle T^{\rm res} \rangle
	\nonumber
\end{equation}
and its product with the overlap integral $S^*$ is proportional to $\exp (-i k_0\delta x)$ and independent of $k_{\rs F}$. The mean number of particles oscillates as a function of the distance $\delta x$:
\begin{multline}
	\langle n \rangle^{\rm res}_\pm =
		2 \langle T^{\rm res} \rangle
	\big( 1+ (\delta x / 2 \xi)^2
		\pm [\cos (k_0 \delta x) + \\ + (\delta x / 2 \xi) \sin(k_0 \delta x)] \big) \times
	\big( 1+ (\delta x / 2 \xi)^2 \pm 1 \big)^{-1}.
	\label{eq:res_q}
\end{multline}
For wave packets with a large delay, $\delta x \gg \xi$, the exchange term decays as $(\delta x)^{-2}$, while the number of transmitted particles is $\langle n \rangle^{\rm res} = 2 \langle T^{\rm res} \rangle$, irrespective of the exchange term sign. On the other hand, for strongly overlapped wave packets, $\delta x \to 0$, and the result $\langle n \rangle^{\rm res}_+ = 2 \langle T^{\rm res} \rangle$ obtained for independent particles in the singlet case is reproduced. In the asymmetric case (a triplet), the number of particles
\begin{equation}
	\langle n \rangle^{\rm res}_- =
	2 \langle T^{\rm res} \rangle (1 - 2 \xi k_0 + 2 \xi^2 k_0^2)
	\label{eq:qminus_res}
\end{equation}
decreases for narrow wave packets with $\xi k_0 < 1$ and increases for packets with $\xi k_0 > 1$. The decrease can reach 50\% for $\xi k_0 = 1/2$,\footnote{It follows from the exponential decrease of $\langle T^{\rm res}\rangle \propto \exp{(-2k_0\xi )}$ that $\langle -Q/e \rangle^{\rm res} \ll 2$ for $\xi k_0 > 1$.} while the (relative) increase for $\xi k_0 > 1$ is unlimited. We note that this increase occurs because $P_{1,-}$ and $P_{2,-}$ almost vanish.

In the case of a QPC, the result is similar: the nondiagonal matrix elements take the form
\begin{equation}
	\langle 1 | T^{\rm qpc} | 2 \rangle \approx
	e^{-i k_0 \delta x} S \langle T^{\rm qpc} \rangle
	\nonumber
\end{equation}
and the interference term in $P_2$ vanishes. The number of transmitted particles
\begin{equation}
	\langle n \rangle^{\rm qpc}_\pm =
	2 \langle T^{\rm qpc} \rangle
	\frac{1+ (\delta x / 2 \xi)^2 \pm \cos (k_0 \delta x) }
	{ 1+ (\delta x/ 2 \xi)^2 \pm 1}
	\label{eq:qpc_q}
\end{equation}
oscillates with $k_0 \delta x$. Different limits discussed above are also reproduced, except for the case of antisymmetric exchange and strongly overlapping packets [see~(\ref{eq:qminus_res})], in which the mean number of transmitted particles is
\begin{equation}
	\langle n \rangle^{\rm qpc}_-
	= 2 \langle T^{\rm qpc} \rangle (1 + 2 \xi^2 k_0^2).
	\label{eq:qminus_qpc}
\end{equation}
Equations~(\ref{eq:qminus_res}) and (\ref{eq:qminus_qpc}) contain the most surprising results: for a large parameter $\xi k_0$, the mean number of particles can increase very strongly compared to that of two independent packets. This is the case when all characteristic lengths are smaller than the dephasing length $L_\varphi$. Finally, we present the expression for the characteristic function in the case of the transfer statistics for two particles:
\begin{align}
	\chi^{\rm res}_\pm =& 1 +
	\langle n \rangle^{\rm res}_\pm (e^{i\lambda} -1 )\\
	&+ \langle T^{\rm res} \rangle^2
	\frac{(1 \pm 1) [ (\delta x /2 \xi)^2 + 1]}
	{ (\delta x /2 \xi)^2 + 1 \pm 1 } (e^{i \lambda} -1)^2, \nonumber\\
	\chi^{\rm qpc}_\pm =& 1 +
	\langle n \rangle^{\rm qpc}_\pm (e^{i\lambda} -1 ) +
	\langle T^{\rm qpc} \rangle^2 (e^{i\lambda} -1 )^2.
	\label{eq:2fcs}
\end{align}

The increase in $\langle n \rangle_-$ is caused completely by the increasing $P_1$. This effect is determined by the Pauli principle and the energy dependence of the transparency. The energy dependence of the transparency leads to the wave packet broadening, which, combined with the Pauli principle, causes a decrease in $P_0$ and $P_2$, and therefore an increase in $P_1$. Nevertheless, situations are possible where $P_2$ also increases.

For example, a high probability $P_{2,-}$ [see~(\ref{eq:fcs_2wp})] is obtained for wave packets with the amplitudes shifted in the $k$-space, $f_2(k) = f_1 (k+\delta k)$, and the large overlap integral $S$ of the transmission amplitude is suppressed for $k$ belonging to the overlap region. A large increase in the transmission probability $P_2$ for two electrons was also observed in Ref.~\cite{Lebedev08b}.

We note that exchange effects in the transfer statistics at constant voltages are considered in Ref.~\cite{Levitov93}, where, in particular, a simple example with a ``Y-joint'' containing three channels is discussed and the characteristic function is found. Pair correlators in this geometry are studied in Refs.~\cite{Martin92,Buttiker92}.

\subsection{$N$ electrons}
\label{sec:fcsNelectrons}

Following~\cite{Hassler07}, we extend the previous analysis to the case of $N$ particles with the incident wave function $\Psi({\bf k})$ defined in the momentum space; the vector ${\bf k} = (k_1, \dots, k_{\s N})$ determines $N$ momenta of particles. We consider independent noninteracting particles scattered from a barrier. The scattered wave function in the asymptotic region (for $t \to \infty $) takes the form
\begin{multline}
	\psi_\text{out}^\pm ({\bf x};t) = \Bigl\{
		\prod_{m=1}^N \int \frac{dk_m}{2\pi} \, [
			r_{k_m} e^{-i k_m (x_m+ v_{\rs F} t)} \Theta(-x_m) \\
			\quad+ t_{k_m} e^{i k_m (x_m- v_{\rs F} t)} e^{\pm i \lambda/2} \Theta(x_m)
		]
	\Bigr\}
	\Psi({\bf k}),
	\label{eq:wf_out2}
\end{multline}
which means that the evolution of the total wave function reduces to the individual evolutions of one-particle wave functions, and we obtain the product of asymptotic states~(\ref{eq:psi_out}). The characteristic function $\chi_{\s N} (\lambda) = \int d{\bf x}\,\psi_\text{out}^-({\bf x};t)^* \psi_\text{out}^+({\bf x};t)$ is expressed as
\begin{equation}
	\chi_{\s N} (\lambda) =
	\Bigl\{
		\prod_{m=1}^N \int \frac{dk_m}{2\pi} \,
		(1 - T_{k_m} + T_{k_m} e^{i\lambda})
	\Bigr\} |\Psi({\bf k})|^2.
	\label{eq:chiN}
\end{equation}

So far we have not specified the exact form of the incident wave function. If we restrict ourselves to the Slater determinant composed of orthonormalized one-particle functions $\phi_m$,
\begin{equation}
	\Psi(k_1,\dots,k_{\s N}) =
	\frac{1}{\sqrt{N!}}\det \phi_m(k_n),
	\label{eq:slater_det}
\end{equation}
then expression~(\ref{eq:chiN}) can be represented as the determinant
\begin{align}
	\chi_{\s N}(\lambda)
	& = \det \int \frac{dk}{2\pi} \, 
	\phi_m^*(k) (1-T_k + T_k e^{i\lambda}) \phi_n(k) \nonumber \\
	& = \det \langle \phi_m | 1 -T + Te^{i\lambda} | \phi_n \rangle
	\label{eq:chiN_det}
\end{align}
containing one-particle matrix elements $\langle\phi_m| \mathcal{O}|\phi_n \rangle$ of the operator $\mathcal{O} = 1 -T + Te^{i\lambda}$.

\begin{figure}[tb]
	\includegraphics[width=8.0cm]{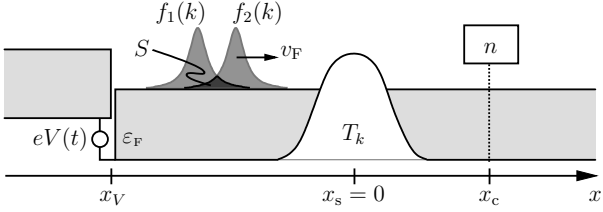}
	\caption{
Quantum wire with a scattering center at $x_{\rm s}$ providing the momentum-dependent transparency $T_k$. The time-dependent potential $eV(t)$ applied at the point $x_{\s V}$ (to the left of the scatterer) gives rise to the incident wave packets $f_1$ and $f_2$ with the overlaps $S = \langle f_2 | f_1 \rangle$. A detector located at the point $x_{\rm c}$ (to the right of the scatterer) measures the statistics of the number $n$ of particles propagated to the right. We consider incident packets with $k > 0$ outside the Fermi sea. As a result, the Fermi sea remains unperturbed in the asymptotic regime. The presence of the Fermi sea at finite times produces additional noise, which we neglect.
	}
	\label{fig:fcs_setup}
\end{figure}

{\it The nonorthogonal basis.} Real situations are typically described using the occupation of orthogonal states in the Slater determinant, as demonstrated above. But, for example, in the case presented in Fig.~\ref{fig:fcs_setup}, electrons fill the states $f_1$ and $f_2$ that have a finite overlap, i.e., are nonorthogonal. However, the $N$-particle Slater determinant can also be composed of nonorthogonal states $|f_m\rangle$ if they are linearly independent, i.e., $\det \langle f_m | f_n \rangle \neq 0$. The correctly antisymmetrized and normalized wave function~(\ref{eq:slater_det}) takes the form
\begin{equation}
	\Psi^f(k_1,\dots,k_{\s N}) =
	\frac{1}{\sqrt{N! \det \langle f_m | f_n \rangle}} \det f_m(k_n).
	\label{eq:slater_det_non}
\end{equation}
Substituting this expression in Eq.~(\ref{eq:chiN}) and repeating the calculations leading to Eq.~(\ref{eq:chiN_det}), we obtain the characteristic function as the ratio of determinants of two $N\times N$ matrices
\begin{align}
	\chi_{\s N}(\lambda)
	& = \frac{ \det \langle f_m | 1 - T + T e^{i \lambda} |
	f_n \rangle}{\det \langle f_m | f_n \rangle} \nonumber\\
	& = \frac{\det (S^f - T^f + T^f e^{i \lambda})}
	{\det S^f},
	\label{eq:chiN_det_non}
\end{align}
where the matrices $S^f$ and $T^f$ are defined by
\begin{equation}
	S^f_{mn} = \langle f_m | f_n \rangle, \qquad
	T^f_{mn} = \langle f_m | T | f_n \rangle.
	\label{eq:matrices}
\end{equation}

\subsection{Invariance of the Slater determinant under linear transformations}
\label{sec:invariance}

Expression~(\ref{eq:chiN_det_non}) for the characteristic function can be considerably simplified and presented in the form describing a generalized binomial distribution~\cite{Hassler08,Abanov08,Abanov09}. A significant feature of such a representation of characteristic function is the fact that the Hilbert space $H_{\s N}$ of dimension $N$ spanned by the set of one-particle states with the wave functions $f_n(k)$ determines unique correctly antisymmetrized wave function (unentangled state) or, in other words, only one Slater determinant exists (up to a phase) for $N$ particles with states from $H_{\s N}$. The antisymmetrized $N$-particle (unentangled) state is thus determined by the Hilbert space $H_{\s N}$, i.e., by all states in the set, and is independent of the particular choice of an orthonormalized basis~\cite{Roothaan51}.

To clarify this, we consider a simple case of the two-particle Slater determinant in the second quantization representation $|\Psi\rangle = a_2^\dag a_1^\dag |0\rangle$ with the vacuum state $|0\rangle$ and fermion operators $a_{1,2}$. Defining the new operators $a_\pm = (a_1 \pm a_2) / \sqrt{2}$, we see that the two-particle state
\begin{equation}
	a_+^\dag a_-^\dag |0\rangle
	= \frac{1}{2} (a_1^\dag + a_2^\dag)
	(a_1^\dag - a_2^\dag) |0\rangle
	= a_2^\dag a_1^\dag |0\rangle
	= |\Psi \rangle
	\label{eq:invariance2}
\end{equation}
remains invariant.

We now consider the $N$-particle Slater determinant in form~(\ref{eq:slater_det_non}). After the passage from basis states $f_m(k)$ to new states $g_m(k)$ via a complex linear transformation
\begin{equation}
	g_m(k) = \sum\limits_n A_{mn} f_n (k), \qquad \det A \neq 0,
	\label{eq:linear}
\end{equation}
the antisymmetric combination
\begin{equation}
	\det g_m(k_n) =
	(\det A) \, \det f_m(k_n)
	\label{eq:f_trans_g}
\end{equation}
remains invariant up to the factor $\det A$; here, we took into account that the determinant of the product of matrices is equal to the product of their determinants. In addition, the normalized $N$-particle determinant states $\Psi^f$ and $\Psi^g$ satisfy the relation
\begin{equation}
	\Psi^g (k_1, \dots, k_{\s N}) =
	{\rm sign}(\det A) \Psi^f (k_1,\dots,k_{\s N}),
	\label{eq:psif_psig}
\end{equation}
where ${\rm sign}(x) = x/|x|$. The only effect of using the new basis is the appearance of the overall factor ${\rm sign}(\det A)$, which does not enter expression~(\ref{eq:chiN}) for the characteristic function. The FCS in bases $f$ and $g$ is therefore the same.

{\it Diagonalization.} The invariance of the determinant can be used to simplify the FCS. Moreover, even without specifying the scatterer type, it is possible to understand the FCS qualitatively. In particular, we can assert that the statistics for states of the Slater determinant type always reduces to a generalized binomial form (which is valid for a single-lead conductor (two-contact wire), but not, generally speaking, for multilead conductors~\cite{Kambly09}).

We first consider how the invariance of determinant~(\ref{eq:f_trans_g}) is manifested in Eq.~(\ref{eq:chiN_det_non}). We note that any one-particle matrix $B$ of form~(\ref{eq:matrices}) is transformed by $A$ as
\begin{equation}
	B^g = A^\dag B^f A, \qquad
	B = S, T.
	\label{eq:trans_M}
\end{equation}
Since $\det (A B)= \det A \det B$, it follows that $\chi_{\s N}$ is invariant under the change of basis
\begin{equation}
	\chi_{\s N} = \frac{\det X^f}{\det S^f}
	= \frac{|\det A\,|^2 \det X^f}
	{|\det A\,|^2 \det S^f}
	= \frac{\det X^g}{\det S^g},
	\label{eq:trans_chi}
\end{equation}
where $X^f \equiv S^f - T^f + T^f e^{i\lambda}$. This invariance can be used to pass to a new orthogonal set $g_m(k)$ with the overlap matrix $S^g_{mn} = \delta_{mn}$ and the transparency matrix taking the diagonal form $T^g_{mn} = \tau_m \delta_{mn}$. The possibility to diagonalize the matrices $T^g_{mn}$ and $S^g_{mn}$ simultaneously follows from transformation law~(\ref{eq:trans_M}) for bilinear forms (in contrast to a linear transformation $\mathsf{L}$, which acts as $\mathsf{L}^g = A^{-1} \mathsf{L}^f A$), taking the positive definiteness of $S^g_{mn}$ into account. The corresponding basis $g_m$ and the eigenvalues $\tau_m$ of the $T^g_{mn}$ matrix are found from the generalized eigenvalue problem
\begin{equation}
	(T^f - \tau_m S^f) a_m =0
	\label{eq:gen_diag}
\end{equation}
with the normalization $a_m^\dag S^f a_m^{\vphantom\dag} = 1$.\footnote{This is used, for example, in the Bogoliubov transformation, where a quadratic Hamiltonian is diagonalized under the condition that the form of commutation relations be preserved (see Sec.~\ref{sec:BdG}).} The eigenvectors $a_m$ compose the columns of the transformation matrix $A = (a_1, \dots, a_{\s N})$. The eigenvalues are determined by the roots of the characteristic polynomial $\det (T^f - \tau\, S^f) = 0$. Expression~(\ref{eq:chiN_det_non}) for $\chi_{\s N}$, written in the basis $g_m(k)$ becomes a generalized binomial function,
\begin{equation}
	\chi_{\s N}(\lambda) =
	\prod_{m=1}^N (1- \tau_m + \tau_m e^{i\lambda}),
	\label{eq:binomial}
\end{equation}
where the determinant is calculated explicitly, and the result depends only on the eigenvalues $\tau_m$.

The generalized eigenvalue problem can be reduced to the usual one by passing to the orthonormalized basis $\phi_m(k)$ with $S^\phi = \openone_{\s N}$, which can be obtained using the Gram-Schmidt diagonalization procedure, with $\phi_m(k) = \sum_n [(S^f)^{-1/2}]_{nm} f_n(k)$.

We see from the foregoing that eigenvalue problem~(\ref{eq:gen_diag}) is independent of the basis, while the eigenvalues and eigenvectors are specified by the transparency operator $T$ acting in the Hilbert space $H_{\s N}$ endowed with the scalar product $\langle f | g \rangle$. Using the language of quadratic forms, these conclusions mean that the eigenvalues and eigenvectors can be found by using the positive definite quadratic form $T(g) = \langle g | T | g \rangle$ and $S(g) = \langle g | g \rangle$, $g \in H_{\s N}$. Representing the bilinear form $T(g)$ with $S(g) = 1$ as a polar plot with the radius $T(g)$, where $g$ determines the direction in $H_{\s N}$, we obtain an ellipsoid in the $N$-dimensional space. The lengths of the major axes of this ellipsoid are given by the eigenvalues, while the corresponding directions are the eigenvectors of our problem~(\ref{eq:gen_diag})~\cite{Courant93Book}. The eigenvalues $\tau_m$ are restricted to the interval $[0, 1]$ and $0 \leqslant T(g) \leqslant S(g)$ by virtue of the unitarity property. Such a description can be used to analyze the general properties of the generalized binomial distribution function~\cite{Hassler07}.

\subsection{Constant voltage}
\label{sec:fcsstatdesc}

The measurement of scattering characteristics of individual electrons in meso- and nanoconductors, as it is performed, for example, for particles in accelerators, is complicated.\footnote{As mentioned in Sec.~\ref{sec:fcs2electrons}, methods for sending isolated electrons~\cite{Feve07,Mahe10} at specified instants (``on request'') in quantum conductors were developed only recently.} It is much simpler to study the mean current or current correlators measured at a constant voltage (in this case, a large number of electrons are involved in transport). At the same time, this case of the FCS is much more complicated than the cases with a fixed number of particles considered in Sec.~\ref{sec:fcsNelectrons} and described by wave packets with the known shape. The problem is that a fermion reservoir emits the number of particles that is unknown beforehand, and we can assume that this number experiences quantum fluctuations. However, these fluctuations are small enough and the transfer statistics (at zero temperature) is almost binomial, as in the case of a fixed number of particles considered in Sec.~\ref{sec:fcsNelectrons}. Since the pair correlator at constant voltage in the quantum case exactly coincides with the pair correlator for the Bernoulli process, the hypothesis that the distribution function for the number of transmitted electrons is binomial appeared soon after the result for noise was obtained in Ref.~\cite{Lesovik89b}.

The confirmation of this hypothesis, however, proved to be not simple~\cite{Levitov93,Levitov96}. We do not derive the binomial statistics rigorously here, although almost all the elements required for this derivation have already been presented above, and only briefly outline the corresponding stages of the derivation in the spirit of~\cite{Hassler08}, which we followed above. Describing a constant voltage requires packets as in expression~(\ref{eq:wavePacket}), which are displaced during the observation time over a distance much smaller than their width, rather than the localized packets with $N$ particles used in Sec.~\ref{sec:fcsNelectrons}, which are all scattered from a barrier after a long time. To obtain the characteristic function, matrices~(\ref{eq:matrices}) still have to be found. The determinant of a Toeplitz matrix obtained as a result can be calculated using the Szego theorem~\cite{Hassler08} or, as in Ref.~\cite{Schonhammer07}, with the help of the relation $\log [\det(1 + M)] = {\rm Tr} [\log(1 + M)]$.

Let us now present the result obtained in Refs.~\cite{Levitov93,Levitov96} using the second quantization representations and other elements used in the subsequent versions of the derivation. Calculations for $t_0 \Theta \gg \hbar$ and $t_0 eV \gg \hbar$ give the characteristic functions
\begin{multline}
	\log\chi(\lambda)=\frac{2t_0}{h}\sum\limits_n\int\limits^{\infty }_{-\infty} dE \,
	\log\bigl[ 1+T_n(E)(e^{i\lambda }-1) \times \\ \times f_{\rs L}(1-f_{\rs R}) +
	T_n(E)(e^{-i\lambda}-1)f_{\rs R}(1-f_{\rs L}) \bigr].
	\label{eq:lnCharF}
\end{multline}
For the distribution function $f_{\rs L/R} = 1/[e^{(E \pm eV/2)/\Theta}+1]$ and the energy-independent transparency, the integral in Eq.~(\ref{eq:lnCharF}) is $-\Theta \log x_1 \log x_2$, where $x^2_{1,2} - ux_{1,2} + w = 0$, $w = e^{eV/\Theta}$, $u = G(\lambda)e^{eV/\Theta} + G(-\lambda)$, and $G(\lambda) = 1 + T(e^{i\lambda}-1)$. In the limit $\Theta \ll eV$, the result is simplified, and we obtain for the shot noise statistics
\begin{equation}
	\chi(\lambda)=
	\big[1+T(e^{i\lambda}-1)\big]^{2eVt_o/h}.
\end{equation}
To find the probability $P_m$ of the transfer of $m$ electrons, it is necessary to perform the Fourier transform of $\chi(\lambda)$ to obtain the binomial distribution $P_{mN} = p^m q^{N-m} C^m_{\s N}$ with $p = T$, $q = 1-T$, and $N = 2eVt_0 / h$.

In the two limit cases $T \to 0$ and $T \to 1$, the binomial distribution reduces to the Poisson distribution. The first case corresponds to the classical shot noise, and the second one to transport in a system almost without reflections. We note that in the second case, the distribution of reflected particles, rather than the transmitted ones, is Poissonian.

For $\Theta =0$ and $eV \neq 0$, the distribution is close to the binomial Bernoulli distribution with the ``success'' probability $p=T$, the ``failure'' probability $q=1-T$, and the number of events $N = 2eV t_0 / h$ linearly increasing in time. This is caused by the almost regular sequence of ``tunneling attempts'' occurring at the frequency $\nu_0 = eV / h$. While the result for the probability of tunneling events is quite clear intuitively, the smallness of fluctuations of the number of events during the measurement time is somewhat unexpected, suggesting the existence of an almost periodic process in the system with the frequency weakly fluctuating about $\omega_0 = eV / \hbar$.\footnote{Such fluctuations are related to the fluctuations, logarithmic in time, in the number of tunneling attempts (see the details in Ref.~\cite{Levitov93} and the discussion of logarithmic corrections to cumulants in Ref.~\cite{Hassler08,Schonhammer07}).}

It is clear that the regularity of tunneling attempts is caused in one way or another by the Pauli principle. But the literal interpretation of the electron transfer process in the spirit of the consideration presented in Sec.~\ref{sec:fcsNelectrons}, where a particle falls on the barrier once during the time interval $\tau_{\s V} = h / eV$, encounters difficulties. The wave packets of such particles should have a size of the order of $\delta k = eV / \hbar v_{\rs F}$ in the $k$-space, which means, as we have seen, that the tunneling probability is the mean of the transparency over the energy interval $\delta E \approx eV$. This picture does not correspond to expression~(\ref{eq:lnCharF}), in which the characteristic function is the product of components for each energy and the charge transfer processes at different energies are independent.\footnote{We note that such a factorization is in fact valid only for energy intervals specified by the inverse observation time $\delta E=\hbar /t$ and, according to Levitov-Lesovik formula~(\ref{eq:lnCharF}), is correct only if transparencies are independent of energy at such scales.} The characteristic frequency $\omega_0 = eV/\hbar $ specified by voltage can be directly manifested only over short times, for example, if we study the corresponding charge fluctuations. We present the general relation useful in this case:
\begin{equation}
	\frac{d^2 \dlangle Q_{x_0}^2(t)\drangle}{dt^2} =
	\dlangle j_{x_0}(t)j_{x_0}(0)\drangle +
	\dlangle j_{x_0}(0)j_{x_0}(t)\drangle,
	\label{eq:new}
\end{equation}
where $x_0$ is the detector coordinate. For the excess current correlator, i.e. for the difference between current coordinates $\Delta x \ll v_{\rs F}/ \omega_0$ and the energy-independent transparency, we have~\cite{Lesovik94b}
\begin{equation}
	\dlangle j(0)j(t)\drangle =
	\frac{2e^2}{\pi ^2} T(1-T) \sin^2(\omega_0 t/2)/t^2.
	\label{eq:cucur}
\end{equation}
Interference between different Fermi surfaces occurs at short times, resulting in current and transferred charge oscillations. Similar oscillations also occur in higher-order correlators. A time-unordered third-order correlator was calculated in Ref.~\cite{Bayandin11} (this result was used in Ref.~\cite{Bachmann00}). The third-order correlator depends on coordinates in a more complicated way because of specific interference in the scatterer region. We present the results for two different cases here. In the first case, if current is measured far from the scattering region, $|x_i|\gg v_{\rs F} \tau_{\s V}$, $|t_i-t_j|\sim \tau_{\s V}$, the correlator with coinciding coordinates and at zero temperature has the form
\begin{multline}
	\dlangle \hat{I}(t_1,x) \hat{I}(t_2,x) \hat{I}(t_3,x) \drangle \\
	=-\frac{e^3}{4\pi^3} \,T(1-T)(1-2T) \\
	\times \frac{
		\sin \omega_0 (t_1-t_2)
		+ \sin \omega_0 (t_3-t_1)
		+ \sin \omega_0 (t_2-t_3)
	}{
		(t_1-t_2)(t_3-t_1)(t_2-t_3)
	}.
\end{multline}
The second case is possible near the scattering region. Formally precisely at the scattering point $x = 0$ at zero temperature, the dependence of the correlator on the transparency differs from that measured away from the scattering region,
\begin{multline}
	\dlangle \hat{I}(t_1,x) \hat{I}(t_2,x) \hat{I}(t_3,x)\drangle_{x=0}
	= \frac{e^3}{2\pi^3} \,T^2(1-T) \\
	\times \frac{
		\sin \omega_0 (t_1-t_2) 
		+ \sin \omega_0 (t_3-t_1) + \sin \omega_0 (t_2-t_3)
	}{
		(t_1-t_2)(t_3-t_1)(t_2-t_3)
	}.
	\label{eq:x0}
\end{multline}
We note that the dependence of this correlator on the transparency $T^2 (1-T)$ coincides with that for the third-order charge cumulant determined from expression~(\ref{eq:fcsll}). The layout of a thought experiment for measuring a symmetrized third-order correlator is considered in Ref.~\cite{Bayandin08}.

Transfer statistics at short times have been poorly investigated to date, although they are no less interesting than those at long times. We know only two papers~\cite{Hassler08,Schonhammer07} in which the short-time statistics were considered.

For the long-time statistics, the third-order charge correlators were measured in Ref.~\cite{Bomze05,Gershon08}. Third-order current correlators were measured in Ref.~\cite{Timofeev07} by detecting variations in the dynamics of a Josephson contact; the voltage correlators were also measured earlier in Ref.~\cite{Reulet10}.

At the same time, more complicated measurements of the total statistics or the characteristic function have been performed so far only for incoherent transport: the authors of~\cite{Gustavsson06} have managed to literally count individual electrons. We note, however, that this situation is not desperate, and qubits available at laboratories today can be used as measuring spins. For example, charge qubits based on double-well potentials with one electron~\cite{Petersson10} can be used for measuring statistics at relatively short times, which, however, can be longer than comparable to $\hbar / \Theta$ and $\hbar / eV$.

As we mentioned above, the presence of a wire near a qubit leads to the qubit decoherence. It is interesting that since the characteristic function is periodic in $\lambda$, the decoherence should also be periodic, or at least its dependence on the coupling constant should be nonmonotonic. From Eq.~(\ref{eq:decoher}), in particular, we can obtain the phase breaking time for a qubit as
\begin{equation}
	\tau^{-1} =
	\left|\frac{eV}{h} \log \left[ 1-4T(1-T)\sin^2\frac{\lambda}{2} \right] \right|.
	\label{eq:taudecoher}
\end{equation}
We see that the phase breaking is especially large at $T = 1/2$, when noise is maximal. Then
\begin{equation}
	\tau^{-1} =
	\left| \frac{eV}{h} \log\left[ \cos^2 \frac{\lambda}{2} \right] \right|
	\label{eq:taudecoher12}
\end{equation}
and the phase breaking is completely absent if $\lambda = 2n\pi$, while for $\lambda = (2n+1)\pi$, the time formally tends to zero. In this case, it is more correct to return to the definitions of the characteristic function presented above, from which it follows that the spin (qubit) phase rotates after the passage of one electron through exactly $\pi$, which can be treated as the complete phase breaking, because the obtained spin (qubit) state is orthogonal to the initial state. In this case, the maximal entanglement of the qubit state with the flying-electron state occurs (if the state of the latter is characterized only by one discrete variable taking ``transmitted'' or ``reflected'' values; see Ref.~\cite{Lesovik10}).

The appearance of singularities of the characteristic function for $T = 1/2$ on the formal level was pointed out in Ref.~\cite{Ivanov10} and interpreted as a ``phase transition'' between thermodynamic phases in the time space. We have described above the physical nature of this phenomenon.

\subsection{Complete description of the FCS for the known transparency statistics}

As we discussed in Sec.~\ref{sec:fcsstatdesc}, if the probability of transmission of electrons through a quantum conductor is known, then the FCS at large times can be described completely. In turn, the transparency, which can be treated as a random quantity (meaning an irregular dependence on the scattering potential parameters), can also be described for some conductors by its own distribution function. Therefore, we can introduce the ``total'' charge distribution function, taking both dynamic fluctuations and transparency fluctuations from sample to sample into account.

We begin with the simple example of a ballistic conductor with a cavity, for which the transparency distribution function in the quasi-one-dimensional case is trivial~\cite{Jalabert94}. In the presence of a weak magnetic field, but such that more than one flux quantum passes through a two-dimensional asymmetric cavity connected with reservoirs by two one-channel leads, the probability $T$ of transmission through this system is uniformly distributed over the segment $[0, 1]$, i.e.,
\begin{equation}
	P(T) = 1.
	\label{eq:PT1}
\end{equation}
The probability of transferring the charge $en$ in time $t$, when the transparency is unknown beforehand but the distribution function $P_{\s T}(T)$ is known, reduces to the integral of the charge distribution function $P_{\s Q}(Q)$ over transparencies with the weight $P_{\s T}(T)$:
\begin{equation}
	\langle P(Q) \rangle =
	\int\limits_0^1 dT P_{\s T}(T) P(Q).
	\label{eq:PQi}
\end{equation}
The characteristic function is averaged similarly.

The characteristic function of the binomial distribution averaged with~(\ref{eq:PT1}) has the form
\begin{align}
	\langle \chi_b (\lambda) \rangle
	= & \int\limits_0^1 dT \bigg[1+T\left(\exp{(i\lambda)}-1\right)\bigg]^{ N } =
	\nonumber \\
	= & \frac{\exp{[i\lambda (N+1)]} -1}{(\exp{(i\lambda)}-1)(N+1)}.
	\label{eq:chi_b}
\end{align}
For an integer $N$, characteristic function~(\ref{eq:chi_b}) can be easily integrated, and we obtain a interesting result for the distribution function
\begin{equation}
	P(k) = \sum\limits_{n=0}^{N}\frac{1}{1+N} \, \delta (k+n),
\end{equation}
which means that the transfer of any number of electrons, beginning from zero and ending with the maximum value $N = 2eVt / h$, is equiprobable. This is caused, in particular, by the boundedness of the binomial distribution function.

The more important case is the dirty conductor considered in Sec.~\ref{sec:dorokhov}, for which the transparency distribution is described by the Dorokhov function. The second cumulant was calculated in Ref.~\cite{Beenakker92b}, while the results for the characteristic function, allowing the calculation of all the moments, were obtained in Ref.~\cite{Lee95}.

For a dirty conductor in which transparencies are described by the Dorokhov distribution function, the noise is three times lower than the Poisson one, or, in terms of the Fano factor, $F = 1/3$, as was shown in Ref.~\cite{Beenakker92b}. This is a consequence of the bimodal nature of the transparency distribution function. Higher moments (cumulants) can also be obtained quite simply by integration; moreover, the generating function for all cumulants, i.e., the mean of the logarithm of the characteristic function, can be obtained as in Ref.~\cite{Lee95},
\begin{equation}
	\overline{\log \chi(\lambda)} =
	\frac{GVt}{e} {\rm arcsinh}^2 \sqrt{e^{i\lambda} - 1}.
	\label{eq:lyl_3}
\end{equation}
In~\cite{Lee95} the first 10 cumulants are also presented explicitly. We present the first four cumulants:
\begin{align}
	& \overline{\dlangle N(t) \drangle} = N_0, \\
	& \overline{\dlangle N^2(t) \drangle} = N_0/3, \\
	& \overline{\dlangle N^3(t) \drangle} = N_0/15, \\
	& \overline{\dlangle N^4(t) \drangle} = -N_0/105, \ldots
	\label{eq:lyl_11}
\end{align}
The first cumulant simply gives the first transferred charge, the second gives the result for noise specifying the dispersion, which we already know, the third characterizes the asymmetry (skewness) of the distribution function (with respect to its top), and the fourth determines the deviation of the distribution function from being Gaussian. The higher cumulants are described by the expressions
\begin{equation}
	\overline{\dlangle N^k(t) \drangle} \sim
	\frac{N_0}{(2\pi)^{k-1}}
	\frac{(k-1)!}{\sqrt{k}}
	\left\{\begin{array}{l}
		(-1)^{(k+2)/2} \mbox{ for even } k, \\
		(-1)^{(k+1)/2} \mbox{ for odd } k. \\
	\end{array}\right.
	\label{eq:lyl_13}
\end{equation}
For comparison, in the case of the Poisson distribution, we have $\overline{\dlangle N^k(t) \drangle} = N_0$ for $k>0$, while for the Gaussian distribution function, as is known, all the cumulants higher than the second are zero.

\subsection{FCS in graphene}

It is surprising that the transparency distribution in {\it pure} graphene (in the case of many conducting channels and a zero doping level) is the same as that for a dirty conductor. This property of the transport ``pseudodiffuseness'' is confirmed by the measurements of noise~\cite{Danneau08}, for which the Fano factor turned out to be $1/3$, as had been predicted in Ref.~\cite{Tworzydlo06}. Scattering in pure graphene occurs at its boundaries, in the absence of doping, the transport being completely provided by decaying modes, which we always neglected above. We do not consider the difference between scattering properties for the Dirac and Schr\"odinger equations here, but we return to this question in Sec.~\ref{sec:SNS}. The Landauer approach to the description of transport in graphene was used in Ref.~\cite{Katsnelson06}.

\subsection{FCS in the presence of interaction}

The problem of the scattering statistics for two electrons can also be solved in the presence of the electron-electron interaction when it is concentrated in the quantum dot region. In this case, the scattering matrix can be found either exactly~\cite{Lebedev08} or by using the perturbation theory~\cite{Goorden07}. This allows describing the result of scattering of two particles in detail, notably, the entanglement appearing in this process~\cite{Lebedev08} and transport in two conductors indirectly interacting via quantum dots~\cite{Goorden07}. It is interesting that in the problem with a constant voltage, the interaction (in the low-voltage limit) does not change the form of characteristic function~(\ref{eq:lnCharF}), although transparencies turn out to be renormalized in a complicated way~\cite{Gogolin07}. (But we note that it is not quite clear at the moment how universal this result can be.)

The description with the help of scattering matrices can be extended to the case of electrons interacting with other degrees of freedom. It has been found that, by developing the theory of emission of photons (or other electromagnetic modes, for example, plasmons) by coherent conductors, it is possible to express photon emission rates~\cite{Lesovik97b} or correlators of the number of photons at different points~\cite{Beenakker01,Lebedev10} in terms of scattering matrices in the conductor. This possibility appears because the wavelengths of emitted photons greatly exceed the characteristic length of a scatterer. Under these conditions, the interaction Hamiltonian
\begin{equation}
	{\hat H}_\mathrm{int}
	= -\frac1c \int d{\bf r} \, \hat{\bf j}({\bf r})
	\hat {\bf A}({\bf r})
	\label{eq:hamAj},
\end{equation}
containing integrals over the coordinates of exact wave functions (scattering states), reduces to integrals of the coefficients of the scattering matrices and second quantized operators. In intermediate calculations, all the quantities for photon correlators are reduced to the convolutions of current correlators at low frequencies, which are independent of coordinates and are expressed in terms of scattering matrices. The same approach (which can be called the perturbation theory in the interaction based on exact scattering wave functions) can be used to describe the electron-phonon interaction if the characteristic wavelengths of phonons are much longer than the characteristic length of the scatterer.

An important case where the interaction can be described by scattering matrices is the contact of a superconductor with a normal quantum conductor or a Josephson contact of two superconductors through a normal interlayer, which can be a barrier, a two-barrier system, a dirty (coherent) conductor, or a graphene film. We consider these cases in detail in Secs.~\ref{sec:NS} and \ref{sec:SNS}.


\section{The Bogoliubov-de Gennes equations}
\label{sec:BdG}

We turn to the description of the quantum transport in superconducting systems. In this section we describe a superconducting system, in general case spatially inhomogeneous, using the Bogoliubov transformations~\cite{deGennes68,deGennes68Book,Svidzinsky82Book,Kittel87}.

We first consider the many-particle effective Bardeen-Cooper-Schrieffer (BCS) Hamiltonian
\begin{align}
	{\hat H} =
	& \sum\limits_\sigma \int d{\bf x} \, {\hat\psi}_\sigma^\dag({\bf x}) \, \biggl[\frac{{\bf\hat P}^2}{2m}
		- \bar{\mu}({\bf x})
	\biggr] \, {\hat\psi}_\sigma^{\phantom \dag}({\bf x})
	\nonumber \\
	& + \int d{\bf x} \, \big[ {\bf\Delta}({\bf x}) \, {\hat\psi}_\uparrow^\dag({\bf x}) \, {\hat\psi}_\downarrow^\dag({\bf x}) + {\rm H.c.} \big],
	\label{eq:BdG_H}
\end{align}
where integrals are taken over the entire volume of the system and ${\bf x} = (x, y, z)$. The first term in the right-hand side of Eq.~(\ref{eq:BdG_H}) is kinetic and contains the operator ${\bf\hat P}^2 / 2m$ determining the quadratic dispersion of the system; here, ${\bf\hat P} = -i\hbar{\bf\nabla} - (e/c) {\bf A}$. The summation is taken over spins $\sigma = \uparrow, \downarrow $. $\bar{\mu}({\bf x}) = \mu - eV({\bf x})$ is the chemical potential in a superconductor or a normal conductor, where $\mu$ is still the electrochemical potential, which is assumed to be a constant defined in the superconductor.\footnote{In Secs.~\ref{sec:landauer}--\ref{sec:fcs} we used the electrochemical potentials~$\mu$ and their differences which determined the deviation of the system from the equilibrium state. Here we additionally consider the chemical potential~$\bar\mu$ related to the local density of the charge involved in the formation of the superconductivity.} The second term in the right-hand side of Eq.~(\ref{eq:BdG_H}) is responsible for the superconductivity, it is associated with the complex order parameter in the superconductor. $\Delta({\bf x}) \equiv |{\bf\Delta}({\bf x})|$ is the superconducting gap. In the general case, the superconducting parameter ${\bf\Delta}({\bf x})$, which is calculated by averaging over phonon degrees of freedom responsible for the superconductivity, depends on the state of the electron system. Below the self-consistent approximation will be used.

We replace the wave functions by a linear combination of new wave functions $u_\nu({\bf x})$ and $v_\nu({\bf x})$:
\begin{equation}
	\left\{\begin{array}{l}
		{\hat\psi}_{\sigma}^{\phantom{\dag}}({\bf x}) =
		\displaystyle\sum\limits_\nu \big\{
			u_\nu^{\phantom{*}}({\bf x}) {\hat a}_{\nu, \sigma}^{\phantom{\dag}} +
			{\rm sign}(\sigma) \, v_\nu^*({\bf x}) {\hat a}_{\nu, -\sigma}^\dag
		\big\}, \\
		{\hat\psi}_{\sigma}^\dag({\bf x}) =
		\displaystyle\sum\limits_\nu \big\{
			u_\nu^*({\bf x}) {\hat a}_{\nu, \sigma}^\dag +
			{\rm sign}(\sigma) \, v_\nu^{\phantom *}({\bf x})
				{\hat a}_{\nu,-\sigma}^{\phantom \dag}
		\big\}.
	\end{array}\right.
	\label{eq:BTrans}
\end{equation}
The summation over states $\nu$ means the summation over the discrete spectrum and the integration over the continuous spectrum. Such a substitution in the Hamiltonian is called the Bogoliubov transformation.

The operators of free electrons satisfy the standard commutation relations for Fermi particles:
\begin{align}
	& [{\hat\psi}_\sigma^\dag({\bf x}),\, {\hat\psi}_{\sigma'}^{\phantom \dag}({\bf x}')] =
	\delta_{\sigma,\sigma'} \delta({\bf x}-{\bf x}'),
	\label{eq:BCanon11}
	\\
	& [{\hat\psi}_\sigma({\bf x}),\, {\hat\psi}_{-\sigma}({\bf x}')] = 0.
	\label{eq:BCanon12}
\end{align}
Let us require that the new operators also satisfy the commutation relations for Fermi particles reflecting the canonical character of the transformation~(\ref{eq:BTrans})~\cite{Svidzinsky82Book}:
\begin{align}
	& [{\hat a}_{\nu,\sigma}^\dag,\, {\hat a}_{\nu',\sigma'}^{\phantom \dag}] =
	\delta_{\sigma,\sigma'} \delta_{\nu,\nu'},
	\label{eq:BCanon00} \\
	& [{\hat a}_{\nu,\sigma},\, {\hat a}_{\nu',\sigma'}] = 0.
	\label{eq:BCanon01}
\end{align}
Then
$\langle {\hat a}_{\nu,\sigma}^\dag {\hat a}_{\nu',\sigma'}^{\phantom \dag} \rangle = \delta_{\sigma,\sigma'} \delta_{\nu,\nu'} f(\varepsilon_\nu)$,
where, as above, $f(\varepsilon)$ is the Fermi distribution function. It can be shown that conditions~(\ref{eq:BCanon11})--(\ref{eq:BCanon01}) lead to relations for the coefficients $u({\bf x})$ and $v({\bf x})$ in Eq.~(\ref{eq:BTrans}):
\begin{align}
	& \sum\limits_\nu \big\{ u_\nu^*({\bf x}) u_\nu^{\phantom *}({\bf x}') +
	v_\nu^{\phantom *}({\bf x}) v_\nu^*({\bf x}') \big\} =
	\delta({\bf x}-{\bf x}'),
	\label{eq:BuvCond1} \\
	& \sum\limits_\nu \big\{ u_\nu^*({\bf x}) v_\nu^{\phantom *}({\bf x}') -
	v_\nu^{\phantom *}({\bf x}) u_\nu^*({\bf x}') \big\} = 0
	\label{eq:BuvCond2}
\end{align}
and
\begin{align}
	& \int d{\bf x} \, \bigl[ u_\nu^\pstar({\bf x}) u_{\nu'}^*({\bf x}) +
	v_\nu^\pstar({\bf x}) v_{\nu'}^*({\bf x}) \bigr] =
	\delta_{\nu,\nu'},
	\label{eq:BuvCond3}
	\\
	& \int d{\bf x} \, \bigl[ u_\nu^\pstar({\bf x}) v_{\nu'}^\pstar({\bf x}) -
	v_\nu^\pstar({\bf x}) u_{\nu'}^\pstar({\bf x}) \bigr] = 0.
	\label{eq:BuvCond4}
\end{align}

Transformation~(\ref{eq:BTrans}) diagonalizes the Hamiltonian~(\ref{eq:BdG_H}) reducing it to the form
\begin{equation}
	{\hat H} =
	U_0 +
	\sum\limits_{\sigma,\nu} \varepsilon_\nu
	{\hat a}_{\nu,\sigma}^\dag {\hat a}_{\nu,\sigma}^{\phantom \dag}
	\label{eq:BdG_Hdiag}
\end{equation}
if the coefficients $u_\nu({\bf x})$ and $v_\nu({\bf x})$ satisfy the second-order differential equations
\begin{equation}
	\left\{\begin{array}{l}
	\biggl[ \cfrac{{\bf\hat P}^2}{2m} - \bar{\mu}({\bf x}) \biggr] \,
	u_\nu^{\phantom *}({\bf x}) -
	{\bf\Delta}({\bf x}) v_\nu^{\phantom *}({\bf x}) =
	\varepsilon_\nu u_\nu^{\phantom *}({\bf x}),
	\vspace{1 mm} \\
	\biggl[ \cfrac{{\bf\hat P}_{\rm c}^2}{2m} - \bar{\mu}({\bf x}) \biggr] \,
	v_\nu^{\phantom *}({\bf x}) +
	{\bf\Delta}^*({\bf x}) u_\nu^{\phantom *}({\bf x}) =
	-\varepsilon_\nu v_\nu^{\phantom *}({\bf x}),
	\end{array}\right.
	\label{eq:BdGeqsFull}
\end{equation}
where ${\bf\hat P}_{\rm c} = {\bf\hat P}|_{e \to -e}$. The energy $U_0$ plays the role of the ground-state energy of the system,
\begin{multline}
	U_0 =
	\int d{\bf x} \sum\limits_\nu
	\biggl\{
		v_\nu^{\phantom *}({\bf x}) \biggl[ \cfrac{{\bf\hat P}^2}{2m} 
		- \bar{\mu}({\bf x}) \biggr] v_\nu^*({\bf x}) \\
		+ u_\nu^*({\bf x}) \biggl[ \cfrac{{\bf\hat P}^2}{2m} 
		- \bar{\mu}({\bf x}) \biggr] u_\nu^{\phantom *}({\bf x})
	\biggl\}
	- \sum\limits_\nu \varepsilon_\nu.
	\label{eq:BdGcTermM}
\end{multline}
Equations~(\ref{eq:BdGeqsFull}), which are called the Bogoliubov-de Gennes (BdG) equations, can be interpreted as the wave equation for the two-component wave function
\begin{equation}
	{\hat\Psi}_\nu({\bf x}) =
	\left[ \!\! \begin{array}{c}
		u_\nu({\bf x}) \\
		v_\nu({\bf x})
	\end{array} \!\! \right]
	\nonumber
\end{equation}
of a quasiparticle with dispersion $\varepsilon_\nu$. The first component $u_\nu({\bf x})$ can be treated as the electron-like part of the wave function and the second component $v_\nu({\bf x})$ as the hole-like component. This interpretation can be useful, for example, in the consideration of the Andreev scattering~\cite{Andreev64a,Andreev64b,Andreev65}.

Eqs.~(\ref{eq:BdGeqsFull}) are written sometimes in the matrix form
\begin{eqnarray}
	\left[\begin{array}{cc}
		{\hat H}_0 ({\bf x})	& {\bf\Delta}({\bf x}) \\
		{\bf\Delta}^* ({\bf x})	& -{\hat H}_0 ({\bf x})
	\end{array}\right] \!
	{\hat\Psi}_\nu({\bf x})
	= \varepsilon_\nu {\hat\Psi}_\nu({\bf x}),
	\label{eq:BdGeqsWest}
\end{eqnarray}
where ${\hat H}_0({\bf x}) = -\hbar^2\partial_x^2 / 2m - \bar{\mu}({\bf x})$ is the Hamiltonian of the system in the normal state with the chemical potential $\bar{\mu}({\bf x})$.

The BdG equations are invariant under the transformations $\varepsilon_\nu \to -\varepsilon_\nu$, $u_\nu \to -v_\nu^*$, $v_\nu \to u_\nu^*$, therefore, the set of solutions of Eqs.~(\ref{eq:BdGeqsFull}) is redundant. It can be simply explained in the case when the superconducting parameter is zero. It is clear that in this case, the same initial electron state in terms of the Bogoliubov quasiparticles can be described either by the creation of an electron-like state or by the annihilation of a hole-like state with the opposite energy. In practice, one of the following variants is usually chosen:
\begin{enumerate}[(i)]
\item $\varepsilon > 0$: in this case, summation is done over spins (taking both the electron-like and hole-like states into account), which is convenient, for example, for the description of Josephson contacts and most natural for the description of excitations above the Fermi sea;
\item $\varepsilon \in \mathbb{R}$: in this case, only the electron-like states are taken into account, which can be convenient for describing contacts of a normal conductor with a superconductor.
\end{enumerate}
These approaches are equivalent and can be chosen in accordance with their practicality. We note that in principle other variants are also possible.

In the general case, the superconducting parameter ${\bf\Delta}({\bf x})$ is not free and depends on the state of the electron system and hence on the solutions of the BdG equations. Therefore, in order to solve the BdG equations, the parameter ${\bf\Delta}({\bf x})$ should be known, which in turn is defined by the same equations. The corresponding self-consistent solution can be obtained, for example, by the iteration method, choosing the initial function ${\bf\Delta}_0({\bf x})$ as the initial approximation. We here present the expression for the superconducting gap in terms of $u_\nu$, $v_\nu$, and $\varepsilon_\nu$ without derivation:\footnote{See the detailed derivation, for example, in Ref.~\cite{Svidzinsky82Book}.}
\begin{equation}
	{\bf\Delta}({\bf x}) =
	-|g| \sum\limits_\nu
	u_\nu^{\phantom *}({\bf x}) v_\nu^*({\bf x})
	\tanh(\varepsilon_\nu / 2\Theta),
	\label{eq:BdGscDelta}
\end{equation}
where $\Theta$ is the system temperature and $g$ is the electron-phonon coupling constant, $g < 0$. The thermodynamic potential of the system (also given without derivation) has the form
\begin{multline}
	\Omega =
	\frac{1}{|g|} \int d{\bf x} \, |{\bf\Delta}({\bf x})|^2 -
	2 \Theta \sum\limits_\nu \log \bigl[ 2\cosh(\varepsilon_\nu / 2\Theta) \bigr] \\
	+ \int d{\bf x} \sum\limits_\nu
	\biggl\{
		v_\nu^{\phantom *}({\bf x}) \biggl[ \cfrac{{\bf\hat P}^2}{2m} 
		- \bar{\mu}({\bf x}) \biggr] v_\nu^*({\bf x}) \\
		+ u_\nu^*({\bf x}) \biggl[ \cfrac{{\bf\hat P}^2}{2m} 
		- \bar{\mu}({\bf x}) \biggr] u_\nu^{\phantom *}({\bf x})
	\biggl\}.
	\label{eq:BdGFreeEn}
\end{multline}

The superconducting gap can sometimes be specified ``manually'' and the problem can be solved quite accurately without resorting to self-consistency\footnote{We consider just these cases.} [in other words, we can select a very good initial function ${\bf\Delta}_0({\bf x})$]. For example, in the case of a small normal contact (an island) connected to a massive superconductor(s) via tunneling junctions we can assume the superconducting gap to be constant in the superconductor and zero in the normal metal. Such island forms a small number of states which cannot considerably affect superconductivity in massive reservoirs with a huge number of states. The same takes place for the contact between a normal (massive) conductor and a superconductor via a quasi-one-dimensional conductor. We note that this situation is quite similar to the problem discussed in Sec.~\ref{sec:landauer} about two massive conductors connected via a quasi-one-dimensional conductor, where the distribution function (density matrix) in the reservoir changes negligibly due to the presence of the second reservoir.

The current density operator is given by
\begin{align}
	{\bf\hat j}({\bf x}) = &
	\frac{ie\hbar}{2m} \sum\limits_\sigma
	\Bigl\{
		\big[\nabla {\hat\psi}_\sigma^\dag({\bf x})\big]
		{\hat\psi}_\sigma^\pdag({\bf x}) -
		{\hat\psi}_\sigma^\dag({\bf x})
		\nabla {\hat\psi}_\sigma^\pdag({\bf x})
	\Bigr\} \nonumber \\
	- & \frac{e^2}{m}{\bf A}({\bf x}) \sum\limits_\sigma
	{\hat\psi}_\sigma^\dag({\bf x}) {\hat\psi}_\sigma^\pdag({\bf x}).
	\label{eq:CurrGen}
\end{align}
To rewrite~(\ref{eq:CurrGen}) in terms of coefficients $u_\nu({\bf x})$ and $v_\nu({\bf x})$ in the Bogoliubov transformation, we average the current operator over the density matrix of the system:
\begin{multline}
	\langle {\bf\hat j}({\bf x}) \rangle =
	\frac{ie\hbar}{m} \sum\limits_\nu \Bigl\{
		\big[
			v_\nu^*({\bf x}) \nabla v_\nu^{\phantom *}({\bf x}) -
			v_\nu^{\phantom *}({\bf x}) \nabla v_\nu^*({\bf x})
		\big] \\
		\times [1-f(\varepsilon_\nu)]
		-\big[
			u_\nu^*({\bf x}) \nabla u_\nu^{\phantom *}({\bf x}) -
			u_\nu^{\phantom *}({\bf x}) \nabla u_\nu^*({\bf x})
		\big] f(\varepsilon_\nu)
	\Bigr\} \\
	-\frac{2e^2}{m}{\bf A}({\bf x})\sum\limits_\nu
	\big\{ |v_\nu({\bf x})|^2(1-f(\varepsilon_\nu)) +
	|u_\nu({\bf x})|^2 f(\varepsilon_\nu)\big\}.
	\label{eq:CurrBdG}
\end{multline}


\section{Electron transport in NS junctions}
\label{sec:NS}

At low temperatures the electron dephasing time in sufficiently pure structures can exceed the traveling time through the normal part of a normal metal-superconductor (NS) system. Therefore, the wave functions can be assumed coherent both in the superconductor and outside it. In this case, the scattering matrix approach is especially convenient.

In the standard theory of the proximity effect, the influence of a superconductor on a normal metal can be described in terms of the condensate wave function penetration from the superconductor to the normal metal over the coherence length. This phenomenon can also be interpreted as an appearance of a coherent coupling between electrons and holes in the normal metal caused by Andreev reflection~\cite{Andreev64a} from the boundary of the NS junction, and can be described by BdG equations~(\ref{eq:BdGeqsFull}). Therefore, due to Andreev reflection the quasi-particle current at the NS interface transforms into the superconducting current~\cite{Andreev64a,Blonder82}.\footnote{According to the recently proposed standpoint, the proximity effect is caused by Cooper electron pairs flying into the normal conductor. The wave functions of electrons in these pairs are entangled in a complicated way. The entanglement is inherent both in spin variables (similarly to the entanglement in the Bohm singlet) and in orbital variables (similarly to the Einstein-Podolsky-Rosen entanglement). The entanglement of Cooper pairs penetrating into the normal conductor was studied in Refs.~\cite{Lesovik01,Recher01,Chtchelkatchev02,Bayandin06}.}

The scattering-matrix approach involves the concept of the Bogoliubov quasiparticles with the wave functions containing both electron and hole components~\cite{Beenakker92,Landauer70,Buttiker85}. Here the Bogoliubov free quasiparticles play the same role as the free electrons in the theory of normal conductors, and all aspects of the theory developed for normal conductors can be extended to hybrid systems. In addition, there are also other effects, for example, the Josephson effect, which can also be successfully described within the scattering matrix method.

The strength of the proximity effect is determined by the normal scattering near the NS junction, in particular, by the quality of the boundary, which affects the shape of the current-voltage ($I$-$V$) characteristic. The $I$-$V$ characteristic in NIS junctions, which differ from NS junctions by the presence of a normal scatterer I reflecting electrons to electrons and holes to holes, has already been studied in Ref.~\cite{Blonder82} (see references to earlier papers therein), where linear transport was considered based on the BdG equations within the quasi-one-dimensional model in the presence of one barrier at the boundary.

The scattering matrix approach allows to take an arbitrary scatterer into account for the systems with superconductors. This approach was used first by Takane and Ebisawa~\cite{Takane91,Takane92} and Lambert~\cite{Lambert91,Lambert93}, while Beenakker~\cite{Beenakker92} derived the formula for the linear conductance of the NS junction using the scattering matrix in the normal metal.\footnote{We also recall Anderson's paper~\cite{Anderson59}, in which the independence of the critical temperature of a weak disorder (the Anderson theorem) was formally proved by using the exact wave functions.} Unlike Green's function methods~\cite{Nazarov96a,Nazarov96b,Hekking94,Volkov93}, which were used to describe the experiments in Refs.~\cite{Petrashov95,Courtois96}, the scattering matrix approach cannot take all inelastic processes into account. However, this approach is rather illustrative for simple scattering potentials while for complex potentials it allows to obtain the result in general form.

Following~\cite{Lesovik97}, we consider the conductivity of a NXS junction, where the region X is a scatterer in the normal part, taken at arbitrary temperatures and voltages. We also introduce some general relations, describe the case of dirty contacts, and analyze systems with one or two scatterers in the region X in greater detail.

\subsection{Current-voltage relation and the spectral conductance}
\label{sec:IVrelation}

\begin{figure}[tb]
	\includegraphics[width=5cm]{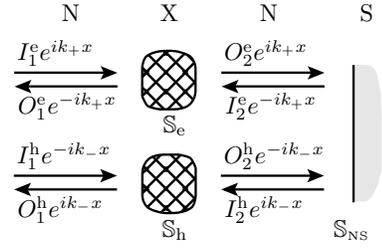}
	\caption{
NS junction scheme. The left normal reservoir has the electrochemical potential $\mu + eV$, the right, superconducting reservoir, has the potential $\mu$; X is an arbitrary normal scatterer, $k_+$ and $k_-$ are the wave vectors of an electron and a hole, respectively.
	}
	\label{fig:disordNS}
\end{figure}

In this section we consider a quasi-one-dimensional multichannel NXS junction, see Fig.~\ref{fig:disordNS}. The structure of the scattering matrix of quasiparticles for such a contact is much more complicated than that for the normal NXN junction. The reason is that in addition to the usual scattering in the normal part there is also Andreev scattering from the NS boundary, where the gap is assumed to jump from zero to the bulk value, and electrons can be reflected to holes and vice versa. To clarify the structure of the scattering processes, we first describe the scattering matrices in the both parts of the contact individually and then consider the full matrix.

The coherent scattering in the normal part is described by the $4N \times 4N$ scattering matrix (we neglect decaying modes in the ballistic segments)
\begin{equation}
	\mathbb{S}_{\rs N} =
	\left[\begin{array}{cccc}
		r_{11}(\varepsilon) & 0 & t_{12}(\varepsilon) & 0 \\
		0 & r_{11}^*(-\varepsilon) & 0 & t_{12}^*(-\varepsilon) \\
		t_{21}(\varepsilon) & 0 & r_{22}(\varepsilon) & 0 \\
		0 & t_{21}^*(-\varepsilon) & 0 & r_{22}^*(-\varepsilon)
	\end{array}\right].
	\label{eq:scattmat}
\end{equation}
Here, $t_{ij}(\varepsilon)$ and $r_{ii}(\varepsilon)$ are the $N \times N$ matrices of transmission and reflection amplitudes in the electron channels. The $\mathbb{S}_{\rs N}$ matrix connects $N$ input electron (hole) channels $I_i^{\rm e}$ ($I_i^{\rm h}$), $i=1,2$, on each side to the output channels of the same energy $O_i^{\rm e}$ $(O_i^{\\,\rm h})$ (see Fig.~\ref{fig:disordNS}):\footnote{Hereafter, it is more convenient to measure all energies in function arguments relative to the electrochemical potential $\mu$ in a massive superconductor, for example, $t_{12}^*(-\varepsilon)$ means $t_{12}^*(\mu -\varepsilon)$. The complex conjugate amplitudes for holes appear because the propagation direction of holes is opposite to the wave vector. The corresponding amplitudes are obtained from the electron ones by complex conjugation, resulting in the appearance of quantities $t^*(-\varepsilon)$ rather than $t(-\varepsilon)$.}
\begin{equation}
	\left[\begin{array}{c}
		O_1^{\rm e} \\ O_1^{\rm h} \\
		O_2^{\rm e} \\ O_2^{\rm h} 
	\end{array}\right] =
	\mathbb{S}_{\rs N}
	\left[\begin{array}{c}
		I_1^{\rm e} \\ I_1^{\rm h} \\
		I_2^{\rm e}\\ I_2^{\rm h}
	\end{array}\right].
	\label{eq:scattmat2}
\end{equation}
The number of channels $N$ is determined by the number of transverse modes; we neglect a change in the number of modes when changing the voltage.\footnote{In principle, the scattering matrix can depend on the applied voltage. For example, it can be used to account for the change of scattering states in the case of voltage-dependent Schottky barrier shape.} Matrix~(\ref{eq:scattmat}) can also be written in the block form
\begin{equation}
	\mathbb{S}_{\rs N} =
	\left[ \begin{array}{cc}\hat{r}_{11}
	(\varepsilon) & \hat{t}_{12}(\varepsilon)\\
	\hat{t}_{21}(\varepsilon) & \hat{r}_{22}(\varepsilon)
	\end{array}\right],
	\label{eq:scattmat3}
\end{equation}
where $\hat {r}_{ii}$ and $\hat {t}_{ij}$ are extended $2N \times 2N$ matrices containing complex conjugate amplitudes for holes. Following the usual procedure, we include the propagation of particles in the ballistic segment between the scatterer X and the NS boundary into the scattering matrix (see Fig.~\ref{fig:disordNS}).

The scattering matrix can often be conveniently represented in the electron-hole parameterization
\begin{equation}
	\bar{\mathbb{S}}_{\rs N} =
	\left[ \begin{array}{cc}
		\mathbb{S}_{\rm e} & 0 \\
		0 & \mathbb{S}_{\rm h}
	\end{array}\right],
	\label{eq:sm_eh1}
\end{equation}
where the $\mathbb{S}_{\rm e}$ and $\mathbb{S}_{\rm h}$ are the submatrices describing the scattering of electrons with energy $\varepsilon$ and holes with energy $-\varepsilon$. These submatrices are composed of the corresponding components of matrix~(\ref{eq:scattmat}), with the slightly modified states in Eq.~(\ref{eq:scattmat2}):
\begin{equation}
	\left[\begin{array}{c}
		O_1^{\rm e} \\ O_2^{\rm e} \\
		O_1^{\rm h} \\ O_2^{\rm h}
	\end{array}\right] =
	\bar{\mathbb{S}}_{\rs N}
	\left[\begin{array}{c}
		I_1^{\rm e} \\ I_2^{\rm e} \\
		I_1^{\rm h} \\ I_2^{\rm h}
	\end{array}\right].
	\label{eq:sm_eh2}
\end{equation}
The scattering described by the $\mathbb{S}_{\rm e}$ and $\mathbb{S}_{\rm h}$ submatrices is shown schematically in Fig.~\ref{fig:disordNS}.

The scattering matrix on the NS interface can be defined in general as
\begin{equation}
	\mathbb{S}_{\rs NS} =
	\left[\begin{array}{cccc}
		r_{\rm ee}(\varepsilon) & r_{\rm eh}(\varepsilon) &
		t'_{\rm ee}(\varepsilon) & t'_{\rm eh}(\varepsilon) \\
		r_{\rm he}(\varepsilon) & r_{\rm hh}(\varepsilon) &
		t'_{\rm he}(\varepsilon) & t'_{\rm hh}(\varepsilon) \\
		t_{\rm ee}(\varepsilon) & t_{\rm eh}(\varepsilon) &
		r'_{\rm ee}(\varepsilon) & r'_{\rm eh}(\varepsilon) \\
		t_{\rm he}(\varepsilon) & t_{\rm hh}(\varepsilon) &
		r'_{\rm he}(\varepsilon) & r'_{\rm hh}(\varepsilon)
	\end{array}\right].
	\label{eq:ascattmat}
\end{equation}
The $\mathbb{S}_{\rs NS}$ matrix relates the wave functions in the normal part and the superconductor,\footnote{In Fig.~\ref{fig:disordNS}, the states in the superconductor are not indicated because we mainly consider scattering amplitudes for those states coming to the superconductor from the normal part.}
\begin{equation}
	\left[\begin{array}{c}
		I_2^{\rm e} \\ I_2^{\rm h} \\
		O_{\rs S}^{\rm e} \\ O_{\rs S}^{\rm h}
	\end{array}\right] =
	\mathbb{S}_{\rs NS}
	\left[\begin{array}{c}
		O_2^{\rm e} \\ O_2^{\rm h} \\
		I_{\rs S}^{\rm e} \\ I_{\rs S}^{\rm h}
	\end{array}\right].
	\label{eq:ascattmat2}
\end{equation}
The input and output channels are labeled in accordance with Fig.~\ref{fig:disordNS}. This matrix can be written in the block form
\begin{equation}
	\mathbb{S}_{\rs NS} =
	\left[\begin{array}{cc}
		\hat{r}_{\rs NS}(\varepsilon) & \hat{t}'_{\rs NS}(\varepsilon) \\
		\hat{t}_{\rs NS}(\varepsilon) & \hat{r}'_{\rs NS}(\varepsilon)
	\end{array}\right],
	\label{eq:ascattmat3}
\end{equation}
where $r$, $r'$, $t$, and $t'$ are the $N \times N$ matrices describing reflection and transmission for the states normalized to the unit flux in normal and superconducting segments and are grouped into the $2N \times 2N$ matrices $\hat {r}_{\rs NS}$, $\hat {r}'_{\rs NS}$, $\hat {t}_{\rs NS}$, and $\hat {t}'_{\rs NS}$.

We calculate the current by considering matrix~(\ref{eq:scattmat}) in the general form and refining it, if necessary, for specific models. We find matrix~(\ref{eq:ascattmat}) explicitly with the help of BdG equation~(\ref{eq:BdGeqsFull}). We temporarily assume that both of these matrices are arbitrary. The result for all types of scattering can be described by the $\mathbb{S}_{\rs NXS}$ matrix like~(\ref{eq:ascattmat}), which is also unitary. We restrict consideration to its $2N \times 2N$ submatrix
\begin{equation}
	\mathbb{R}_{\rs NXS} =
	\left[ \begin{array}{cc}
		R_{\rm ee} & R_{\rm eh} \\
		R_{\rm he} & R_{\rm hh}
	\end{array} \right]
	\label{eq:globalR}
\end{equation}
describing reflection to the normal region,
\begin{equation}
	\left[\begin{array}{c}
		O_1^{\rm e} \\ O_1^{\rm h}
	\end{array}\right] =
	\mathbb{R}_{\rs NXS}
	\left[\begin{array}{c}
		I_1^{\rm e} \\ I_1^{\rm h}
	\end{array}\right].
	\label{eq:globalR1}
\end{equation}
Here, $R_{\rm ee}$, $R_{\rm eh}$, $R_{\rm he}$, and $R_{\rm hh}$ are $N \times N$ reflection matrices. Below, we calculate the matrix $\mathbb{R}_{\rm NXS} = \mathbb{R}_{\rm NXS}(\varepsilon,V)$ for the scattering matrices given in Eqs.~(\ref{eq:scattmat}) and (\ref{eq:ascattmat}).

We now derive an expression for the current using total scattering matrix~(\ref{eq:ascattmat}). The contribution to the current from the state coming from the normal conductor with a given energy $\varepsilon$ is
\begin{align}
	I_n(\varepsilon, V) =
	\frac{e\hbar k_n}{m}
	\bigg\{
		1 - & \sum\limits_m |R_{{\rm ee},m n}(\varepsilon,V)|^2 \nonumber \\
		 + & \sum\limits_m |R_{{\rm he},m n}(\varepsilon,V)|^2
	\bigg\}.
\end{align}
This contribution depends on the voltage because a change in the electrostatic potential causes a change in the scattering state. But the deformation of the scattering state caused by the applied voltage does not itself lead to the appearance of a nonzero total current.\footnote{This can be shown by using the total scattering matrix that takes Andreev scattering into account.}

It is important that the applied voltage produces a difference of electrochemical potentials in the normal part and the superconductor, resulting in the finite total current
\begin{equation}
	I = -\int d\varepsilon \, \frac{G_{\rm s}(\varepsilon,V)}{e} \,
	\big[f(\varepsilon)-f(\varepsilon-eV)\big],
	\label{eq:NScurrent}
\end{equation}
where the spectral conductance
\begin{align}
	G_{\rm s}(\varepsilon,V) =
	\frac{2e^2}{h}{\rm Tr}
	\Big\{
		1 - & R_{\rm ee}^\dag(\varepsilon,V)
		R_{\rm ee}^{\phantom\dag}(\varepsilon,V) \nonumber \\
		+ & R_{\rm he}^\dag(\varepsilon,V)
		R_{\rm he}^{\phantom\dag}(\varepsilon,V)
	\Big\}
	\label{eq:cond}
\end{align}
describes the contribution to the current from the input states with the energy $\varepsilon$ for a specified voltage $V$ (the energy is measured from the electrochemical potential in the superconductor). The factor 2 in the right-hand side of Eq.~(\ref{eq:cond}) takes the spin degeneracy into account.

Expressions~(\ref{eq:NScurrent}) and (\ref{eq:cond}) determine the differential conductivity
\begin{align}
	\left.\frac{dI}{dV}\right|_{\s V} =
	- & \int d\varepsilon\, f'(\varepsilon -eV)\, G_{\rm s}(\varepsilon,V)
	\nonumber \\
	- & \int d\varepsilon\, \frac{1}{e} 
		\frac{\partial G_{\rm s}(\varepsilon,V)}{\partial V} \,
	\big[f(\varepsilon)-f(\varepsilon -eV)\big].
	\label{eq:diff_cond_dIdV}
\end{align}

At zero temperature, Eq.~(\ref{eq:diff_cond_dIdV}) can be conveniently represented as a series
\begin{equation}
	\left.\frac{dI}{dV}\right|_{\s V} =
	G_{\rm s}\left(eV,0\right) +
	2V \left. \frac{\partial G_{\rm s}(\varepsilon,V)}{\partial V} \right|_{\varepsilon=eV,\,V=0} +
	\ldots
	\label{eq:diff_cond_2}
\end{equation}
Unlike definition of the differential conductance $dI/dV = G_{\rm s}(eV,0)$ in Ref.~\cite{Blonder82}, the expression~(\ref{eq:diff_cond_2}) takes the change in transparency into account.

To complete the general derivation, the matrix $\mathbb{R}_{\rm NXS}$ in Eq.~(\ref{eq:globalR}) must be expressed in terms of the scattering matrices~(\ref{eq:scattmat}) and (\ref{eq:ascattmat}):
\begin{multline}
	\mathbb{R}_{\rs NXS}(\varepsilon, V) =
	\hat{r}_{11}(\varepsilon) \\
	+ \hat{t}_{12}(\varepsilon)
	\big[
		1 - \hat{r}_{\rs NS}(\varepsilon)\hat{r}_{22}(\varepsilon)
	\big]^{-1}
	\hat{r}_{\rs NS}(\varepsilon) \hat{t}_{21}(\varepsilon).
	\label{eq:Rhat}
\end{multline}

The simplest process contributing to the resistance, apart from the direct scattering in the normal part, is the propagation through the normal part $(\hat{t}_{21})$, reflection from the NS interface $(\hat {r}_{\rs NS})$, and propagation back through the normal segment $(\hat{t}_{12})$. All subsequent processes can be interpreted as multiple reflections from the normal scatterer and the NS interface. Expressions~(\ref{eq:NScurrent}), (\ref{eq:cond}), and (\ref{eq:Rhat}) determine the general form of the $I$-$V$ characteristic without any assumptions about the scattering characteristics; for example, the shape of $\Delta(x)$ at the NS interface can be arbitrary.

We now calculate the spectral conductance~(\ref{eq:cond}) using the Andreev approximation and assuming that $\Delta(x)$ is a step function.

We briefly consider $I$-$V$ characteristic symmetry under a change of the $V$ sign. In doing so we take into account that for $|eV| < \Delta$, the incoming quasiparticles cannot penetrate into a massive superconductor. The probability flow in the states with $|\varepsilon| < \Delta$ is completely reflected, and, therefore, the total scattering matrix $\mathbb{R}_{\rm NXS}(\varepsilon,V)$ in Eq.~(\ref{eq:globalR}) is unitary. The unitarity leads to the relations $R_{\rm ee}^\dag R_{\rm ee}^{\phantom *} + R_{\rm he}^\dag R_{\rm he}^{\phantom *} = 1$ and $R_{\rm ee}^{\phantom *} R_{\rm ee}^\dag + R_{\rm eh}^{\phantom *} R_{\rm eh}^\dag = 1$. Symmetry of the solutions of the electron and hole BdG equations guarantees that $R_{\rm eh}^{\phantom *}(\varepsilon,V) = -R_{\rm he}^*(-\varepsilon,V)$. Hence, the conductivity for $|eV| < \Delta$ can be written as
\begin{align}
	G_{\rm s}(\varepsilon,V) & =
	\frac{4e^2}{h}{\rm Tr} \Big\{R_{\rm he}^\dag(\varepsilon,V) R_{\rm he}^{\phantom\dag}(\varepsilon,V) \Big\} =
	\nonumber \\
	& = \frac{4e^2}{h}{\rm Tr} \Big\{R_{\rm eh}^\dag(\varepsilon,V) R_{\rm eh}^{\phantom\dag}(\varepsilon,V) \Big\} =
	\nonumber \\
	& = \frac{4e^2}{h}{\rm Tr} \Big\{R_{\rm he}^\dag(-\varepsilon,V) R_{\rm he}^{\phantom\dag}(-\varepsilon,V) \Big\} =
	\nonumber \\
	& = G_{\rm s}\left(-\varepsilon,V\right).
\end{align}
However, this symmetry does not lead to $I$-$V$ characteristic symmetry under the change of the bias voltage sign~\cite{Leadbeater96}. Such $I$-$V$ characteristic symmetry would mean fulfillment of the condition $G_{\rm s}(\varepsilon,V) = G_{\rm s}(-\varepsilon,-V)$, which requires that $G_{\rm s}(\varepsilon,V)$ be independent of voltage. In this case, we would obtain $G_{\rm s}(\varepsilon)|_{\varepsilon = eV} = dI/dV|_{\s V}$, and the differential conductivity would therefore be symmetric with respect to voltage.

In reality, however, experiments with SNS junctions~\cite{Magnee94,PoirierUnpublished} revealed $I$-$V$ characteristic asymmetry for $|eV| > \Delta$, which can be explained in the context of the previous discussion, taking into account the voltage dependence of the Schottky barrier on the SN boundaries. The asymmetry degree is determined by a quantity of the order of $eV/\mu$ or $eV/U$, where $U$ is the measure of the scattering potential height. To account for the voltage dependance of $G_{\rm s}$ explicitly, it is necessary to calculate the scattering matrix $\mathbb{S}_{\rs N}$ at the applied electrostatic potential. In principle, this problem requires a self-consistent solution of the scattering problem and the Poisson equation~\cite{Christen96}. In many practically interesting cases, it is possible to account for the voltage dependence of the scattering matrix only approximately.

\subsection{Conductance in the Andreev approximation}

We use expression~(\ref{eq:cond}) and evaluate it under boundary conditions for a pure NS interface in the Andreev approximation. The stationary states in the ballistic segment are plane-wave solutions of the BdG equations~\cite{deGennes68Book,Svidzinsky82Book}. We assume that ${\bf\Delta}(x) = \Delta e^{i\varphi}$ for $x>0$ and ${\bf\Delta}(x) = 0$ for $x<0$, which means that gap suppression in the contact region in the superconductor is neglected.

The NS boundary couples holes and electrons from one spatial channel with the scattering amplitude depending on the excitation energy and the effective chemical potential. Taking transverse quantization into account, the effective chemical potential has the form ${\bar\mu}_n = {\bar\mu} - \hbar^2{\bf k}_{{\s\bot},n}^2/2m$. In the limit $\varepsilon,\Delta \ll \bar\mu_n$, the BdG equations are reduced to linear equations by linearizing the dispersion law $k_n^{(0)} = \sqrt{2m\bar\mu_n}/\hbar$.

\begin{figure}[tb]
	\includegraphics[width=8.5cm]{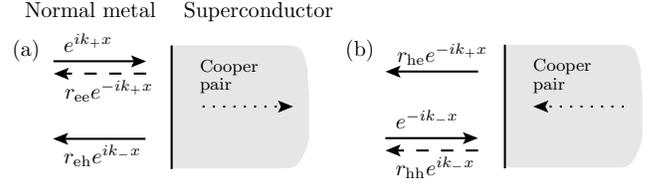}
	\caption{
Scattering at the NS interface. (a)~Scattering of an electron to a hole. (b)~Scattering of a hole to an electron. The dashed lines indicate zero amplitudes in case of the ideal NS boundary.
	}
	\label{fig:ns-scattering}
\end{figure}

The matrix of scattering from an ideal NS boundary has the form (Fig.~\ref{fig:ns-scattering})
\begin{equation}
	\hat{r}_{\rs NS}=
	\left[\begin{array}
		{cc} 0 & r_{\rm eh} \\
		r_{\rm he} & 0
	\end{array}\right] =
	\left[\begin{array}{cc}
		0 & e^{-i\varphi}\Lambda(\varepsilon) \\
		e^{i\varphi}\Lambda(\varepsilon) & 0
	\end{array}\right],
	\label{eq:rNS}
\end{equation}
where
\begin{equation}
	\Lambda(\varepsilon)=
	\left\{\begin{array}{ll}
		\cfrac{
			\varepsilon -
			{\rm sign}(\varepsilon)\sqrt{\varepsilon^2-\Delta^2}
		}{
			\Delta
		}
		\sim \cfrac{\Delta}{2|\varepsilon|}, &
		|\varepsilon| > \Delta, \\
		\cfrac{
			\varepsilon - i\sqrt{\Delta^2-\varepsilon^2}
		}{
			\Delta
		}
		= \exp\Bigl(
			-i\arccos\cfrac{\varepsilon}{\Delta}
		\Bigr), &
		|\varepsilon| < \Delta.
	\end{array}\right.
	\label{eq:Lambda}
\end{equation}

The total $N \times N$ reflection matrices $R_{\rm ee}$ and $R_{\rm he}$ can be determined from Eq.~(\ref{eq:Rhat}). Using Eq.~(\ref{eq:cond}) we obtain the expression for the spectral conductance for all energies:
\begin{align}
	G_{\rm s}(\varepsilon,V) & =
	\frac{2e^2}{h} 
	\left(1+|\Lambda(\varepsilon)|^2\right) \nonumber \\
	& \times {\rm Tr}\bigg\{
		t_{21}^\dag(\varepsilon)
		\Big[
			1 - \big[\Lambda^*(\varepsilon)\big]^2
			r_{22}^\top(-\varepsilon) r_{22}^\dag(\varepsilon)
		\Big]^{-1} \nonumber \\
	& \quad \times \Big[
		1-|\Lambda(\varepsilon)|^2
		r_{22}^\top(-\varepsilon) r_{22}^*(-\varepsilon)
	\Big] \nonumber \\
	& \quad \times \Big[
		1-\Lambda^2(\varepsilon)
		r_{22}^{\phantom *}(\varepsilon)r_{22}^*(-\varepsilon)
	\Big]^{-1}
	t_{21}(\varepsilon)
	\bigg\}.
	\label{eq:speccond}
\end{align}
Here the superscript ``$\top$'' denotes transposition.

Equations~(\ref{eq:NScurrent}) and (\ref{eq:speccond}) specify the $I$-$V$ characteristic in the Andreev approximation. The spectral conductance depends on the electron scattering matrix at energies $\pm\varepsilon$, indicating presence of Andreev reflection. The dependence of conductance~(\ref{eq:speccond}) on the phases of transmission and reflection amplitudes is extremely important for determining resonance peaks in the conductance. Elementary processes contributing to these phases are propagations of electrons and holes between the NS interface and the normal scatterer.

If the channels do not mix and the matrices $t_{ij}$ and $r_{ij}$ are diagonal the conductance reduces to the quasi-one-dimensional conductance
\begin{equation}
	G_{\rm s}(\varepsilon,V) =
	\sum\limits_{n=1}^{N} G_n(\varepsilon,V),
	\label{eq:onechannel}
\end{equation}
where
\begin{align}
	G_n(\varepsilon,V)
	& = \frac{2e^2}{h}
	\left[1+|\Lambda(\varepsilon)|^2\right] \, T_n(\varepsilon,V) \nonumber\\
	& \times \left[
		1-|\Lambda(\varepsilon)|^2R_n(-\varepsilon,V)
	\right] \nonumber\\
	& \times \Big\{
		1+|\Lambda(\varepsilon)|^4
		R_n(\varepsilon,V) R_n(-\varepsilon,V) \nonumber\\
		& \quad - 2{\rm Re} \left[
			\Lambda^2(\varepsilon)
			r_n(\varepsilon,V) r_n^*(-\varepsilon,V)
		\right]
	\Big\}^{-1},
	\label{eq:onechannel_n}
\end{align}
$r_n \equiv (r_{22})_{nn}$ are the amplitudes of normal reflection on the superconductor side, and $R_n = |r_n|^2$ and $T_n = 1 - R_n$ are the reflection and transmission probabilities in the $n$th channel. The last term in curly brackets in the right-hand side of Eq.~(\ref{eq:onechannel_n}) describes the important scattering process involving the propagation through a sector between the superconductor and the normal scatterer twice: once by an electron and once by a hole.

For high energies, $|\varepsilon|\gg \Delta$ (ad $|\varepsilon|\ll {\bar\mu}$), Andreev scattering is strongly suppressed, decaying as $\Lambda(\varepsilon)\sim $ $\Delta / 2|\varepsilon| \to 0$. In this case, the spectral conductance~(\ref{eq:speccond}) tends to the normal limit (the usual Landauer formula)
\begin{equation}
	G_{\rm s}(\varepsilon,V) = \frac{2e^2}{h}
	{\rm Tr}\left\{
		t_{21}^\dag(\varepsilon,V) \,
		t_{21}^{\phantom\dag}(\varepsilon,V)
	\right\}.
	\label{eq:land}
\end{equation}
We note that conductance~(\ref{eq:land}) is not necessarily symmetric under the voltage sign change.

For voltages smaller than the gap width, $|\varepsilon| < \Delta$, reflections of an electron to a hole and conversely occur with the probability one, $|\Lambda(\varepsilon)| = 1$, and then expression~(\ref{eq:onechannel_n}) reduces to the form~\cite{Lesovik97}
\begin{align}
	G_n(\varepsilon,V)
	& = \frac{4e^2}{h}
	T_n(\varepsilon,V) \, T_n(-\varepsilon,V) \nonumber\\
	& \times \Big\{
		1+ R_n(\varepsilon,V) R_n(-\varepsilon,V) \nonumber\\
		& \quad -2{\rm Re}\big[
			\Lambda^2(\varepsilon)
			r_n(\varepsilon,V) r_n^*(-\varepsilon,V)
		\big]
	\Big\}^{-1}.
	\label{eq:subonechannel}
\end{align}
The reflection and transmission amplitudes at energies~$\pm\varepsilon$ enter~(\ref{eq:subonechannel}) symmetrically, providing $I$-$V$ characteristic symmetry (voltage dependence of the scattering potential is neglected).

By contrast, for voltages exceeding the gap, the spectral conductance~(\ref{eq:onechannel}) becomes asymmetric in general. An important difference between the conductance of the NS junction~(\ref{eq:subonechannel}) and normal conductance~(\ref{eq:land}) is a dependence~(\ref{eq:subonechannel}) on phases of the scattering amplitude in the normal part.

In the linear response limit ($\varepsilon, eV \to 0$), which can be obtained by setting $\Lambda^2(0) = 1$ in Eq.~(\ref{eq:subonechannel}), the conductance takes the remarkably simple form~\cite{Beenakker92}
\begin{equation}
	G(0) =
	\frac{4e^2}{h} \sum\limits_n
	\frac{T_n^2(0)}{\big[ 2-T_n(0) \big]^2}.
	\label{eq:been}
\end{equation}
This expression is also valid for mixed channels: in this case, $T_n(0)$ are the transparency eigenvalues [see~(\ref{eq:Gdiag})].

\subsection{Linear conductance in special cases}

We analyze expression~(\ref{eq:been}) in limit cases. The best known limit is the weak tunneling limit for $T \ll 1$, in which
\begin{equation}
	G(0) =
	\frac{e^2}{h} \sum\limits_n
	T_n^2(0).
	\label{eq:beenBTK}
\end{equation}
In this case, the subgap conductivity is strongly suppressed and a current appears either at high voltages, as in the experiments in Refs.~\cite{Giaever60,Giaever62,Giaever74}, or at finite temperatures and voltages comparable to the gap. Until recently, only such NS junctions could be studied experimentally.

In the opposite limit, when the NS boundary is ideal, we obtain
\begin{equation}
	G(0) =
	\sum\limits_n \frac{4e^2}{h}.
	\label{eq:beenT1}
\end{equation}
We see that in the last case, the conductivity per channel is twice the normal conductivity. This result is sometimes interpreted in the following way: due to electron pairing into Cooper pairs and spin degeneracy the factor 2 in the expression for the conductance disappears. At the same time the factor 4 appears since the charge of the elementary carrier doubles. Such an interpretation is possible, however, we believe that the situation here is most likely as follows: the spin degeneracy does not disappear at all (which can be seen, for example, from the analysis of single electron injections from the normal region to the superconductor); in this case, a pair in the superconductor can be found for each electron with any spin direction (in other words, a hole is reflected). But in contrast to a normal contact, no electrons with energies in the interval from $\mu - |eV|$ to $\mu$ escape from the superconductor. This can be explained by the fact that electrons escaping from the normal reservoir below the Fermi level are paired with electrons above the Fermi level and absorbed in the superconductor, resulting in the appearance of an uncompensated current in the energy interval $2|eV|$, which leads to the doubled total current.

We finally consider a contact between a dirty normal conductor and a superconductor. In this case, we know the transparency distribution function~\cite{Dorokhov82,Dorokhov84}, and, as for other quantities, we can obtain the mean conductance of the NS junction. If the transparency of the normal part is described by the Dorokhov function, then the conductance accidentally coincides with the normal conductance~\cite{Beenakker97}
\begin{equation}
	G_{\rs NS}=
	\frac{4e^2}{h}\sum\limits_n
	\bigg\langle	\frac{T_n^2(0)}{\big[ 2-T_n(0) \big]^2} \bigg\rangle =
	G_{\rs N}=
	\frac{2e^2}{h} \Big\langle \sum\limits_n T_n \Big\rangle.
	\label{eq:beenDorokh}
\end{equation}
This result was already obtained by the Green's function method in Ref.~\cite{Artemenko79}.

\subsection{Conductance of NINIS junction}
\label{sec:condNINIS}

In the 1990s, several very interesting experiments~\cite{Pothier94,Petrashov95,Courtois96} were performed in which the dependences of the NS junction conductance on temperature, voltage, and magnetic fluxes were studied. It is interesting that the ratio of scattering intensities at the contact boundary and in the normal part determines the $I$-$V$ characteristic profile. This ratio determines whether a peak in the conductivity appears at zero temperature or at small but finite voltages~\cite{Volkov93,Marmokos93,Yip95}. Such peaks, which are called zero anomaly and finite-voltage anomaly, were studied in a number of interesting experiments~\cite{Kastalsky91,Nguyen92,Nitta94,Bakker94,Magnee94,PoirierUnpublished}.

We consider a model NINIS junction, which analysis is useful for understanding $I$-$V$ characteristic anomalies in dirty NS junctions. In addition, this system is of interest as an example of rather complicated scattering in the normal part, which can be used as a model for studying the interaction of smeared normal levels in I$_1$NI$_2$ interferometer and Andreev levels in INS Fabry-Per\'ot interferometer. We describe mechanisms responsible for zero and finite-voltage anomalies~\cite{Lesovik97} under certain conditions imposed on scattering intensity in barriers, which allows a qualitative understanding of these anomalies nature.

At first, we discuss conductance structure in a single-channel NI$_1$NI$_2$S junction and then present numerical results for a multichannel case in which the resonance structure does not disappear after averaging over channels, in contrast to INI junctions~\cite{Lesovik97}.

As before, we assume that channels are separated\footnote{In case of one channel we omit the subscript~$n$.} and the result in Eq.~(\ref{eq:onechannel}) can be used for the conductance $G_{\rm s}$, which depends on the phases $\chi_\pm^r$ of reflection amplitudes $r(\pm\varepsilon)$ and the complex amplitude $\Lambda(\varepsilon)$ of Andreev reflection. We use the notation $r(\pm\varepsilon) = \sqrt{R_{\pm}}e^{i\chi^r_\pm}$ for the reflection amplitude, where the phase factors $\chi^r_\pm $ are determined by the barriers I$_1$ and I$_2$ and propagation between them (for simplicity, voltage dependence of scattering is neglected).

We represent the Andreev reflection amplitudes as $\Lambda(\varepsilon) = |\Lambda| \exp[-i\vartheta(\varepsilon)]$ with the phase $\vartheta(\varepsilon) = \arccos(\varepsilon/\Delta)$ for $\varepsilon < \Delta$ and $\vartheta (\varepsilon) = 0$ for $\varepsilon > \Delta$. The expression for the conductance then reduces to the form
\begin{align}
	G_{\rm s}(\varepsilon)
	& = \frac{2e^2}{h}
	\left(1+|\Lambda|^2\right) T_+ \left(1-|\Lambda|^2 R_-\right) \nonumber\\
	& \times \Big\{
		1 + |\Lambda|^4 R_+ R_- 
		- 2|\Lambda|^2\sqrt{R_+R_-} \nonumber\\
		& \quad\quad \times \cos\big[
			\chi^r_+ -\chi^r_- -2\vartheta(\varepsilon)
		\big]
	\Big\}^{-1}.
	\label{eq:advert}
\end{align}
It follows from Eq.~(\ref{eq:advert}) that the conductance is always less than or equal to $4e^2/h$. We note that for $\varepsilon > \Delta$, Andreev scattering is suppressed, $|\Lambda | < 1$. For $\varepsilon < \Delta$, the phase $\vartheta (\varepsilon)$ is defined for resonances. Conductance~(\ref{eq:advert}) reaches the maximum value $4e^2/h$, which is twice the normal value, when the reflection probabilities $R_+$ and $R_-$ are equal and the phases $\chi^r_\pm$ satisfy the resonance condition
\begin{equation}
	\cos\left[
		\chi^r_+ - \chi^r_- 
		- 2\vartheta(\varepsilon)
	\right] = 1.
	\label{eq:res}
\end{equation}
A similar condition is known for a normal two-barrier system NI$_1$NI$_2$N, in which the transmission probability $T = 1$ and the maximum conductance $2e^2 / h$ can be achieved if the probabilities of reflection from barriers at the resonance energy are equal.

Given expression~(\ref{eq:advert}) for the conductance, we consider a single-channel NINS junction consisting of a ballistic NS junction containing one barrier at a distance $d$ from the ideal NS boundary. In the high-barrier limit, $R_+$ and $R_-$ are approximately equal. The reflection amplitudes $r(\pm\varepsilon)$ describing the propagation of electrons and holes have almost constant absolute values, while the phases are $\chi^r_\pm = \pi + 2k_\pm d$. Substituting the wave numbers $k_\pm = mv_{\rs F} \pm \varepsilon / v_{\rs F}$ (where $v_{\rs F}$ is the Fermi velocity in the channel) in Eq.~(\ref{eq:res}), we obtain positions of the Andreev resonances:
\begin{equation}
	\varepsilon_n =
	\frac{v_{\rs F}}{2d}
	\left(
		n\pi + \arccos\frac{\varepsilon_n}{\Delta}
	\right).
	\label{eq:alevel}
\end{equation}
Expression~(\ref{eq:alevel}) predicts the resonances in the conductance with a typical width proportional to the barrier transparency $T$ [similar Rowell-Macmillan resonances with a width of the order of $T / \Lambda(\varepsilon)$ are located at $\varepsilon_n = n\pi v_{\rs F} / 2d$ at voltages exceeding the gap width]. The phase $\vartheta(\varepsilon)$, changing from $\pi/2$ to 0 as $\varepsilon$ changes from 0 to $\Delta$, ensures the existence of at least one Andreev resonance for an arbitrarily small $d$. In the limit $d \to 0$, the resonance position coincides with the gap voltage, which is in accordance with the result obtained for the NIS junction in Ref.~\cite{Blonder82}. Hence the peak that was assigned in Ref.~\cite{Blonder82} to a singularity in the density of states near the gap can be interpreted as the Andreev resonance shifted to the gap at $d \to 0$.

We now introduce an additional barrier at the NS boundary and analyze an obtained two-barrier NI$_1$NI$_2$S junction, still using expression~(\ref{eq:advert}).

According to the adopted definitions, $\chi^r_\pm$ are the reflection phases of an electron incident on a two-barrier potential from the superconductor. The corresponding reflection amplitudes are given by
\begin{equation}
	r(\pm\varepsilon) = r_2 + \frac{t_2^2 r_1 e^{2ik_{\pm}d}}
	{1-r_1 r_2 e^{2ik_{\pm}d}},
	\label{eq:rpm}
\end{equation}
where $r_i$ and $t_i$ are the amplitudes of the left ($i = 1$) and right ($i = 2$) (on the NS interface) barriers [also see expression~(\ref{eq:ra1})]. The phases of these reflection amplitudes play an important role in the formation of the conductance structure since they control the existence of resonances according to Eq.~(\ref{eq:res}). We fix the barrier I$_1$ and gradually increase I$_2$, keeping the inequality $R_1 > R_2$. In this case, the INI interferometer produces noticeable Andreev resonances. For $r_1 \gg r_2$, the phases $\chi_\pm^r$ of the reflection amplitudes $r(\pm\varepsilon) \approx t_2^2r_1e^{2ik_{\pm}d}$ depend linearly on energy and change by $2\pi$ on the scale $hv_{\rs F}/d$, giving rise to equidistant resonances, in accordance with~(\ref{eq:alevel}).

The resonance positions can be found from the known phase function $\chi^r(\varepsilon)$: they are determined by the energies $\pm\varepsilon$ for which the phase difference is $\delta\chi^r(\varepsilon) = \chi^r_+ - \chi^r_- = \pi + 2n\pi$. The doubled period of $\delta\chi^r(\varepsilon)$, compared with the period of $\chi^r_\pm (\varepsilon)$, takes the pairing of resonances into account.

When barrier strengths are equal, $R_1 \approx R_2$, due to a large phase gradient near the normal resonance with energy $\varepsilon$, Andreev resonances tend to be pinned by normal resonances at the energies $+\varepsilon$ or $-\varepsilon$. This rule is violated when the normal resonance coincides with the electrochemical potential. In this case, Andreev resonances are separated from the electrochemical potential by a finite value.

As the strength of the second barrier increases further, $R_2 > R_1$, Andreev resonances become weaker and finally disappear. Although normal resonances are still present in this regime in the normal INI interferometer, only weak Andreev resonances can be seen in the conductance. The phase function becomes almost constant for $R_2 \gg R_1$ [see~(\ref{eq:rpm})] and condition~(\ref{eq:res}) for the resonance phases cannot be satisfied.

We now compare transport in two-barrier systems NI$_1$NI$_2$S and NI$_2$NI$_1$S, i.e., in the systems with the reversed sequence of barriers I$_1$ and I$_2$. We note that the transparency $T(\varepsilon)$ is the same in both cases. Hence, unlike the nonlinear conductance of the NINIS junction, the nonlinear conductance~(\ref{eq:land}) of the normal NININ junction as well as the linear conductance~(\ref{eq:been}) of the superconducting NINIS junction are independent of the sequence of barriers I$_1$ and I$_2$. We assume that $R_1\gg R_2$. For the direct barrier sequence (NI$_1$NI$_2$S), the energy dependence of the phase $\chi^r(\varepsilon)$ is strong, resulting in the appearance of Andreev resonances at a finite voltage. Electrons entering the INI interferometer on the normal side have enough time to form Andreev resonances and escape, typically to the superconductor. For the reversed barrier sequence (NI$_2$NI$_1$S), the barrier I$_1$ on the NS boundary becomes the main one. The weak energy dependence of the scattering phase $\chi^r(\varepsilon)$ prevents the formation of narrow resonances. This reflects the fact that electrons entering the INI interferometer escape through I$_2$ to the normal part without forming Andreev resonances.

We now analyze a multichannel NINIS junction quantitatively using the expression~(\ref{eq:onechannel}). This expression allows the analysis to be done for finite voltages and temperatures beyond the scope of a linear response studied in Ref.~\cite{Melson94}. We consider an NI$_1$NI$_2$S junction with two delta barriers with scattering probabilities $R_i$ ranging from 0.2 to 1 ($i = 1, 2$). We change the relative strength of the barriers to cover the interval between the two limits $R_1 > R_2$ and $R_1 < R_2$, which were discussed in Sec.~\ref{sec:condNINIS}. The distance $L$ between the barriers was chosen to be of the order of the coherence length $L \approx \xi = v_{\rs F}\hbar/\Delta$ in the superconductor, such that one or several Andreev resonances can be formed in the first channel (with the maximal longitudinal velocity). The number of resonances increases upon increasing the angle of incidence (counted from the normal to the NS interface) or, equivalently, upon increasing the channel's number. The contact cross section was set equal to $(100/k_{\rs F})^2$ and the ratio of the gap to the chemical potential was $\Delta / \bar\mu = 0.002$.

Each channel yields the typical structure of paired Andreev resonances discussed at the beginning of this section. Their position and width depend on the ratio of barriers strength I$_1$ and I$_2$ and the longitudinal kinetic energy in each channel. We note that the total conductance obtained by summation over the channels still has the structure produced by Andreev resonances. In contrast, the total conductance of the corresponding normal NI$_1$NI$_2$N junction is almost constant, i.e., normal resonances cancel each other.

The numerical study of three-dimensional NINS junctions shows that position and number of resonances in the total conductivity coincide with those in the first channel~\cite{Chaudhuri95}.\footnote{This occurs due to a decrease in the transparency with an increase in the angle of incidence and the nonuniform distribution of the angle of incidence over channels~\cite{Chaudhuri95}.} In NINIS junctions, such direct dependence has not been found.

\begin{figure}[tb]
	\includegraphics[width=7cm]{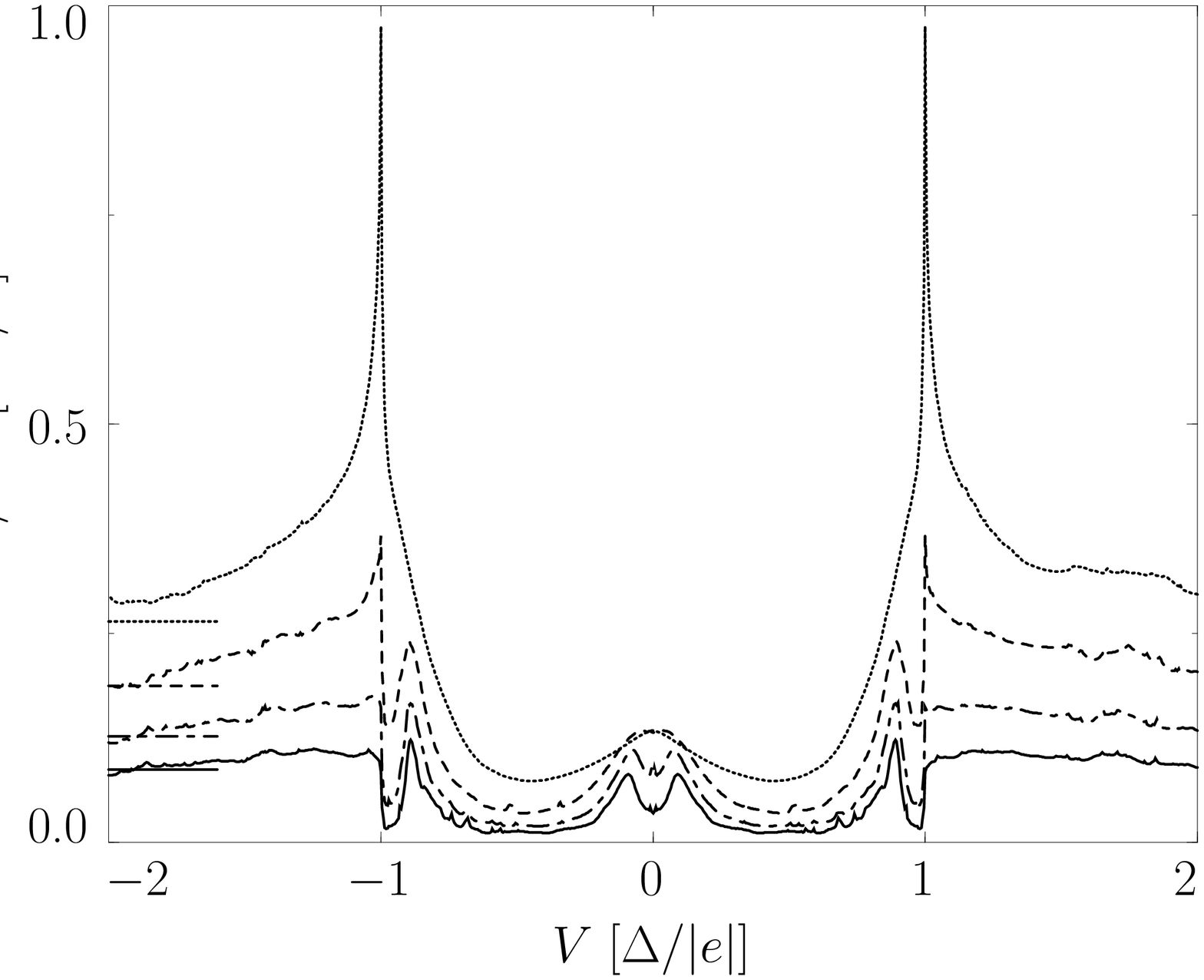}
	\caption{
Differential conductance (averaged per channel) in a multichannel NINIS junction of the width $d = 2v_{\rs F} / \Delta = 2\pi\xi$ as a function of the applied voltage at the temperature $\Theta = 0$. The curves (top down) correspond to the probabilities of reflection from the first barrier $R_1 = 0.2$, $0.5$, $0.7$, and $0.8$ at the constant reflection probability $R_2 = 0.5$ for the second barrier. The corresponding conductances in the normal state, indicated by the horizontal segments on the left part of the figure, are virtually independent of voltage in the limits indicated. As the strength I$_1$ of the first barrier increases, an anomaly appears at zero voltage due to the appearance of a new Andreev resonance at $R_1 > R_2$. (Figure from Ref.~\cite{Lesovik97}.)
	}
	\label{fig:vach1}
\end{figure}

\begin{figure}[tb]
	\includegraphics[width=7cm]{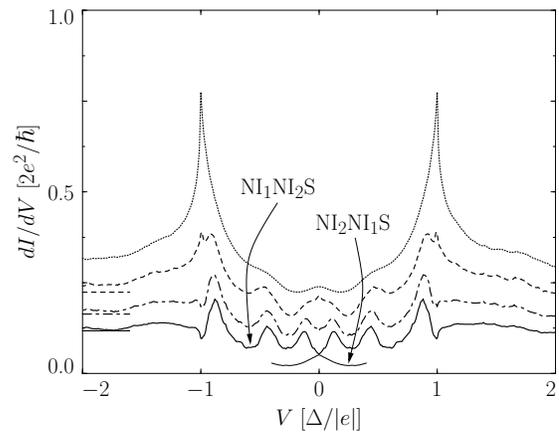}
	\caption{
Differential conductance (averaged per channel) in a multichannel NINIS junction of the width $d = 4v_{\rs F} / \Delta = 4\pi\xi$ as a function of the applied voltage at the temperature $\Theta =0$. The curves (top down) correspond to the probabilities of reflection from the first barrier $R_1 = 0.04$, $0.2$, $0.4$, and $0.54$ at the constant reflection probability $R_2 = 0.2$ from the second barrier. The corresponding conductances in the normal state are indicated by the horizontal segments on the left part of the figure. As the strength of the first barrier increases, the zero-voltage anomaly transforms into a finite-voltage anomaly and several Andreev resonances appear. For $R_1=0.54$ and $R_2=0.2$, we interchanged scatterers I$_1$ and I$_2$ in the INI part ($R_1 = 0.2$ and $R_2 = 0.54$) (the lower short curve); in this case, the conductivity at zero voltage remains the same, but the local minimum transforms into a local maximum. (Figure from Ref.~\cite{Lesovik97}.)
	}
	\label{fig:vach2}
\end{figure}

We now consider expression~(\ref{eq:subonechannel}) for the conductance, which is valid for voltages lower than the gap, and the properties of the $I$-$V$ characteristic near zero voltage. For $R_1 > R_2$, the denominator in Eq.~(\ref{eq:subonechannel}) changes rapidly because of a strong energy dependence of the phase of the reflection amplitude $r_n(\varepsilon)$, which is responsible for the appearance of a peak in the $I$-$V$ characteristic at nonzero voltage. The $I$-$V$ characteristic structure after summation is shown in Figs.~\ref{fig:vach1} and \ref{fig:vach2} (solid curves). The repulsion of Andreev levels from zero energy gives a minimum of $dI/dV$ for zero voltage. For $R_1 < R_2$, the phase of the reflection amplitude $r_n(\varepsilon)$ is almost independent of the energy, and the conductance structure is determined by the numerator in expression~(\ref{eq:subonechannel}). The expansion of the product $T_n(\varepsilon)T_n(-\varepsilon)=T_n^2 - w_n^2 \varepsilon^2$ near zero energy indicates the presence of a maximum at zero (zero anomaly).\footnote{The dominator cannot affect this property as long as the total transparency of a two-barrier system is not too large, $T_n < 0.55$.}

The manifestation of the zero-voltage anomaly is shown in Figs.~\ref{fig:vach1} and \ref{fig:vach2} (dashed curves). These figures illustrate crossovers from the zero to finite anomaly for two different distances $d$ between barriers upon increasing the barrier strength I$_1$ at a constant barrier strength I$_2$. If $d$ exceeds the coherence length in the superconductor, several resonances appear (see Fig.~\ref{fig:vach2}). In case of the reversed barrier sequence, the local minimum of the conductance at zero voltage transforms into a local maximum, however value of the conductance at zero does not change. This is shown in Fig.~\ref{fig:vach2} by two solid curves near the zero voltage: the upper curve corresponds to the direct barrier sequence (NI$_1$NI$_2$S) and the short lower curve to the reversed sequence (NI$_2$NI$_1$S). To understand what determines width of the peaks and position of the finite anomaly, we compare them with the Thouless energy. The Thouless energy $E_{\rs Th}$~\cite{Edwards72,Altland96} in a disordered system can be defined as a dimensionless conductance $g$ times a distance $\delta E$ between levels (in a closed system), $E_{\rs Th} = g \, \delta E$. In a system with weakly transparent barriers there is a good correspondence between such energy, the width, and the position of the finite bias anomaly. The width of the peak at zero voltage coincides with the characteristic correlation energy $\langle G(E+\varepsilon)G(E)\rangle_{\s E}$ in the conductance correlator and with the Thouless energy.\footnote{Angular brackets $\langle \cdot \rangle_{\s E}$ denote averaging over energies.} In this case, the transparency distribution function of the two-barrier system is bimodal and resembles that of a dirty system~\cite{Melson94}. But as the total transparency approaches unity, the width of the resonances no longer coincides with the energy $E_{\rs Th}$. In this limit, the two-barrier system poorly simulates the bimodal distribution for the dirty system and the Thouless energy is no longer the characteristic energy of the problem. At finite temperature, the anomaly at a finite bias voltage is smeared to form the zero bias anomaly.

The behavior of the zero and finite bias anomalies in disordered NS junctions has been studied in many experiments~\cite{Kastalsky91,Nguyen92,Nitta94,Bakker94,Magnee94}. The theoretical consideration of dirty NS systems show that this behavior is determined by the relation between the scattering strength on the NS interface and in the normal part~\cite{Volkov93,Marmokos93,Yip95,vanWees92}. In case of a small disorder, zero bias anomaly appears, while in case of a strong disorder, a peak appears in the normal part at a finite bias~\cite{Yip95} of the order of the Thouless energy $E_{\rs Th}$, which has been confirmed experimentally~\cite{PoirierUnpublished}. Such a behavior of the anomaly is similar to the behavior in a ballistic two-barrier NINIS junction, described in Ref.~\cite{Lesovik97}. The ``ballistic'' model of a dirty NS junction considered above therefore assumes the interpretation of the peak at a finite bias as a superposition of the smeared Andreev levels appearing between superconductor and the strongly reflecting normal part.


\section{Electron transport in SNS junctions}
\label{sec:SNS}

In this section, we consider nondissipative transport in superconductor-normal metal-superconductor (SNS) junctions, i.e., the Josephson effect~\cite{Josephson62}. Recently, it has become possible to make such contacts at meso- and nanoscales, for example, based on two-dimensional electron gas in heterostructures~\cite{Wees88,Wharam88,vanWees91,Takayanagi95,Akazaki96,Schraepers98}, using electron tunneling microscope~\cite{Sutton96}, or lithography~\cite{Morpurgo98,Baselmans99}, or atomic contacts~\cite{Krans93,Mueller92,Scheer98}, carbon nanotubes~\cite{Kasumov99,JarilloHerrero06,Cleuziou06,Eichler09}, single molecules~\cite{Roch08,Winkelmann09}, or graphene~\cite{Heersche07,Du08,OjedaAristizabal09}. The possibility of using such contacts in different applications appears to be extremely interesting~\cite{Ohta99Book}.

The specific features of these systems are mainly manifested in the regime when the conductivity is determined by several conducting channels (or even a single channel). As the number of conducting channels changes, quantization of the superconducting critical current~\cite{Takayanagi95,Mueller92,Beenakker91,Furusaki91,Furusaki92,Chtchelkatchev00,Kuhn01} and charge~\cite{Sadovskyy07a,Sadovskyy07b,Sadovskyy10} can be observed. The current is quantized in units of $e\Delta/\hbar$~\cite{Beenakker91b,Furusaki91,Furusaki92} and the charge in units of $2e$~\cite{Sadovskyy07a}. Usually, the gate potential is used as a control parameter in an experiment. By changing this potential the effective chemical potential in a two-dimensional gas can be varied. By varying the gate potential in structures with resonances, it is possible to shift resonances with respect to the electrochemical potential, thereby opening and closing conducting channels. Interesting phenomena also occur in the intermediate state with the partially opened channel. In this case, the current and charge strongly depend on the phase and lie between their quantized values.

The interest in such structures is additionally stimulated by the possibility of their practical application in superconducting quantum interference devices (SQUIDs)~\cite{Shmidt00Book,Clarke77,Clarke04BookV1,Clarke04BookV2,Kleiner04}, in which Josephson junctions inserted into a superconducting ring act as sensitive elements converting magnetic flux into current. SQUIDs are fabricated based on well known multichannel macroscopic Josephson contacts~\cite{Kulik70,Ishii70}. Study of Josephson nanocontacts can help decrease the size and increase the sensitivity of such devices. Other applications are also possible, such as a Josephson transistor~\cite{Houton91,Chrestin94,Wendin96,Wendin99a,Wendin99b,Kuhn01}.

Below, we consider the SXS junction with the nonsuperconducting part X of an arbitrary structure and derive the equation, in terms of the scattering matrix of part X, for the energy levels that carry almost all the current, and analyze this equation in most interesting cases.

\subsection{Energy levels and current in an SXS junction}

\begin{figure}[tb]
	\includegraphics[width=5.5cm]{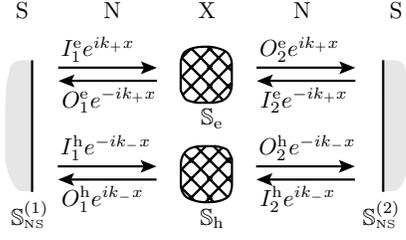}
	\caption{
States of Bogoliubov quasiparticles in SXS junction. Calculations are performed by using the model S$\tilde{\rm N}$X$\tilde{\rm N}$S system with the $\tilde{\rm N}$ region length tending to zero. Andreev reflection, occurring on the NS boundaries, is described by the respective scattering matrices $\mathbb{S}_{\rs NS}^{(1)}$ and $\mathbb{S}_{\rs NS}^{(2)}$ on the left and right boundaries. The normal part X is characterized by a scattering matrix $\mathbb{S}_{\rs N}$, which is separated for clarity into two parts, the electron matrix $\mathbb{S}_{\rm e}$ and the hole matrix $\mathbb{S}_{\rm h}$.
	}
	\label{fig:sxs}
\end{figure}

We consider the problem of two superconductors separated by a distance $L$ (Fig.~\ref{fig:sxs}). We assume that a normal scatterer X with the scattering matrix $\mathbb{S}_{\rs N}$ given by expression~(\ref{eq:scattmat3}) is located between NS interfaces. Since we are going to look for the quantization conditions, we consider only one conducting channel.\footnote{Assuming that a normal metal is connected to a superconductor adiabatically, we believe that transverse modes are well defined, and we solve the one-dimensional BdG equations for each channel.} Let us write the scattering matrix $\mathbb{S}_{\rs N}$ in more convenient form
\begin{eqnarray}
	\mathbb{S}_{\rs N} =
	\! \left[\begin{array}{cccc}
		\!\! \sqrt{R_+} e^{i\chi^r_+} \!\! & 0 & \!\! \sqrt{T_+} e^{i\chi^t_+} \!\! & 0 \!\! \\
		\!\! 0 & \!\! \sqrt{R_-} e^{i\chi^r_-} \!\! & 0 & \!\! \sqrt{T_-} e^{i\chi^t_-} \!\! \\
		\!\! \sqrt{T_+} e^{i\chi^t_+} \!\! & 0 & \!\! \sqrt{R_+} e^{i\chi^r_+} \!\! & 0 \!\! \\
		\!\! 0 & \!\! \sqrt{T_-} e^{i\chi^t_-} \!\! & 0 & \!\! \sqrt{R_-} e^{i\chi^r_-} \!\!
	\end{array}\right] \!, \;\;
	\label{eq:ns:x-matrix}
\end{eqnarray}
where $T$ and $R$ are the transmission and reflection probabilities of the scatterer X, and $\chi^t$ and $\chi^r$ are the corresponding phases. The subscripts ``$\pm$'' correspond to the energies $\pm\varepsilon$. For convenience of calculations, we assume that normal metal regions with an infinitely small length are located between the normal scatterer and the NS interfaces.

We now define matrices similar to Eq.~(\ref{eq:ascattmat3}), describing scattering on the left and right NS boundaries in the Andreev approximation. On the left and right NS interfaces, expression~(\ref{eq:rNS}) takes the respective forms
\begin{align}
	\hat{r}_{\rs NS}^{(1)} & =
	\left[\begin{array}{cc}
		0 & r_{\rm eh}^{(1)} \\ r_{\rm he}^{(1)} & 0
	\end{array}\right] =
	\left[\begin{array}{cc}
		0 & \! e^{-i\varphi_1} \! \\ \! e^{i\varphi_1} \! & 0
	\end{array}\right]
	\Lambda(\varepsilon), \;
	\label{eq:rNS_L} \\
	\hat{r}_{\rs NS}^{(2)} & =
	\left[\begin{array}{cc}
		0 & r_{\rm eh}^{(2)} \\ r_{\rm he}^{(2)} & 0
	\end{array}\right] =
	\left[\begin{array}{cc}
		0 & \! e^{-i\varphi_2} \! \\ \! e^{i\varphi_2} \! & 0
	\end{array}\right]
	\Lambda(\varepsilon).
	\label{eq:rNS_R}
\end{align}

The states below the gap, $|\varepsilon| < \Delta$, form a discrete spectrum, while the states above the gap, $|\varepsilon| > \Delta$, form a continuous spectrum. We consider the first case and write the quantization condition for discrete Andreev levels:
\begin{equation}
	{\rm det}\bigl[
		1 -
		\mathbb{S}_{\rm e}
		{\hat r}_{\rm eh}^{(1)}
		\mathbb{S}_{\rm h}
		{\hat r}_{\rm he}^{(2)}
	\bigr] = 0,
	\label{eq:ccSXS}
\end{equation}
where we again use the electron-hole parameterization in Eqs.~(\ref{eq:sm_eh1}) and (\ref{eq:sm_eh2}). Processes described by this equation are illustrated in Fig.~\ref{fig:sxs}. As mentioned in Sec.~\ref{sec:BdG}, in solving problems of this type, it is convenient to set $\varepsilon > 0$, taking both electron-like and hole-like states into account. As previously, $\Lambda(\varepsilon) = \exp[-i\vartheta(\varepsilon)]$ and $\vartheta(\varepsilon) = \arccos(\varepsilon/\Delta)$.

Substituting expressions~(\ref{eq:ns:x-matrix})--(\ref{eq:rNS_R}) in Eq.~(\ref{eq:ccSXS}), we obtain the quantization condition
\begin{multline}
	\cos(S_+ - S_- - 2 \vartheta) \\
	= \sqrt{R_+ R_-}\cos\beta + \!\sqrt{T_+ T_-} \cos\varphi,
	\label{eq:sxs:qc}
\end{multline}
determining the excitation spectrum $\varepsilon_\nu$ of Hamiltonian~(\ref{eq:BdG_Hdiag}) in the SXS system. Here, $\varphi = \varphi_2-\varphi_1$ is the difference between the superconducting phases in the left and right superconductors and $S_\pm = \chi_\pm^t + k_\pm L$ is the phase gained by electrons and holes in the normal region, where $k_\pm = \sqrt{2m({\bar\mu} \pm \varepsilon)}/\hbar$ are the corresponding wave vectors. In the case of symmetric barriers, the phase $\beta=(\chi^t_+ -\chi^r_+) - (\chi^t_- - \chi^r_-)$ is an integer multiple of~$\pi$ and gives rise to a continuous function $\sqrt{R_+R_-} \cos \beta $ changing its sign at each resonance~\cite{Kuhn01,Chtchelkatchev00}.\footnote{Equation~(\ref{eq:sxs:qc}) was obtained in a somewhat simplified form in Ref.~\cite{Wendin96b}.}

The total current in the ground state is $(2e/\hbar) \partial_\varphi U_0$. The ground-state energy $U_0$ of the system is given by expression~(\ref{eq:BdGcTermM}). We note that the last term in the right-hand side of Eq.~(\ref{eq:BdGcTermM}) is the sum of all excitation energies taken with the opposite sign. It is interesting that only this term depends on the superconducting phase difference $\varphi$, allowing the calculation of the ground-state Josephson current if the excitation spectrum $\varepsilon_\nu$ is known. Using this specific feature in the phase dependence of the ground-state energy $U_0$, we obtain the Josephson current equal to $I = \sum_\nu I_\nu$, where $I_\nu$ can be found by differentiating the energy $\varepsilon_\nu$ with respect to the superconducting phase, taken with the opposite sign, $I_\nu = - (2e/\hbar) \partial_\varphi \varepsilon_\nu$.

The total nondissipative current flowing through the SXS junction consists of two parts, one of them originates from the discrete levels below the gap and the other one from the continuous spectrum above the gap. We consider the contribution from the discrete component only, because it typically dominates~\cite{Chtchelkatchev02Thesis,Chtchelkatchev00}. After straightforward differentiation in Eq.~(\ref{eq:sxs:qc}), we obtain
\begin{equation}
	I_\nu = - \frac{2e}{{\cal T}_\nu} \sqrt{T_+T_-} \sin \varphi,
	\label{eq:J_gen}
\end{equation}
where the factor 2 is due to the double spin degeneracy,
\begin{align}
	{\cal T}_\nu
	& = \sin(\delta S - 2\vartheta)
		\hbar\partial_\varepsilon[\delta S - 2\vartheta]
	\nonumber \\
	& + \hbar\partial_\varepsilon [
		\sqrt{T_+T_-} \cos\varphi + \sqrt{R_+R_-} \cos\beta
	] \nonumber
\end{align}
has the dimension of time and represents the generalized quasiparticle traveling time in the normal part of the contact, and $\delta S \equiv S_+ - S_-$.

Expressions~(\ref{eq:sxs:qc}) and (\ref{eq:J_gen}) are valid for any scattering matrix. We use them below to describe particular systems.

\subsection{SNS junction: constriction in a two-dimensional gas}
\label{sec:sns_constriction}

We first consider a multichannel SNS junction with ideal NS boundaries. The simplest (and best known) case is a short SNS junction without inner scatterers. In this case, substituting $T_+ = T_- = 1$, $R_+ = R_- = 0$, and $\delta S = 0$ in Eq.~(\ref{eq:sxs:qc}), we obtain the energy $\varepsilon = \Delta \cos {(\varphi/2)}$. If the transparency $T$ of the normal part is not equal to unity but weakly depends on the energy in the interval $[{\bar\mu}-\Delta,\, {\bar\mu}+\Delta]$, then the electron and hole transparencies coincide, $T_+ = T_- = T$, and expression~(\ref{eq:sxs:qc}) gives one level per channel~\cite{Beenakker91}:
\begin{equation}
	\varepsilon_n = \Delta \sqrt{1-T_n \sin^2(\varphi/2)},
	\label{eq:E_T}
\end{equation}
where $T_n$ is the transparency of the $n$th channel.

For example, for a rectangular barrier of length~$L$ specifying the effective chemical potential ${\bar\mu}_{x,n}$ (with the chemical potential $\bar\mu$ at infinity), the transparency has the form
\begin{equation}
	T_n = \frac{
		4 {\bar\mu} {\bar\mu}_{x,n}
	}{
		4 {\bar\mu} {\bar\mu}_{x,n} +
		({\bar\mu} - {\bar\mu}_{x,n})^2
		\sin^2 \bigl[ \sqrt{2m{\bar\mu}_{x,n}/\hbar^2} L\bigr]
	},
	\label{eq:T_sns_rect}
\end{equation}
where $n$ is the transverse mode number. Below, we omit the subscript $n$ for simplicity.

\begin{figure}[tb]
	\includegraphics[width=7cm]{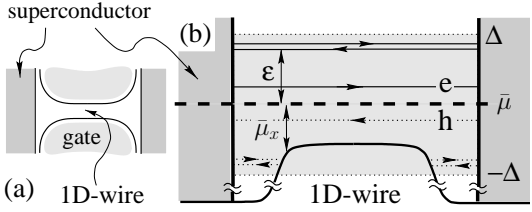}
	\caption{
SNS junction. (a)~Adiabatic constriction in a normal metal between two superconductors. (b)~The corresponding one-dimensional smooth effective potential. (Figure from Ref.~\cite{Kuhn01}.)
	}
	\label{fig:sns_setup}
\end{figure}

We consider an SNS junction based on a QPC in a two-dimensional electron gas [Fig.~\hyperref[fig:sns_setup]{\ref{fig:sns_setup}(a)}]. A one-dimensional contact is formed by two gates suppressing the electron density of the two-dimensional electron gas. Figure~\hyperref[fig:sns_setup]{\ref{fig:sns_setup}(b)} shows the effective one-dimensional chemical potential ${\bar\mu}_x$ corresponding to a channel $n$. As the ``top'' ${\bar\mu}_x(0)$ of this potential increases, the channel $n$ gradually closes: first for holes and then for electrons.

Expression~(\ref{eq:T_sns_rect}) describes a potential with breaks and is rarely realized in practice. The system outlined in Fig.~\hyperref[fig:sns_setup]{\ref{fig:sns_setup}(a)} can be described by the parabolic potential
\begin{equation}
	{\bar\mu}_x(x) = 
	{\bar\mu}_x(0) + m \Omega^2 x^2/2,
	\nonumber
\end{equation}
where $\hbar\Omega = (4/\pi) \sqrt{\varepsilon_{\s L} [\bar\mu -{\bar\mu}_x(0)]}$ describes the ``curvature'' of the potential at $x = 0$ and $\varepsilon_{\s L} = \hbar^2\pi^2/2mL^2$ is the quantization energy over the contact length. The value of $\Omega $ is selected such that the relation ${\bar\mu}_x(\pm L/2) = \bar\mu$ is satisfied. In this case, the transparency $T$ is given by Kemble formula~(\ref{eq:kemble}) and depends only on the effective chemical potential in the maximum and its curvature:
\begin{equation}
	T = \frac{1}{1 + \exp(-2\pi {\bar\mu}_x(0) / \hbar\Omega)}.
	\label{eq:Kemble}
\end{equation}
In the typical case, the expansion of the potential near its maximum can be restricted to the quadratic term because, as the transparency changes according to the Kemble formula from small $(T \ll 1)$ to large $(T \approx 1)$, scattering is determined by the potential in a rather small vicinity of the potential maximum (see Sec.~\ref{sec:condsmearing}).

\begin{figure}[tb]
	\includegraphics[width=6cm]{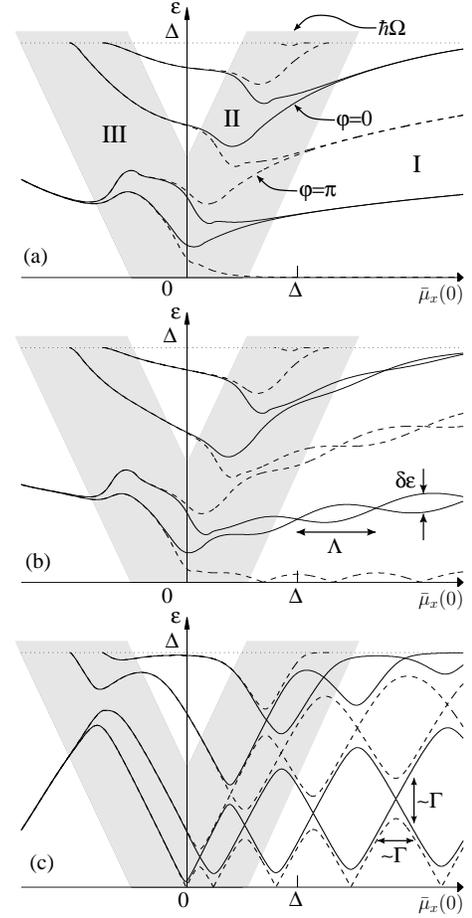}
	\caption{
Subgap spectrum for a parabolic potential in the normal part and scatterers with the strength $Z$ on NS boundaries. (a)~Ideal NS boundaries ($Z = 0$). (b)~Weak scattering ($Z = 0.1$) leads to the appearance of weak resonances and Andreev spectrum splitting. In the case of strong scattering ($Z = 1$), an Andreev quantum dot is formed. In regions I and II, the spectrum depends on the difference between superconducting phases $\varphi$ (solid curves correspond to the phase $\varphi = 0$ and dashed curves to the phase $\varphi = \pi$). (Figure from Ref.~\cite{Kuhn01}.)
	}
	\label{fig:qc}
\end{figure}

For a channel of an arbitrary length, it is necessary to calculate dimensionless actions $S_\pm$ involved in quantization condition~(\ref{eq:sxs:qc}). The action for a parabolic barrier can be calculated explicitly~\cite{Chtchelkatchev00}:
\begin{equation}
	\frac{S(E)}{\hbar} =
	\frac{2E}{\hbar\Omega}
	\left\{
		\! \kappa^2 \sqrt{1+\frac{1}{\kappa^2}} +
		\log \! \left[ |\kappa| \! \left( \! 1+\sqrt{1+\frac{1}{\kappa^2}} \right) \right]
	\right\},
	\label{eq:Sparab}
\end{equation}
where
\begin{equation}
	\kappa^2 =
	Q\, \frac{\hbar\Omega}{E} =
	\frac{\pi^2 \hbar^2\Omega^2}{16 E \varepsilon_{\s L}}.
	\label{eq:Kparab}
\end{equation}
Here $S_\pm = S(E = {\bar\mu}_x(0) \pm \varepsilon)$ and $Q \equiv (\pi/4)^2\hbar\Omega/\varepsilon_{\s L}$ is a dimensionless parameter, typically, $Q \gg 1$. Note that the additional change in the action $S(E)$ by $\pi$ after passing through zero energy in the interval $\hbar\Omega$ cannot be obtained in the WKB approximation.

Figure~\hyperref[fig:qc]{\ref{fig:qc}(a)} shows the numerical solution of quantization equations~(\ref{eq:sxs:qc}) for the model described above. For a large positive chemical potential ${\bar\mu}_x(0) > \varepsilon + \hbar\Omega$ (region I), the system can be described by expressions~(\ref{eq:E_T}) and (\ref{eq:Kemble}). As ${\bar\mu}_x$ decreases (region II), only the electron levels remain, as shown in Fig.~\hyperref[fig:sns_setup]{\ref{fig:sns_setup}(b)}; in this case, the energy levels are split, even for $\varphi =0$. In region III, the channel produced by an electron-like level is closed. In regions with ${\bar\mu}_x(0) < -\varepsilon - \hbar\Omega$, the energy levels no longer depend on the phase and hence represent closed channels. The behavior of the system is described in more detail in Ref.~\cite{Kuhn01}.

In semiclassical region I, each Andreev level produces a nondissipative current with the amplitude $2|e|/[\tau_++\tau_-+2\hbar/\sqrt{\Delta^2- \varepsilon^2}]$, where $\tau_\pm = \hbar \partial_\varepsilon S_\pm$ are the traveling times of electron-like and hole-like quasiparticles in the normal part. For the parabolic potential,
\begin{equation}
	\tau(E) = \Omega^{-1} [2\log[|\kappa|(1+\sqrt{1+\kappa^{-2}})].
	\nonumber
\end{equation}
For small energies, the traveling time increases logarithmically, $\tau (E) \approx \Omega^{-1} \log(4Q\hbar\Omega/E)$, in the interval $\hbar\Omega < E < Q \hbar\Omega$ and is saturated at $\tau_0 \approx \Omega^{-1} \log(4Q)$ and energy $E < \hbar\Omega$, at which the system can no longer be considered in the WKB approximation.

\begin{figure}[tb]
	\includegraphics[width=6.2cm]{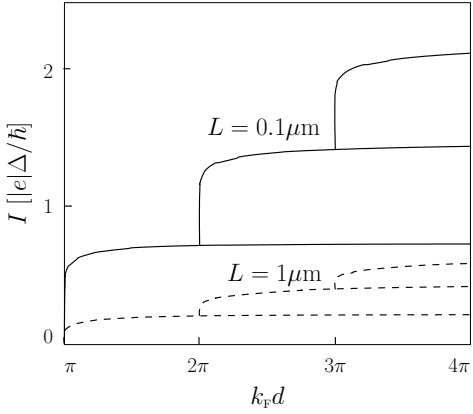}
	\caption{
Quantization of the superconducting current. As the width $d$ of the normal channel increases, the superconducting current increases by quanta $|e|/(\tau_0 + \hbar/\Delta)$. (Figure from Ref.~\cite{Chtchelkatchev00}.)
	}
	\label{fig:supr_quant}
\end{figure}

Let us analyze the dependence of the current on the phase difference~$\varphi$. For $\varphi = 0$, double degeneracy occurs, and levels in a pair make contributions to the current with the same modulus but opposite signs. Therefore, for $\varphi = 0$, the discrete spectrum makes no contribution to the current. As $\varphi$ increases, the degeneracy is lifted and each split pair contributes to the nondissipative current, monotonically increasing with increasing $\varphi$ and reaching the maximum at $\varphi = \pi - 0$. This means that the current takes the critical value at the point $\varphi = \pi - 0$.\footnote{It is quite difficult to prove that the contribution of the continuous spectrum to the critical current is insignificant. The discrete and continuous spectra can be simultaneously taken into account by using the Krein's theorem, as in Refs.~\cite{Krichevsky00,Akkermans91}. At the same time, it can be shown relatively easily that the contribution from the continuous spectrum vanishes at $\varphi = \pi$~\cite{Chtchelkatchev00}. However, this does not mean that the critical current through the contact, determined by both the discrete and continuous spectra, is then caused only by the discrete spectrum at $\varphi = \pi - 0$. To prove this statement, it is necessary to show that the sum of discrete and continuous spectra reaches a maximum at $\varphi = \pi -0$. For example, this can be done for the contact length $L$ and the chemical potential ${\bar\mu}_x(0)$ in the middle of the contact distant from the point $(\sqrt{\xi_0 / k_{\rs F}},\Delta)$ in the $(L,{\bar\mu}_x(0))$ coordinates~\cite{Chtchelkatchev00}.} For $\varphi = \pi - 0$, all the levels except the lowest one become degenerate again and none of the degenerate pairs of levels contributes to the current.

Finally, we obtain a simple expression for the critical current $I_{\rm c} \equiv \max_\varphi \{I\}$ in semiclassical region I:
\begin{equation}
	I_{\rm c} = \frac{|e|}{\tau_0 + \hbar/\Delta},
	\label{eq:crit_current}
\end{equation}
where $\tau_0$ is the traveling time calculated for a parabolic potential, which is constant in an opening channel, and $\tau_0 = \Omega^{-1} \log 4Q$ decays as $\tau_0 = \Omega^{-1}\log 4Q\hbar\Omega/{\bar\mu}_x(0)$ for ${\bar\mu}_x(0) > \hbar\Omega$, and becomes equal to the free-traveling time, $\tau_0 = L/v_{{\rs F},x}$ for ${\bar\mu}_x(0) > \hbar\Omega$. In the last case, for a channel opened at large energies, we obtain the known formula $I_{\rm c} = |e|v_{\rs F} /(L+\pi \xi_0)$ for the critical current. The critical current increases by $I_{\rm c}$ with the appearance of each new open channel (Fig.~\ref{fig:supr_quant}).

The basic qualitative features of the critical current quantization predicted theoretically were confirmed experimentally in Ref.~\cite{Takayanagi95}.

\subsection{SINIS junction: the Andreev quantum dot}

\begin{figure}[tb]
	\includegraphics[width=5.5cm]{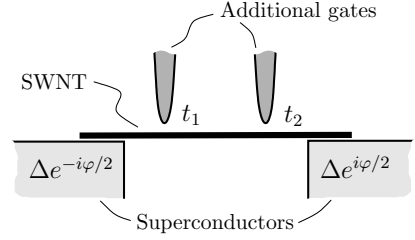}
	\caption{
Outline of the experimental realization of the Andreev quantum dot based on a single-wall nanotube (SWNT). Two additional gates form the electron density in certain regions and produce effective barriers with transmission amplitudes $t_1$ and $t_2$.
	}
	\label{fig:andreev_dot_2gates}
\end{figure}

In Sec.~\ref{sec:sns_constriction} we considered quantum constriction, assuming that only Andreev reflection occurs on NS boundaries ($Z = 0$). We now consider the case where the boundaries contain scatterers with $Z > 0$. Figure~\hyperref[fig:qc]{\ref{fig:qc}(b)}, showing an intermediate case with $Z = 0.1$, demonstrates oscillations with a period $\lambda$ and an amplitude $\delta \varepsilon$ caused by weak resonances of double barrier. In Fig.~\hyperref[fig:qc]{\ref{fig:qc}(c)} is depicted the case of strong normal resonances, $Z = 1$ (the corresponding system based on a carbon nanotube is shown in Fig.~\ref{fig:andreev_dot_2gates}). We can see the distinct resonance structure of Andreev levels, which is determined by the resonance structure of the normal part. An important parameter is the resonance width $\Gamma$ given by Eq.~(\ref{eq:Taa_G1}). We restrict our analysis to the case ${\bar\mu}_x(0) \gg Q \hbar\Omega$, where the curvature $\Omega$ of the potential no longer plays any role, and the potential just shifts normal resonances.

The excitation spectrum in the Josephson contact with the normal part of any structure can be determined from expression~(\ref{eq:sxs:qc}). Let us derive the quantization condition for double barrier with $\Omega = 0$. Substituting transmission coefficients~(\ref{eq:Taa_T}) and phases~(\ref{eq:ra1}) in Eq.~(\ref{eq:sxs:qc}), we obtain
\begin{multline}
	(R_1 + R_2)
		\cos\Bigl( 2\pi \frac{\varepsilon}{\delta} \Bigr)
	- 4\sqrt{R_1 R_2} \,
		\cos\Bigl( 2\pi \frac{\varepsilon_{\rs D}}{\delta} \Bigr) \,
		\sin^2 \vartheta \\
	+T_1 T_2 \cos\varphi
	= \cos\Bigl(2\vartheta - 2\pi \frac{\varepsilon}{\delta}\Bigr) +
	R_1 R_2 \cos\Bigl(
		2\vartheta + 2\pi\frac{\varepsilon}{\delta}
	\Bigr).
	\label{eq:sinis:qc1}
\end{multline}
Here, we choose the resonance with some number $n$ and energy $E_n$ and omit the index $n$ for simplicity. The energy $\varepsilon_{\rs D} = E_n - {\bar\mu}(0)$ [where ${\bar\mu}(0) = \mu - eV_{\rm g}$] determines the resonance position $E_n = \varepsilon_{\s L}[n-(\chi_1^r+\chi_2^r)/2\pi]^2$ with respect to the chemical potential in the normal part, which is in turn controlled by the external gate potential $V_{\rm g}$. We assume that the distance to neighboring resonances $\delta \equiv (E_{n+1} - E_{n-1})/2$ significantly exceeds the superconducting gap $\Delta$.

The dependences of the energy states on the effective chemical potential ${\bar\mu}_x(0)$ are shown in Fig.~\ref{fig:qc}. The parameter $Z$ [see~(\ref{eq:tDelta})--(\ref{eq:rDelta})] specifies the ``strength'' of normal scatterers. For a symmetric contact, we have $T_1 = T_2 = 1/(1+Z^2)$ and $R_1 = R_2 = Z^2 / (1+Z^2)$. For $Z = 0$, expression~(\ref{eq:sinis:qc1}) describes an SNS junction; for $Z = 0.1$, it describes a contact with weak scatterers at the NS boundaries, and for $Z = 1$, it describes a contact with quite strong scatterers at the NS boundaries.

Figure~\ref{fig:qc} shows that as the scatterer strength increases, a resonance structure appears. The most interesting case is that of a strong phase dependence of energy, which occurs when some normal resonance passes through the chemical potential. Expression~(\ref{eq:sinis:qc1}) can be simplified in the vicinity of this point and an analytic expression for the Andreev level can be obtained~\cite{Beenakker91Conf,Wendin96,Sadovskyy07b} as
\begin{equation}
	\varepsilon =
	\sqrt{\varepsilon_{\rs D}^2 +
	{\tilde\Gamma}^2},
	\label{eq:Andreev_en0}
\end{equation}
where
\begin{equation}
	{\tilde\Gamma} =
	\Gamma\sqrt{ \cos^2 \frac{\varphi}{2} + A^2 }, \quad
	A = \frac{|T_1 - T_2|}{2\sqrt{T_1 T_2}},
	\label{eq:E_Gamma}
\end{equation}
and $\Gamma = (T_1+T_2)\delta / 4\pi$ is the half-width of the normal resonance. Expression~(\ref{eq:E_Gamma}) is valid when the resonance $E_n$ approaches the chemical potential $\bar\mu$ by a distance of the order of the normal resonance half-width $|\varepsilon_{\rs D}| \lesssim \Gamma$, while the half-width itself is much smaller than the superconducting gap, $\Gamma \ll \Delta$ [Fig.~\hyperref[fig:qc]{\ref{fig:qc}(c)}]. In this case, the current through the Andreev quantum dot is
\begin{equation}
	I =
	\frac{2e}{\hbar} \,
	\frac{
		\Gamma^2 \sin\varphi
	}{
		4 \sqrt{\varepsilon_{\rs D}^2 + {\tilde\Gamma}^2}
	}.
	\label{eq:J_m2p}
\end{equation}
The critical current is given by
\begin{equation}
	I_{\rm c} =
	\frac{|e|\Gamma}{\hbar}
	\biggl\{
		\sqrt{1+A^2 + \frac{\varepsilon_{\rs D}^2}{\Gamma^2}} -
		\sqrt{A^2 + \frac{\varepsilon_{\rs D}^2}{\Gamma^2}}
	\biggr\}.
	\label{eq:J_m2pCrit}
\end{equation}

\subsection{SGS junction and the Dirac-Bogoliubov-de Gennes equations}

\begin{figure}[tb]
	\includegraphics[width=5.0cm]{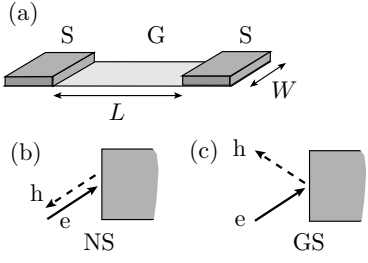}
	\caption{
(a)~SGS junction: a graphene sheet G of length $L$ and width $W$ connected to two superconductors S. (b)~Retro Andreev reflection of an electron to a hole from the ideal NS boundary in the Andreev approximation. The hole repeats the electron trajectory in the opposite direction. (c)~Specular reflection from the ideal GS boundary followed by a change in the normal component of the velocity.
	}
	\label{fig:sgs_setup}
\end{figure}

This section is devoted to the Josephson current in a superconductor-graphene-superconductor (SGS) junctions [Fig.~\hyperref[fig:sgs_setup]{\ref{fig:sgs_setup}(a)}]. The method for preparing a graphite (graphene) monolayer was developed a few years ago~\cite{Novoselov04}. Later, a current through a graphene Josephson contact was measured~\cite{Heersche07}. Below, we describe the electron transport and calculate the critical current in the SGS system.

Graphene is described by the relativistic Dirac wave equation. In this case, low-energy quasiparticles have the linear dispersion $\varepsilon = kv$ and zero mass, while the velocity~$v$ is constant and independent of energy. This leads to a number of interesting physical phenomena such as the Klein tunneling~\cite{Katsnelson06b,Cheianov06,Beenakker08}.

Calculations for the SGS junction are quite similar to those for the SNS junction; however, in BdG equations~(\ref{eq:BdGeqsWest}), we must now substitute the Dirac Hamiltonian~\cite{Beenakker06,Titov06,Cuevas06}
\begin{equation}
	{\hat H}_{0}=-i \hbar v(\sigma_x\partial_x + \sigma_y\partial_y),
	\label{eq:Ham_dirac}
\end{equation}
which describes graphene in the absence of superconductors. The resulting system of four first-order differential equations is called the Dirac-Bogoliubov-de Gennes (DBdG) equations.

As previously, the coefficients $u = [u_1, u_2]^\top$ and $v = [v_1, v_2]^\top$ describe the electron and hole parts of the wave function, but now each of them consists of two components. These components have the opposite spin and valley indices related to the two sublattices in the hexagonal graphene lattice. Solving the DBdG equations on ideal GS boundaries, we obtain the coefficients of the scattering matrix responsible for reflection on the graphene side. The coefficients for the left and right boundaries are respectively given by
\begin{align}
	{\hat r}_{\rs GS}^{(1)} & =
	\left[\begin{array}{cc} 0 & r_{\rm eh}^{(1)} \\ r_{\rm he}^{(1)} & 0 \end{array}\right] =
	\left[ \begin{array}{cc}
		0 & \! e^{-i\varphi_1 - i\vartheta{\hat\sigma}_x} \! \\
		\! e^{i\varphi_1 + i\vartheta{\hat\sigma}_x} \! & 0 \\
	\end{array} \right],
	\label{eq:rGS_L}
	\\
	{\hat r}_{\rs GS}^{(2)} & =
	\left[\begin{array}{cc} 0 & r_{\rm eh}^{(2)} \\ r_{\rm he}^{(2)} & 0 \end{array}\right] =
	\left[ \begin{array}{cccc}
		0 & \! e^{-i\varphi_2 + i\vartheta{\hat\sigma}_x} \! \\
		\! e^{i\varphi_2 - i\vartheta{\hat\sigma}_x} \! & 0 \\
	\end{array} \right].
	\label{eq:rGS_R}
\end{align}
Similarly to matrix~(\ref{eq:globalR1}), reflection matrices~(\ref{eq:rGS_L}) and (\ref{eq:rGS_R}) relate the wave functions incident on the GS boundary and reflected wave functions. However due to the presence of the valleys, the matrix size doubles compared to the size of the analogous matrix for the NS boundary. Expressions~(\ref{eq:rGS_L}) and (\ref{eq:rGS_R}) describe the subgap scattering, with $\vartheta = \arccos(\varepsilon/\Delta)$ as previously.

We note that Andreev reflection from the ideal NS boundary transforms an electron to a hole (or a hole to an electron) with the opposite velocity vector. This means that the reflected hole (electron) propagates along the same path as the incident electron (or hole), but in the opposite direction [Fig.~\hyperref[fig:sgs_setup]{\ref{fig:sgs_setup}(b)}] \cite{Andreev64a,Andreev64b,Shmidt00Book}. After reflection from the ideal GS boundary, only the normal component of the velocity changes, i.e., specular reflection occurs [Fig.~\hyperref[fig:sgs_setup]{\ref{fig:sgs_setup}(c)}] \cite{Beenakker06}, which is called specular Andreev reflection to distinguish it from the well known usual Andreev scattering from the NS boundary (retro Andreev reflection).\footnote{These statements are valid in the Andreev approximation on the NS boundary and a low graphene doping level on the GS boundary.}

We assume that a graphene sheet is ideally rectangular and use the boundary condition for transverse quantization~\cite{Tworzydlo06}
\begin{equation}
	k_{y,n} = (n+1/2)\pi/W,
	\label{eq:gr_tr_cc}
\end{equation}
where $k_{y,n}$ is the transverse component of the wave vector in the $n$th channel. The effective chemical potential ${\bar\mu}_{x,n}$ in the $n$th channel is determined by the relation ${\bar\mu}_{x,n}^2 = {\bar\mu}^2 - (\hbar v k_{y,n})^2$, where $\bar\mu$ is the chemical potential measured with respect to the Dirac point, i.e., the graphene doping level. A particle with an energy $\varepsilon$ has the wave vector $k = ({\bar\mu} + \varepsilon)/\hbar v$ and the corresponding longitudinal component $k_{x,n} = (k^2 - k_{y,n}^2)^{1/2}$. Solving the Dirac equation ($\Delta = 0$) for a rectangular potential of length $L$ (with the wave vector equal to $k_{x,n}$ in the potential region and $k$ outside that region), we obtain the scattering matrices describing this potential
\begin{equation}
	\mathbb{S}_{\rm e} =
	\left[ \begin{array}{cc}
		0 & {\tilde t}_{\rm ee} \\
		{\tilde t}_{\rm ee} & 0 \\
	\end{array} \right],
	\quad
	\mathbb{S}_{\rm h} =
	\left[ \begin{array}{cc}
		0 & {\tilde t}_{\rm hh} \\
		{\tilde t}_{\rm hh} & 0 \\
	\end{array} \right],
	\label{eq:Gsm0}
\end{equation}
where
\begin{equation}
	{\tilde t} =
	\left[ \begin{array}{cc}
		\! \cos(k_xL) + \frac{k_y}{k_x} \sin(k_xL) \! & \frac{ik}{k_x} \sin(k_xL) \\
		\frac{ik}{k_x} \sin(k_xL) & \! \cos(k_xL) - \frac{k_y}{k_x} \sin(k_xL) \! \\
	\end{array} \right].
	\label{eq:Gsm}
\end{equation}
The opposite energy signs correspond to electrons and holes, ${\tilde t}_{\rm ee} = {\tilde t}(\varepsilon)$ and ${\tilde t}_{\rm hh} = {\tilde t}(-\varepsilon)$. In the general form, the quantization condition for the SGS junction has the form
\begin{equation}
	{\rm det}\bigl[
		1 -
		{\tilde t}_{\rm ee}
		{\hat r}_{\rm eh}^{(1)}
		{\tilde t}_{\rm hh}
		{\hat r}_{\rm he}^{(2)}
	\bigr] = 0.
	\label{eq:ccSGS}
\end{equation}
Substituting (\ref{eq:rGS_L}), (\ref{eq:rGS_R}), and (\ref{eq:Gsm}) in Eq.~(\ref{eq:ccSGS}), we can obtain the quantization condition determining energy levels in the system under consideration~\cite{Titov06,Cuevas06}.

In the case of a short contact ($L \ll \Delta, \xi$), the energy levels are described by the simple expression
\begin{eqnarray}
	\varepsilon_n = \Delta \sqrt{1-T_n\sin^2(\varphi/2)},
	\label{eq:E_sgs_short}
\end{eqnarray}
which completely coincides with expression~(\ref{eq:E_T}) for a short SNS junction; the only difference being in transparency definition
\begin{equation}
	T_n =
	\frac{
		k_{x,n}^2
	}{
		k_{x,n}^2\cos^2(k_{x,n}L)+({\bar\mu}/\hbar v)^2\sin^2(k_{x,n}L)
	}.
	\label{eq:T_graphene}
\end{equation}
It follows that transparencies in this case differ considerably from those in Eq.~(\ref{eq:T_sns_rect}) for quadratic dispersion.

Each channel makes a contribution to the current, which can be found by differentiating Eq.~(\ref{eq:E_sgs_short}) with respect to $\varphi$. The total current has the form
\begin{equation}
	I =
	\frac{e\Delta}{\hbar}\sum\limits_{n=0}^{\infty}
	\frac{T_n\sin\varphi}{\sqrt{1-T_n\sin^2(\varphi/2)}}.
	\label{eq:sgs_curr_dir}
\end{equation}
We note that unlike the summation in the case of an SNS junction, the summation in Eq.~(\ref{eq:sgs_curr_dir}) extends to infinity over the propagating (real $k_{x,n}$) and evanescent (imaginary $k_{x,n}$) modes. For $L \ll W, \xi $, summation can be replaced by integration.

\begin{figure}[tb]
	\includegraphics[width=6.7cm]{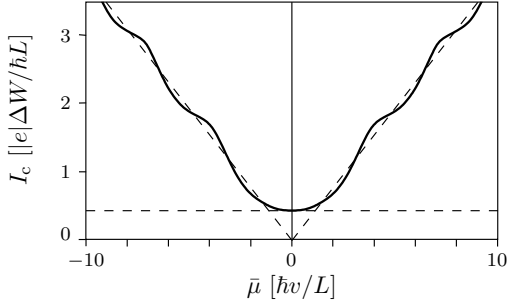}
	\caption{
Critical current $I_{\rm c}$ in an SGS junction as a function of the graphene doping level $\bar\mu$. (Figure from Ref.~\cite{Titov06}.)
	}
	\label{fig:sgs_crit_curr}
\end{figure}

The numerical results obtained for the critical current are presented in Fig.~\ref{fig:sgs_crit_curr}. The main feature is that the critical current does not vanish at the zero doping level, which is confirmed experimentally~\cite{Heersche07}. For ${\bar\mu} = 0$, the total current is determined by the first nonvanishing term in the expansion of current~(\ref{eq:sgs_curr_dir}) in the small parameter $|{\bar\mu}| \ll \hbar v / L$~\cite{Titov06}:
\begin{equation}
	I = \frac{e\Delta }{\hbar}\,\frac{2W}{\pi L}\cos(\varphi/2)\,
	{\rm arctanh}[\sin(\varphi/2)].
	\label{eq:curr_dir}
\end{equation}
In this case, the critical current (shown by the horizontal dashed straight line in Fig.~\ref{fig:sgs_crit_curr}) is described by the expression
\begin{equation}
	I_{\rm c} = 1.33\,\frac{e\Delta }{\hbar}\,\frac{W}{\pi L}.
	\label{eq:crit_curr_dir}
\end{equation}

Away from the Dirac point (${\bar\mu}\gg\hbar v/L$), the critical current is proportional to the doping level (see the inclined dashed asymptotes in Fig.~\ref{fig:sgs_crit_curr}):
\begin{equation}
	I_{\rm c}=1.22\,\frac{|e|\Delta }{\hbar}\,\frac{|{\bar\mu}| W}{\pi\hbar v}.
	\label{eq:crit_curr_dop}
\end{equation}

The case of finite temperature is considered in Refs.~\cite{Gonzalez08,Hagymasi10}.


\section{Shot noise in NS systems at a finite voltage}
\label{sec:noiseNS}

In this section, we present general expressions for the differential shot noise in a nonideal NS junction in terms of the scattering matrix of the normal part. As mentioned in Sec.~\ref{sec:secquant}, shot noise appears due to the discreteness of the charge carried by a particle and the probabilistic nature of scattering. Shot noise in nonideal NS junctions is produced both by normal scattering processes and by nonideal Andreev scattering~\cite{Khlus87,Muzykantskii94,deJong94,Hessling96}. At low temperatures the latter process, caused only by the transfer of electron pairs, can be represented as tunneling of Cooper pairs as a whole, similarly to the tunneling of regular particles. Thus, the Andreev scattering leads to fluctuations with amplitude proportional to the double electron charge. In this case, the fluctuation amplitude in SNS junctions with applied voltage can increase to a value proportional to the larger number of charge quanta~\cite{Averin96,Dieleman97}.

Now let us consider an NXS junction. As above, X denotes a region with an arbitrary normal scattering matrix. The current fluctuation power spectrum at low frequencies is determined by irreducible current-current correlator~(\ref{eq:asymmcorr}) for $\omega \to 0$. The time-dependent current operator is defined as
\begin{equation}
	\hat I(\tau)=
	e^{i(\hat H-\mu \hat N) \tau} \, {\hat I} \,
	e^{-i(\hat H - \mu \hat N) \tau},
	\label{eq:currentop0}
\end{equation}
where $\hat N$ is the particle number operator. The latter expression differs from previously used Eq.~(\ref{eq:curr4noise}). Here the Hamiltonian determining time evolution is replaced by the effective Hamiltonian, which can be diagonalized in the mean-field approximation using the Bogoliubov transformation. This approach neglects order parameter fluctuations in the superconducting region and assumes that Bogoliubov quasiparticles coherently propagate through the entire system, not changing their energy.

Time-dependent current operator~(\ref{eq:currentop0}) can be expressed in terms of solutions of BdG equations~(\ref{eq:BdGeqsFull}) and the Bogoliubov creation and annihilation operators,
\begin{multline}
	\hat I(t) = -\frac{ie}{m}\sum\limits_{{\nu'},{\nu}}
	\int\! dydz
	\big(
		u_{\nu'}^* \hat{\partial_x} u_{\nu}^\pstar
		\hat a_{\nu'}^\dag \hat a_{\nu}^\pdag \\
		- v_{\nu'}^* \hat{\partial_x} v_{\nu}^\pstar
		\hat a_{\nu}^\pdag \hat a_{\nu'}^\dag
	\big) \,
	e^{i\left(\varepsilon_{\nu'}-\varepsilon_{\nu}\right)t},
	\label{eq:currentop}
\end{multline}
where we introduce the new operator $u{\hat\partial_x}v \equiv u \partial_x v - v \partial_x u$. As before, we take into account states with the positive energy $\varepsilon_\nu > 0$ only and omit the channel number $n$ where it is not important. For simplicity, we calculate expression~(\ref{eq:currentop}) in the normal region.

For voltages smaller than superconducting gap ($|eV| < \Delta$) quasiparticles can appear from normal reservoir only and the wave functions are still dependent only on the parts of the scattering matrix responsible for reflection ($r_{\rm ee}$, $r_{\rm he}$, $r_{\rm eh}$, and $r_{\rm hh}$). Each of these $N \times N$ matrices describes the whole NXS junction.

The noise power is determined by transitions (due to the operator $\hat I$) between the states
\begin{align}
	& |{\rm s}\rangle = |f_{\beta,\nu,n} = 1, f_{\beta',\nu',m} = 0\rangle,
	\nonumber \\
	& |{\rm i}\rangle = |f_{\beta,\nu,n} = 0, f_{\beta',\nu',m} = 1\rangle
	\nonumber
\end{align}
that differ only by the occupation of two single-particle states with the energy subscripts $\nu$ and $\nu'$ in the respective $n$th and $m$th channel. The subscripts $\beta$ and $\beta'$ indicate the electron (e) of hole (h) states. For example, a transition between the incident electron ($\beta = {\rm e}$) from the $n$th channel and the incident hole ($\beta = {\rm h}$) from the $m$th channel occurs due to the presence of a nonzero matrix element describing the interaction between reflected electrons (and holes), $\langle {\rm s} | \hat I | {\rm i} \rangle \propto f_{\rm e}(1-f_{\rm h}) (r_{\rm eh}^\dag r_{\rm ee}^\pdag - r_{\rm hh}^\dag r_{\rm he}^\pdag)_{mn}$. Here, the occupation numbers $f_\beta$ for electrons and holes are given by the Fermi distribution $f_{\rm e} = f(\varepsilon - eV)$ and $f_{\rm h} = f(\varepsilon + eV)$; the voltage is measured relative to the electrostatic potential in the superconductor. Summation over the channels gives the contribution to fluctuations
\begin{align}
	\sum\limits_{m,n} |\langle {\rm s} | I | {\rm i} \rangle|^2
	& \propto f_{\rm e}(1-f_{\rm h}) \nonumber \\ \vspace{-5mm}
	& \times {\rm Tr}\Bigl\{
		(r_{\rm ee}^\dag r_{\rm eh}^\pdag -
			r_{\rm he}^\dag r_{\rm hh}^\pdag) 
		(r_{\rm eh}^\dag r_{\rm ee}^\pdag -
			r_{\rm hh}^\dag r_{\rm he}^\pdag)
	\Bigr\} \nonumber \\
	& = f_{\rm e}(1-f_{\rm h}) \, {\rm Tr}\Bigl\{
		r_{\rm he}^\dag r_{\rm he}^\pdag
		(1-r_{\rm he}^\dag r_{\rm he}^\pdag)
	\Bigr\}. \nonumber
\end{align}
Considering similar processes, we obtain the expression for the low-frequency power spectrum~\cite{Anantram96,Martin96} valid for $\Theta \ll |eV| < \Delta$:
\begin{align}
	S & = \frac{4e^2}{h} \int\limits_{0}^{\Delta} d\varepsilon
	\biggl(
		\big[ f_{\rm e}(1-f_{\rm h}) + f_{\rm h}(1-f_{\rm e}) \big] \nonumber \\
		& \quad \times {\rm Tr}\Bigl\{
			r_{\rm he}^\dag r_{\rm he}^\pdag
			\bigl(1-r_{\rm he}^\dag r_{\rm he}^\pdag \bigr) 
		\Bigr\} \nonumber \\
		& \quad + \left[ f_{\rm e}(1-f_{\rm e}) + f_{\rm h}(1-f_{\rm h}) \right]
		{\rm Tr}\Bigl\{
			\bigl(r_{\rm he}^\dag r_{\rm he}^\pdag \bigr)^2
		\Bigr\}
	\biggr).
	\label{eq:pgeneral}
\end{align}
The first term in the right-hand side of Eq.~(\ref{eq:pgeneral}) describes transitions between the states making contributions to the current with opposite signs, while the second term describes transitions making contributions with the same signs. At zero temperature, the second term vanishes and the shot noise is determined by the first term. At finite temperature, both terms contribute, in particular, to the Johnson-Nyquist noise~\cite{Johnson27,Nyquist28} appearing due to the thermal fluctuations.

As in Sec.~\ref{sec:IVrelation}, we express the reflection matrix~$r_{\rm he}(\varepsilon)$ in terms of the reflection matrix of the normal part~X and the Andreev reflection amplitude. Substituting the result in Eq.~(\ref{eq:pgeneral}), we obtain the general expression for shot noise in a multichannel NXS junction. We restrict our consideration by the case of contact with constant cross section and with the differential conductance $G_n(\varepsilon) = (4e^2/h) [r_{\rm he}^\dag r_{\rm he}^\pdag]_{nn}$ described by expression~(\ref{eq:subonechannel}). At zero temperature the shot noise power spectrum takes the form
\begin{equation}
	S = (1/|e|) \int\limits_0^{|eV|} d\varepsilon\, \sum\limits_n \zeta_n(\varepsilon),
	\nonumber
\end{equation}
where $\zeta_n$ is the differential shot noise in the $n$th channel
\begin{align}
	\zeta_n(\varepsilon)
	& = \frac{2|e|^3}{h}
	4\big[r_{\rm he}^\dag r_{\rm he}^\pdag \big(1-r_{\rm he}^{\dag}r_{\rm he}\big)\big]_{nn} \nonumber \\
	& = \frac{2|e|^3}{h}
	4 T_n(\varepsilon) T_n(-\varepsilon) \nonumber \\
	& \quad \times \Bigl\{
		(R_n(\varepsilon) + R_n(-\varepsilon) 
		- 2 {\rm Re} \bigl[
			\Lambda(\varepsilon)^2 r_n^\pstar(\varepsilon)
			r_n^*(-\varepsilon)
		\bigr]
	\Bigr\} \nonumber \\
	& \quad \times \Bigl\{
		T_n(\varepsilon) T_n(-\varepsilon) +
		R_n(\varepsilon) 
		+ R_n(-\varepsilon) \nonumber \\
		& \qquad - 2 {\rm Re} \bigl[
			\Lambda(\varepsilon)^2 r_n^\pstar(\varepsilon)
			r_n^*(-\varepsilon)
		\bigr]
	\Bigr\}^{-2}.
	\label{eq:diffnoise}
\end{align}
In the limit $\varepsilon \to 0$ $(\Lambda \to -i)$ the dependence on the phases gained by quasiparticles between the scatterer X and the NS boundary disappears and we obtain the linear response~\cite{deJong97}
\begin{equation}
	\zeta_n(0) =
	\frac{2|e|^3}{h}
	\frac{16 \, T_n(0)^2 \, [1-T_n(0)]}{[2-T_n(0)]^4}.
\end{equation}
Using the Dorokhov distribution function once again
one can find the total noise in a dirty conductor as~\cite{Beenakker97}
\begin{equation}
	S = (2/3) |e|I.
\end{equation}
The corresponding Fano factor $F = 2/3$ is twice the Fano factor for noise in a normal dirty conductor (see Sec.~\ref{sec:noisedesc}). This fact indicates the presence of charge $2e$ in the system. It was observed experimentally in Refs.~\cite{Jehl99,Jehl00}.

\subsection{Noise in NINS junctions}

\begin{figure}[tb]
	\includegraphics[width=7.0cm]{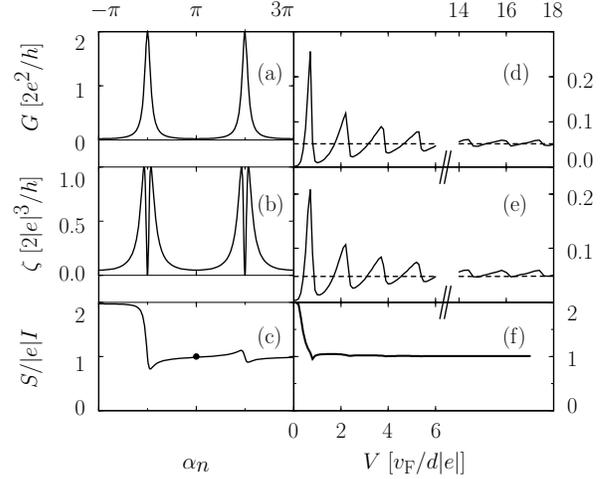}
	\caption{
(a) and (d) Conductance at zero temperature, (b) and (e) shot noise, and (c) and (f) noise power spectrum for the NINS junction (solid curves) with $\int dx\, V(x) = 3\hbar v_{\rs F}$, mean transparency $T = 0.05$, $E_{\rs F} = 500\Delta$, and $d = 20 v_{\rs F} / \Delta$, $v_{\rs F} / d \ll \Delta \ll E_{\rs F}$. Plots in Figs.~(a)--(c) are determined by Eqs.~(\ref{eq:ninsnoise}) and (\ref{eq:ninscond}) for one channel, and in Figs.~(d)--(f) by Eqs.~(\ref{eq:subonechannel}) and (\ref{eq:diffnoise}) averaged over $8 \times 10^4$ channels. The conductance (d) and noise (e) asymptotically approach the corresponding values (dashed straight lines) in the NIN junction. The ratio $S/|e|I$ shown in Figs.~(c) and (f) approaches the classical value $S/|e|I = 1$ at high voltages. (Figure from Ref.~\cite{Fauchere98}.)
	}
	\label{fig:shot2}
\end{figure}

Let us consider a strong scatterer I ($T \ll 1$) without the internal resonance structure, located at a distance $d$ from the NS boundary, ${\rm X} = {\rm NIN}$.\footnote{This model describes, for example, a thin NS film and a metal needle brought near it~--- the system, where Rowell-McMillan resonances were observed~\cite{Rowell66}.} At the energy scale considered here the reflection probability $R$ weakly depends on energy. Therefore, the energy dependence being completely determined by the phase $\chi^r(\varepsilon) \approx 2(k + \varepsilon/v)d + \chi_0$ gained between the scatterer I and the NS boundary. Here $k$ and $v$ are the wave vector and velocity in a quasi-one-dimensional channel. Differential shot noise~(\ref{eq:diffnoise}) in the $n$th channel
\begin{equation}
	\zeta_n(\varepsilon) =
	\frac{2|e|^3}{h}
	\frac{
		4 T^2 2R \left[1-\cos \alpha_n(\varepsilon)\right]
	}{
		\left\{T^2 + 2R \left[1-\cos \alpha_n(\varepsilon)\right] \right\}^2
	}
	\label{eq:ninsnoise}
\end{equation}
depends only on the phase difference $\alpha_n(\varepsilon) = \chi_n^r(\varepsilon) - \chi_n^r(-\varepsilon) - 2\vartheta(\varepsilon)$. The resonance structure of $\zeta_n$ is reflected in the differential conductance
\begin{equation}
	G_n(\varepsilon) =
	\frac{4e^2}{h}
	\frac{ T^2}{T^2 + 2R \left[1-\cos \alpha_n(\varepsilon)\right]}.
	\label{eq:ninscond}
\end{equation}
Figures~\hyperref[fig:shot2]{\ref{fig:shot2}(a)} and~\hyperref[fig:shot2]{\ref{fig:shot2}(b)} show the dependences of the conductance and spectral noise on the phase difference~$\alpha_n$. The minimal value of the denominator in Eqs.~(\ref{eq:ninsnoise}) and (\ref{eq:ninscond}) reaches at $\varepsilon_{\nu,n} = v_n / 2d [n\pi +\arccos(\varepsilon_{\nu,n}/\Delta)]$ and corresponds to the resonances shown in these figures. Differential shot noise $\zeta_n$ in Eq.~(\ref{eq:ninsnoise}) vanishes at these resonances and reaches a maximum at $\cos\alpha_n = (2R-T^2)/2R$, when energies are still close to resonances.

It is remarkable that such a nontrivial structure is preserved even in the multichannel case,\footnote{For example, the resonance structure of normal double barrier is not preserved in the multichannel case.} which is shown in Figs.~\hyperref[fig:shot2]{\ref{fig:shot2}(d)} and \hyperref[fig:shot2]{\ref{fig:shot2}(e)}. This can be explained by the fact that levels ``adhere'' to the electrochemical potential of the superconductor. Comparing Figs.~\hyperref[fig:shot2]{\ref{fig:shot2}(b)} and \hyperref[fig:shot2]{\ref{fig:shot2}(e)} we see that a sharp double peak in the noise [Fig.~\hyperref[fig:shot2]{\ref{fig:shot2}(b)}] disappears in the multichannel case [Fig.~\hyperref[fig:shot2]{\ref{fig:shot2}(e)}], while the noise $S$ takes the maximum value (instead of the minimal) at the resonance, repeating the corresponding conductance curve [Fig.~\hyperref[fig:shot2]{\ref{fig:shot2}(d)}].

For large voltages $|eV| \gg v_{\rs F}/d$ the noise and conductance can be estimated by averaging the phase $\alpha_n$ in Eqs.~(\ref{eq:ninsnoise}) and (\ref{eq:ninscond}),
\begin{equation}
	\bar{\zeta}=
		\frac{1}{2\pi}\int\limits_0^{2\pi} d\alpha\, \zeta(\alpha) =
		\frac{2|e|^3}{h} T, \quad
	\bar{G} = \frac{2e^2}{h} T.
	\label{eq:ninscond_av}
\end{equation}
Here we replace the sum over channels by the integral, $(1/N)\sum_n \to (1/2\pi) \int_0^{2\pi} d\alpha$.
This means that both the noise and the conductance reach their normal values (in the corresponding NIN junction) at the voltages $v_{\rs F}/d \ll |eV| \ll \Delta$.

Let us now consider the Fano factor $F = S/|e|I$, where $S = \int_0^{|eV|} d\varepsilon \, \zeta(\varepsilon)$, $I = \int_0^{|eV|} d\varepsilon \, G(\varepsilon)$. At low voltages, $F = 2$, which reflects the fact that charge carriers are Cooper pairs. At high voltages ($v_{\rs F}/d \ll |eV| \ll \Delta$), the Fano factor decreases to the normal value $F = 1$. Such a decrease in $F$, which occurs immediately after the first Andreev resonance [as shown in Figs.~\hyperref[fig:shot2]{\ref{fig:shot2}(c)} and \hyperref[fig:shot2]{\ref{fig:shot2}(f)}], is caused by the noise suppression in the resonance region. A similar noise suppression was observed in Ref.~\cite{Lefloch03}.

\subsection{Noise in NINIS junctions}
\label{sec:noiseNININS}

\begin{figure}[tb]
	\includegraphics[width=7.3cm]{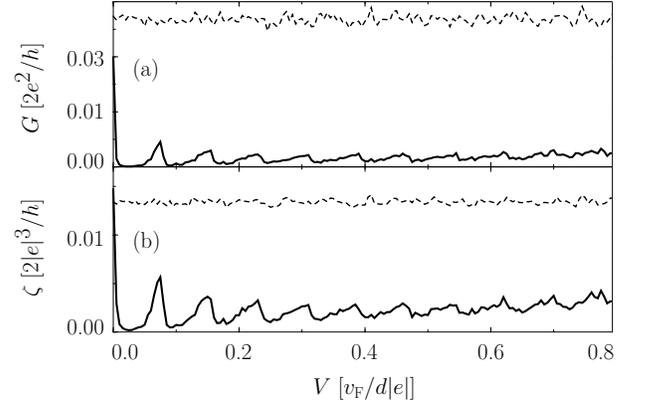}
	\caption{
(a) Differential conductance and (b) shot noise in the NINIS junction with many channels $(8\times 10^4)$. The transparencies of scatterers are the same $T_1 = T_2 \approx 0.05$, the distance between scatterers is $d = 20 v_{\rs F} / \Delta$, and $\int dx \, V(x) = 3\hbar v_{\rs F}$. Averaged shot noise per channel~(\ref{eq:diffnoise}) has local maxima at conductance resonances (solid curves). The conductance and shot noise for the corresponding nonsuperconducting NININ structure are shown by dashed curves. (Figure from Ref.~\cite{Fauchere98}.)
	}
	\label{fig:shot3}
\end{figure}

As mentioned in Sec.~\ref{sec:NS}, the NINIS junction can be interpreted as a qualitative model of a disordered conductor~\cite{Dorokhov82,Melson94}.

The resonance structure of the INI scatterer leads to the bimodal distribution of the transparency $T$. For a symmetric INI scatterer $(T_1 = T_2 \ll 1)$ the transparency distribution is given by~\cite{deJong97}
\begin{equation}
	\rho(T)=\frac{1}{\pi}\frac{T_1}{2}\frac{1}{\sqrt{T^3(1-T)}}, \quad
	T \in \left[\frac{T_1^2}{\pi^2}, 1\right].
	\label{eq:bimodal}
\end{equation}
Expression~(\ref{eq:bimodal}) has an analogue in the case of a disordered conductor~\cite{Dorokhov82,Melson94}. The structure of $\rho (T)$ in the NININ junction does not affect macroscopic transport properties. The differential shot noise and linear conductance can be calculated for bimodal transparency distribution~(\ref{eq:bimodal}),
\begin{align}
	& \zeta =
	\frac{2|e|^3}{h} \int dT\, \rho(T) T(1-T) =
	\frac{2|e|^3}{h} \frac{T_1}{4}, \\
	& G =
	\frac{2e^2}{h} \int dT\, \rho(T) T =
	\frac{2e^2}{h} \frac{T_1}{2}
\end{align}
and describe coherent transport~\cite{Chen91}. In the noncoherent case the resistance is a sum of two resistances determined by barriers I connected in series. In this case $\zeta / |e|G = 1/2$ as a consequence of charge conservation~\cite{Beenakker92b}. In the coherent case the linear response of the NINIS junction can be found from expressions~(\ref{eq:subonechannel}), (\ref{eq:diffnoise}), and (\ref{eq:bimodal}):
\begin{align}
	& \zeta(0) =
	\frac{2|e|^3}{h}
	\int dT\, \rho(T) \frac{16T^2(1-T)}{(2-T)^4} =
	\frac{2|e|^3}{h}
	\frac{T_1}{2} \frac{3}{4\sqrt{2}}, \\
	& G(0) =
	\frac{2e^2}{h}
	\int dT\, \rho(T) \frac{2T^2}{(2-T)^2} =
	\frac{2e^2}{h}
	\frac{T_1}{2} \frac{1}{\sqrt{2}}.
\end{align}
The ratio of the spectral noise density to the conductance in this case is $\zeta(0) / |e|G(0) = 3/4$~\cite{deJong97}. The noise to conductance ratios in the coherent case for ${\rm X} = {\rm I}$ and ${\rm X} = {\rm IN}$ and for a contact with ${\rm X} = {\rm D}$ (where D is a diffusion conductor) are compared in Tab.~\ref{tab:noises_ratios}. The transparency distribution for a unit barrier has a maximum at $T \ll 1$. The ratios $\zeta_{\rs N} / |e|G_{\rs N} = 1$ in the NIN junction and $\zeta_{\rs S} / |e|G_{\rs S} = 2$ in the NIS junction characterize the carrier as an electron or a Cooper pair, correspondingly. In the presence of disorder, superconductivity in the NDS junction produces twice the noise determined by the normal NDN junction. However, in the case of a double barrier noise in the NINIS junction increases $3/2$ times compared with that in the NININ junction (see Tab.~\ref{tab:noises_ratios}). Therefore, we can conclude that noise doubling is not universal and can depend on the features of the transparency distribution function. This is explained by the fact that noise is caused by channels with mean transparencies $0 \lesssim T \lesssim 1$, while the current is related to open channels with $T \to 1$.

\begin{table}[htb]
	\caption{
The ratio $\zeta /|e|G$ of the spectral noise density to the conductance for NXN and NXS junctions. Results are valid~\cite{Khlus87,deJong97,Chen91,Beenakker92b} for a weak transparency $T \ll 1$ and many channels $N \to \infty$.
	}
	\begin{tabular}{|l|c|c|c|}
		\hline
		& Single barrier, \vspace{-0.1mm}
		& Double barrier, \vspace{-0.1mm}
		& Disorder, \vspace{-0.1mm} \\
		& ${\rm X} = {\rm I}$
		& ${\rm X} = {\rm INI}$
		& ${\rm X} = {\rm D}$ \\ \hline
		NXN, $\zeta_{\rs N}/|e|G_{\rs N}$	&	1	&	1/2 	&	1/3 \\ \hline
		NXS, $\zeta_{\rs S}/|e|G_{\rs S}$	&	2	&	3/4	&	2/3 \\ \hline
	\end{tabular}
	\label{tab:noises_ratios}
\end{table}

At finite voltages the differential noise is determined by Eq.~(\ref{eq:diffnoise}) with reflection amplitudes~(\ref{eq:rpm}) of the double INI barrier. This noise for a symmetric barrier $T_1 = T_2 = 0.05$ is shown in Fig.~\ref{fig:shot3}. For voltages of the order of the Andreev level $|eV| \sim v_{\rs F}/d$ the resonance structure is independent of the number of channels, while the resonance peaks are reflected in both the noise and conductance. At higher voltages $|eV| \gg v_{\rs F}/d$ resonances disappear due to dephasing of electrons and holes. In this case the difference between NIN and NIS junctions also disappears.


\section{Conclusion}

We have considered almost all basic aspects of the scattering matrix approach for the description of electron transport. However, since the review size is limited, topics such as multichannel cases, transparency statistics for dirty conductors, the integer quantum Hall effect, etc., have only barely been discussed or just mentioned, as, for example, problems with time-dependent fields. Although the fundamentals of the method described here are presented in books (see, e.g., Refs.~\cite{Datta95,Dem00} and handbook~\cite{Chtchelkatchev11Book}), a number of issues have been mentioned only in articles, while other issues have not been considered at all. We hope that this review compensates at least partially for this deficiency, especially as regards the Russian literature. We note in conclusion that the possibilities inherent in the approach initiated by Landauer in 1957~\cite{Landauer57} are far from being exhausted, especially concerning the description of systems with interacting particles, while at the same time this approach has already become a convenient tool in solving electron transport problems in noninteracting case.

This work was supported by the RFBR grant No. 14-02-01287 (G.B.L.) and grants NSF ECS-0608842, ARO W911NF-09-1-0395, and DARPA HR0011-09-1-0009 (I.A.S.).

We thank M.V.~Suslov, I.S.~Burmistrov, L.I.~Glazman, V.I.~Fal'ko, L.E.~Fedichkin, V.~Bouchiat, T.~Martin, D.~Ivanov, A.~Akhmerov, and especially A.V.~Ilyin for reading the manuscript and for their useful remarks.

We also thank our co-authors D.E.~Khmel'nitskii, L.B.~Ioffe, L.S.~Levitov, V.I.~Fal'ko, C.~Presilla, T.~Martin, G.~Blatter, F.~Hassler, M.V.~Suslov, G.M.~Graf, N.M.~Chtchelkatchev, and A.V.~Lebedev and colleagues A.I.~Larkin, B.L.~Al'tshuller, D.A.~Ivanov, K.A.~Matveev, R.~Landauer, V.V.~Ryazanov, C.~Glattli, M.~Sanquer, V.~Bouchiat, and M.~Reznikov, who helped us in discussions to understand many aspects of quantum electron transport. The review is partially based on the lecture course read by G.B.L. at the Eidgen\"ossische Technische Hochschule Z\"urich in 2008, and the lecture notes prepared by I.A.S. and F.~Hassler. The plan of lectures proposed by G.~Blatter also affected the structure of this review.


\appendix

\section{Properties of scattering matrices}
\label{sec:scattmatr}

\subsection{Properties of scattering states}
\label{sec:scattstates}

The basic properties of scattering states both for purely one-dimensional and for multichannel and multilead cases are obtained by the same methods as in the three-dimensional case, which is usually considered in textbooks. However, the orthonormalization and completeness of a set of scattering states can also be found from the following (not very rigorous) considerations: we create scattering states from plane waves, e.g., at an instant $t = t_{\rm in}$ [the set of wave functions $\psi_k(x,t_{\rm in}) = \exp{(ikx)}$ at that instant] by adiabatically increasing the scattering potential (localized in some region) up to a specified value. Then both the orthonormalization and the completeness of the set $\{\psi_k(x,t)\}$ in the subsequent instants automatically follow from the unitarity of the evolution of the wave packets because the initial set of plane waves had these properties (for plane waves, this can be proved by direct explicit calculations). Because the states $\psi_k(x,t)$ transform into Lippmann-Schwinger scattering states as $t \to \infty$ (it is this statement that should be proved more rigorously), we have obtained the orthonormalization and completeness for them.

We note that these properties can be used only in the region of wave packets that have already scattered from a potential with the required accuracy close to the specified potential, i.e., $|x| \ll v_k \tau$, where $\tau$ is the time elapsed from the instant at which the potential was close to the specified potential, with the required accuracy.\footnote{The possible appearance of coupled states requires some modification of our arguments; however, we do not consider this question here because such states rarely contribute to the transport phenomena under study.} It also follows from these considerations that the type of normalization (for example, to the energy delta function) of scattering states can be determined from their asymptotic forms: the density flow in the incident wave can be related to a plane wave, for which the normalization can be done easily.

\subsection{Unitarity of the scattering matrix}
\label{sec:unitarity}

The scattering matrix $\mathbb{S}$ is parameterized as
\begin{equation}
	\mathbb{S}=
	\left[ \! \begin{array}{cc}
		r & t' \\
		t & r' \\
	\end{array} \! \right].
	\label{eq:Sdef}
\end{equation}
The scattering matrix is unitary,
\begin{equation}
	\mathbb{S} \, \mathbb{S}^\dag = 1,
	\label{eq:S_unitary}
\end{equation}
which means that the amplitudes $t$, $t'$, $r$, and $r'$ are not independent quantities:
\begin{align}
	& r^\dag r+t'^\dag t' = t^\dag t+r'^\dag r' = 1,
	\label{eq:LT} \\
	& tr^\dag + r't'^\dag = 0.
	\label{eq:rrp}
\end{align}
For example, if $t$, $t'$, and $r$ are known, we can find from Eq.~(\ref{eq:rrp}) that
\begin{equation}
	r' = -tr^\dag[t'^\dag]^{-1}.
	\label{eq:rp}
\end{equation}
The Hermitian conjugation symbol ``${\dag}$'' is used here instead of the complex conjugation symbol ``$*$'' to emphasize that scattering amplitudes can be matrices in the multichannel case.\footnote{In addition, scattering amplitudes are not always described by scattering matrices. For example, the numbers of channels on the left and right sides of the barriers can be different. The scattering amplitudes are also matrices in the spin space in general. These matrices become nondiagonal in the case of spin-flip scattering, which can be caused by the spin-orbital interaction in the barrier or the action of an inhomogeneous exchange field (caused by ferromagnetic barriers) on the electron spin.} In the one-dimensional spatially symmetric case (with the time reversal symmetry assumed), it follows from Eq.~(\ref{eq:rrp}) that
\begin{equation}
	tr^*= \pm i\sqrt{TR}.
	\label{eq:tr}
\end{equation}
In the case of scattering from the delta function, the plus sign in Eq.~(\ref{eq:tr}) corresponds to the attractive potential in which a bound state exists, while the minus sign corresponds to the repulsive potential I in which only a continuous spectrum exists [see~(\ref{eq:tDelta})]. In the case of scattering from a double barrier, the sign changes after passing through the (ideal) resonance. Such a behavior of the phase affects the general interference pattern in the INIS structure (see Sec.~\ref{sec:condNINIS}).

\subsection{Symmetry of the Hamiltonian under time reversal}
\label{sec:timerev}

When the system Hamiltonian is invariant under a symmetry transformation, this invariance is extended in a certain way to the scattering matrix. If the Hamiltonian in invariant under time reversal, then the scattering matrix satisfies the relation
\begin{equation}
	\mathbb{S} = \mathbb{S}^\top.
	\label{eq:S_sym}
\end{equation}
It follows from Eq.~(\ref{eq:S_sym}) that $t = t'$. Using relations~(\ref{eq:LT}) and (\ref{eq:rrp}), we can also find that $|r|=|r'|$ and that the usual relation exists between reflection $(R = |r|^2 = |r'|^2)$ and transmission $(T = |t|^2 = |t'|^2)$ probabilities:
\begin{equation}
	R+T=1.
\end{equation}

If we take a magnetic field into account, then, because time reversal changes the direction of the magnetic field $\bf B$ to the opposite~\cite{Landau06BookV2,Datta95}, we have
\begin{equation}
	\mathbb{S}_{\s -{\bf B}}^{\phantom\top} =
	\mathbb{S}_{\s{\bf B}}^\top.
	\label{eq:SB}
\end{equation}
For the transmission amplitude for a scatterer inside which the magnetic field acts nontrivially on the orbital (one-dimensional) motion of particles, we have $t^{\phantom '}_{\s{\bf B}}=t'_{\s -{\bf B}}$.

\newpage

\raggedright

\end{document}